\documentstyle[epsfig, subfigure, lscape, onecolumn]{mn}

\def\ltapprox{\raise 2pt \hbox {$<$} \kern-1.1em \lower 5pt \hbox {$\approx$}}
\def\ltsim{\raise 2pt \hbox {$<$} \kern-1.1em \lower 4pt \hbox {$\sim$}}
\def\gtsim{\raise 2pt \hbox {$>$} \kern-1.1em \lower 4pt \hbox {$\sim$}}

\title{Compressible Turbulence in Galaxy Clusters: Physics and 
Stochastic Particle Re-acceleration}
\author[G. Brunetti, A. Lazarian]
      {G. Brunetti,$^1$ A. Lazarian$^2$\\ 
       $^1$ INAF - Istituto di Radioastronomia, via Gobetti 101,
       I--40129 Bologna, Italy \\
       $^2$ Department of Astronomy, University of Wisconsin 
       at Madison, 5534 Sterling Hall, 475 North Charter Street, 
       Madison, WI 53706, USA \\
} 

\begin{document}
\maketitle



\begin{abstract}
We attempt to explain  
the non-thermal emission arising from galaxy clusters
as a result of the re--acceleration of electrons by compressible 
turbulence induced by cluster mergers.
On the basis of the available observational facts we put forward a 
simplified model of
turbulence in clusters of galaxies focusing our attention on the 
compressible motions.
In our model intracluster medium (ICM) is represented by 
a high beta plasma in 
which turbulent motions are driven at large scales. The corresponding injection
velocities 
are higher than the Alfv\'en velocity. As a result, the turbulence is 
approximately isotropic up to the scale at which the turbulent velocity gets 
comparable with the Alfv\'en velocity. 
These motions are most important for the energetic particle acceleration, 
but at the same time they are subjected to most of the plasma damping.
Under the hypothesis that turbulence in the ICM is highly super-- Alfv\'enic 
the magnetic field is passively advected and the field lines are bended
on scales smaller than that of the classical, unmagnetized, 
ion--ion mean free path. This affects ion diffusion and 
the strength of the effective viscosity.
Under these conditions the bulk of turbulence in hot (5--10 keV temperature)
galaxy clusters is likely to be dissipated 
at collisionless scales via resonant coupling with thermal and 
fast particles.
We use collisionless physics to derive the amplitude 
of the different components of 
the energy of the compressible modes,
and review and extend the treatment
of plasma damping in the ICM.
We calculate the acceleration of both protons and electrons taking 
into account both Transit Time Damping acceleration and non-resonant 
acceleration by large scale compressions.
We find that relativistic electrons can be re--accelerated 
in the ICM up to energies of several GeV provided that the rms velocity of
the compressible turbulent--eddies is $(V_L/c_s)^2 \approx 0.15-0.3$; 
$c_s$ is the sound speed in the ICM.
We find that under typical conditions $\approx$ 2--5 \% of the energy flux 
of the cascading of compressible motions injected at large scales goes 
into the acceleration of fast particles and that this may explain 
the observed non--thermal emission from merging galaxy clusters.
\end{abstract}


\maketitle

\section{Introduction}

In the last years observations of galaxy clusters 
have shown that non--thermal components are mixed together with
the thermal component of the intracluster medium (ICM).
A fraction of massive galaxy clusters
hosts diffuse radio emission in the form of radio
halos, Mpc--sized diffuse synchrotron radio sources at the cluster
center, and radio relics, elongated diffuse synchrotron radio sources
at the cluster periphery. This directly proves the presence of GeV 
relativistic electrons
(and/or positrons) and of $\mu$G magnetic fields diffused 
on Mpc scales through the cluster volume (e.g., Feretti 2005,
for a review). A related issue is the discovery of non--thermal
emission in the hard X--ray band detected in a few galaxy
clusters (e.g., Fusco--Femiano et al.~2004; Rephaeli, Gruber,
Arieli ~2006).

The most spectacular example of non--thermal emission from galaxy clusters
is that of giant radio halos. 
These are very extended (Mpc) synchrotron radio emissions,  
not connected with cluster radio sources, at the 
center of clusters, with a steep spectrum and
a typical synchrotron luminosity in the
range $\approx 10^{40}-10^{42}$erg s$^{-1}$.
A remarkable point is that the emitting 
particles have a life--time considerably shorther than that necessary 
to diffuse over the scales of these radio halos, and this poses a theoretical
problem on their origin (e.g., Jaffe 1977).
In principle if the content of cosmic ray hadrons in the ICM is sufficiently
large, fast electrons and positrons may be continuously injected
in the ICM via hadronic collisions between cosmic--rays 
and thermal protons (Dennison 1980; 
Blasi \& Colafrancesco 1999), alternatively different forms of
{\it in situ}--stochastic acceleration and re--acceleration 
operating for a small fraction of the cluster life  
may provide a viable source for high energy emitting particles
(Schlickeiser et al.~1987; Brunetti et al.~2001; Petrosian 2001).
Future gamma ray observations (GLAST, {\it Cerenkov} telescopes) 
will constrain the content of cosmic--ray hadrons in galaxy clusters
and provide an important tool to better understand the origin of the
relativistic particles in the ICM (e.g. Reimer 2004 \& 
Blasi et al.~2007 for recent reviews).

\noindent
It is believed by several authors 
that the re--acceleration scenario may provide 
a promising picture to explain the bulk of present--day radio
data (e.g., reviews by: Brunetti 2003,04; Petrosian 2003;
Blasi 2004; Hwang 2004; Feretti 2005).
This model essentially relies
on the hypothesis that a fraction of the kinetic energy associated
with cluster--cluster mergers is channelled into turbulence and
re--acceleration of relativistic particles in the ICM.

\noindent
MHD turbulence is known to be an important agent for particle acceleration
since Fermi (1949) first pointed this out.
Second order Fermi acceleration by MHD turbulence
was appealed for  acceleration of particles in many astrophysical
environments, e.g. Solar wind,  Solar 
flares, ICM, gamma-ray bursts (see Schlickeiser
\& Miller 1998; Chandran 2003; Brunetti et al. 2004; 
Petrosian \& Liu 2004; Cho \& Lazarian 2006; 
Petrosian et al. 2006; Becker, Le \& Dermer 2006; Dogiel et
al.~2007).
Naturally, properties of compressible MHD turbulence (see 
Shebalin, Matthaeus, \& Montgomery 1983; 
Higdon 1984; Montgomery, Brown, \& Matthaeus 1987;
Shebalin \& Montgomery 1988, Zank \& Matthaeus 1992; 
Cho \& Lazarian 2003 and references
therein) are essential for 
understanding the acceleration mechanisms.

Among the advances in understanding MHD turbulence, we would like to mention
the Goldreich \& Sridhar (1995; henceforth GS95) model of turbulence. 
GS95 dealt with
incompressible MHD turbulence and showed
that Alfv\'en and pseudo-Alfv\'en modes
follow the scale-dependent anisotropy of $l_{\|} \sim l_{\perp}^{2/3}$,
where $l_{\|}$ is the size of the eddy along the local mean magnetic
field and $l_{\perp}$ that of the eddy perpendicular to it.
Lithwick \& Goldreich (2001) conjectured that this scaling of incompressible
modes is also true for Alfv\'en modes and slow modes in the
presence of compressibility. 
In Cho \& Lazarian (2002; 2003, henceforth CL03) the rational
for considering separately the evolution of
slow, fast and Alfv\'en mode cascades was justified. 
The numerical simulations in CL03 and Kowal \& Lazarian (2006) verified
that Alfv\'en and slow mode velocity fluctuations  are indeed consistent 
with the GS95 scaling, while fast modes exhibit isotropy
in both gas-pressure (high $\beta_{pl}$) and 
magnetic-pressure (low $\beta_{pl}$) dominated plasmas. The former case is
the most appropriate for clusters of galaxies that we deal in this paper.

A single most important change in the paradigm that has become obvious
recently is that if the turbulent--energy injection happens at large scales, 
the cascading Alfv\'enic mode is presented at sufficiently
small scales by very elongated eddies. Thus the interactions 
of these mode with cosmic rays 
differs considerably from that of the isotropic eddies that 
earlier researchers dealt with. 
Under these conditions nearly isotropic fast modes were identified as 
the dominant agent for scattering and resonance 
acceleration (Yan \& Lazarian 2002).
As a consequence of this, our
understanding of energetic particle-turbulence interactions via
gyroresonance and the Transit-Time Damping (TTD) (Chandran 2000;
Yan \& Lazarian 2002, 2004; Farmer \& Goldreich 2004; Cassano \&
Brunetti 2005) as well as the non-resonance acceleration of cosmic
rays by 
large scale compressible motions (see Ptuskin 1988, Chandran 2003,
Cho \& Lazarian 2006) has been altered.
This calls for the corresponding advances in the treatment of cosmic--ray 
acceleration in the environment of
clusters of galaxies (see e.g. Brunetti 2006 \& Lazarian 2006a).

\section{Outline of the paper}

In this paper we proceed in three main steps :

I) As a first point we discuss the problem of turbulence in the
ICM and work up a simplified but physical picture of the 
properties and relevant scales of turbulence in galaxy clusters.
As we discuss below (see \S 3.1) the plasma in clusters of galaxies is
expected to be both magnetized and turbulent. Its Reynolds numbers are
expected to vary as the magnetic field grow (\S 3.2), but they are expected
to be high enough to allow turbulence to be excited.
Finally turbulence in hot galaxy clusters is expected to be dissipated
via collisionless dampings and this makes the particle acceleration
process a natural consequence (\S 3.4).
This part of the paper is mainly designed to provide a reference
picture for observers and a viable astrophysical 
starting point for theoretical developments.

II) As a second point we discuss the physics of compressible motions
in the collisionless regime.
This part of the paper is a necessary extension of previous
seminal studies of collisionless turbulence and is 
aimed at the presentation of necessary general 
equations to use in the paper.
In particular to characterize the plasma-cosmic rays 
interactions we characterize
the compressible motions using dielectric tensor (\S 4.1), 
give the expression for the energy spectrum of the 
fast modes in \S 4.2 and describe the TTD damping in
intracluster plasma in \S 4.3; complex expressions and calculations
are reported in Appendices A--C .

III) Finally we discuss the issue of stochastic particle acceleration
in galaxy clusters by compressible turbulence.
The resonant TTD acceleration is discussed in \S 5.1,
while the effect of non-resonant acceleration is discussed in \S 5.2.
In \S 6 we discuss the results in the framework of the particle
re--acceleration model in galaxy clusters: 
in \S 6.1 we briefly review the basic physics of cosmic
rays in galaxy clusters,
and in \S 6.2 we present detailed calculations on particle re--acceleration
in the ICM.
Here we claim that compressible
turbulence may drive efficient particle acceleration in the ICM.
This is an extension of recent studies on the argument and provides 
a view of the process of particle re--acceleration in the ICM which 
is additional (or alternative) to that of Alfv\'enic acceleration.

In \S 7 we discuss the most relevant findings and simplifications, 
and in \S 8 we provide a short summary.

\section{Turbulence in the ICM}

\subsection{Injection of turbulence in the ICM}

Cluster mergers and accretion of matter at the virial radius may
induce large--scale motions with 
$V_L \sim 1000$ km s$^{-1}$ in massive clusters.
Numerical simulations suggest that turbulent motions 
may store an appreciable
fraction, 5--30\%, of the thermal energy of the ICM
(e.g., Sunyaev, Bryan \& Norman 2003; 
Dolag et al.~2005; Vazza et al.~2006).
Simulations of merging clusters 
provide an insight into the gas dynamics during
a merger event (e.g., Roettiger, Burns, Loken 1996;
Roettiger, Loken, Burns 1997; Ricker \& Sarazin 2001): 
sub--clusters generate laminar bulk
flows through the sweeped volume of the main clusters which inject
turbulence via e.g. Kelvin--Helmholtz instabilities at the interface of the
bulk flows and the primary cluster gas.
The largest turbulent eddies 
decay into smaller and 
turbulent velocity fields developing a turbulent cascade.

A simple, but well motivated by physical arguments, 
semi--analytical approach allows to follow 
the cosmological injection of merger--turbulence.
Calculations from Cassano \& Brunetti (2005) 
suggest that turbulence in the ICM is transient
being mostly injected during the most massive mergers.
However, since more frequent minor mergers may also contribute to
the injection of such turbulence, some minimum 
level of turbulence should be rather ubiquitous in the ICM.
In these calculations turbulence is assumed to be
injected in the cluster volume swept by the sub-clusters,
which is bound by the effect of the ram pressure stripping,
and the turbulent energy is calculated as a fraction of the 
${\rm PdV}$ work done by the sub-clusters infalling onto the main cluster.
Essentially merger--driven 
turbulence is powered by the gravitational potential well
and thus the energy of this turbulence should
approximatively scale 
with the cluster thermal energy (Cassano \& Brunetti 2005).
Support to this scaling 
comes from a recent analysis of a sample of galaxy clusters from
cosmological numerical simulations (Vazza et al.~2006). 

Turbulence is an important ingredient in the physics of the
ICM as it is necessary to understand the amplification of 
magnetic fields in clusters (Dolag et al.~2002;
Schekochihin et al.~2005; Subramanian et al.~2006), an issue
closely related to the non--thermal emission from clusters but
that we will not address in this paper.
Turbulence might provide a source of
heating to balance the cooling of cluster cores
(Fujita, Matsumoto \& Wada 2004), and
the knowledge of the basic aspects of turbulence in galaxy
clusters is also crucial to model the transport of heat and
metals in the ICM (Narayan \& Medvedev 2001;
Cho et al., 2003; Voigt \& Fabian 2004; Lazarian 2006b).

In spite of obvious observational challenges, 
indications of some level (at least 10--20\%
of the thermal energy) of turbulence in the ICM
comes from gas--pressure maps in the X--rays  
(Schuecker et al. 2004), 
and also from the lack of resonant scattering from
X--ray spectra 
(Churazov et al. 2004; Gastaldello \& Molendi 2004).

\noindent
Interestingly enough, also 
upper limits to the turbulent--energy content in the
ICM were 
obtained in a few nearby galaxy clusters from kinematical arguments 
related to the properties of H$\alpha$ and X--ray filaments 
(e.g. Fabian et al.~2003; Crawford et al.~2005; Sun et al.~2006).
Assuming that 
turbulence is driven at hundred--kpc scales 
the above upper limits actually can be used to
place upper limits on the intensity of strong turbulence in the 
ICM (supersonic or trans--sonic turbulence).

\subsection{Reynolds Number and
developing of turbulence in the ICM}

In this Section we discuss the important issue of the 
Reynolds number of the fluid in the ICM, and derive 
its value by assuming a simple, but physically motivated,
scenario.

A fluid becomes turbulent when the rate of
viscous dissipation at the injection scale, $L_o$, is much
smaller than the energy transfer rate, i.e. when
the Reynolds number is ${\it Re} = V_L L_o /\nu_K \gg 1$,
where $V_L$ is the injection velocity and
$\nu_K$ is the kinetic fluid viscosity.
The main source of uncertainty here comes from our ignorance 
of $\nu_K$ in the ICM.

\noindent
If the ICM were not magnetized 
$\nu_K \sim l_{mfp} v_i/3$, were $v_i$ is the velocity of
thermal ions,
and $l_{mfp}$ is the ion--ion mean free path {\it
in case of pure Coulomb interactions} (e.g., Braginskii 1965):

\begin{equation}
l_{mfp} \sim 15000 \, \left( {{n_{e}}\over{10^{-3}{\rm
cm}^{-3}}}
\right)^{-1}
\left( {{T}\over{\rm 8\,  keV}} \right)^2
\left( {{40}\over{\ln \Lambda}} \right)
\,\,\,\,\,\, (\rm pc)
\label{lmfp}
\end{equation}

\noindent
where $\ln \Lambda$ is the Coulomb logarithm.

\noindent
Thus the corresponding Reynolds number would be:

\begin{equation}
{\it R}e \sim
52 \, \Big( {{V_L}\over{1000 \, {\rm km/s}}} \Big)
\Big( {{L_o}\over{300 \, {\rm kpc}}} \Big)
\Big( {{n_{\rm th}}\over{10^{-3}{\rm cm}^{-3}}} \Big)
\Big( {{T}\over{\rm 8 \, keV}} \Big)^{-5/2}
\left( {{\ln \Lambda}\over{40}} \right)
\label{reynolds_standard}
\end{equation}

\noindent
which is formally just sufficient for initiating the 
developing of turbulence.

However,
in the presence of (even a small) {\it stationary}
magnetic field the Reynolds
number for motions in the direction perpendicular to the magnetic field
gets extremely high essentially because the perpendicular
mean free path of particles is limited to the Larmor gyroradius--scale
(e.g., Braginskii 1965).
Potentially, diffusion along the wandering turbulent--magnetic 
field lines
could significantly increase the particle diffusivity
and the plasma viscosity. For instance, estimates in 
Narayan \& Medvedev (2001)
suggest that electron diffusivity 
in a turbulent medium can be of the order of $1/5$ of
the classical Spitzer value for unmagnetized medium, provided that the
injection velocity, $V_L$, is equal to the Alv\'en velocity.

\noindent
More general calculations (Lazarian 2006a, and ref. therein) 
show that things could be more complicated and that
the effective viscosity depends on
the super-- or sub-- Alfv\'enic nature of the
turbulence\footnote{The super- and sub-Alfv\'enic are determined in 
terms of the total magnetic field.}. 
Turbulence in the ICM is super-Alfv\'enic, i.e. 
turbulence with the injection velocity larger than the Alfv\'en one.
In this case the turbulent hydrodynamic motions 
can easily bend the magnetic field lines. The trajectory of the particle 
that follows such a field line gets diffusive even in the absence of 
collisions. The effective mean free path of a particle is determined by 
the scale at which magnetic tension can withstand the hydrodynamic forces, 
i.e. the scale at which the turbulent velocity, $V_{l_A}$,  
gets equal to the Alfv\'en one, $v_A =  B/\sqrt{4 \pi \rho}$, where $B\sim
(B_{o}^2+B_{rms}^2)^{1//2}$, (Lazarian 2006b). 
This scale, at which turbulence becomes MHD, is 
\footnote{In deriving Eq.(\ref{la})
we use the hydro-- scaling $V_l \propto l^{1/3}$}
$l_A \sim L_o \left( {{v_A}\over{V_L}} \right)^3$ and
assuming typical conditions in Mpc regions at the center
of massive (and hot) galaxy clusters where radio halos are found,
it is :

\begin{equation}
l_A 
\sim 100 \,  \left( {{B}\over{\mu G}} \right)^3
\left({{L_o}\over{300\, {\rm kpc}}}\right) 
\left({{V_L}\over{10^3{\rm km/s}}}\right)^{-3}
\left({{n_{th}}\over{10^{-3}{\rm cm}^{-3}}}\right)^{-{3\over 2}}
\,\,\,\,\,\, (\rm pc)
\label{la}
\end{equation}

\noindent
which is $< l_{mfp}$.
This implies the important point that, 
even for motions along the magnetic field, 
the Reynolds number in a turbulent ICM is 
larger than that estimated with the classical 
formula (Eq.\ref{reynolds_standard}).
Actually one finds 
${\it Re} >$ few times $10^3$ which ensures that the
ICM gets turbulent.

\noindent
The main uncertainty in the evaluation of the Reynolds number
comes from the value of the effective mean free path of
particles.
Eq.(\ref{la}) accounts for the effect of the turbulent magnetic
field, however additional mechanisms may affect the value
of the particle mean free path in the ICM, for example 
plasma instabilities.
Plasma instabilities could be at work in the ICM, 
e.g. turbulent compressions themselves
may drive instabilities.
These instabilities in the ICM may induce
scatterings of thermal ions {\it which reduce the effective
mean free path and further increase the value of the
Reynolds number} (e.g.,  Schekochihin \& Cowley 2006; 
Lazarian \& Beresnyak 2006).
In what follows to be conservative and with the aim to simplify the
overall picture, we disregard this effect, so that
our estimates of the Reynolds number in the ICM
would be considered as a lower limit.

\subsection{Turbulent Modes}

\subsubsection{Basic properties of the turbulent modes}

Turbulence discussed in the previous Sections is a complex mixture
of several turbulent modes.
The ICM is a compressible high--beta plasma.
At large scales, where magnetic fields are not dynamically important
($V_L >> v_A$), the turbulence is essentially 
in the hydro-- regime, and we shall assume that
turbulence in the ICM is done by solenoidal and compressible 
(essentially sound waves) motions.
At smaller scales, in the MHD regime, it is $V_l \leq v_A$ and 
three types of modes should exist in a compressible
magnetized plasma: Alfv\'en, slow and fast modes.
Slow and fast modes may be roughly thought as the MHD counterpart of
the compressible modes, while Alfv\'en modes may be thought
as the MHD counterpart of solenoidal Kolmogorov eddies
(a more extended discussion can be found in 
Cho, Lazarian \& Vishniac 2002, and ref. therein).
Sound modes at large scales have propagation
properties
similar to that of the fast modes in the MHD-- regime.
For this reason in this paper 
we shall use the properties of these
modes for describing compressible turbulence, 
i.e. hydro-- modes (sound waves) at large scales and 
fast modes themselves at small scales.
Fast modes are compressive waves which
propagate across or at an angle to the local magnetic
field.
The fast mode branch in a plasma extends from low frequencies
up to the electron cyclotron frequency. At frequencies below
the ion cyclotron frequency, and in the weak
damping limit, the dispersion relation
of these modes is given by $\omega = V_{\rm ph} k$, where
the phase velocity is given by
(e.g., Krall \& Trivelpiece 1973) :

\begin{equation}
V_{\rm ph}^2 =
{{c_s^2 + v_A^2}\over{2}}
\left\{
1 +
\sqrt{
1 - 4 \left( {{k_{\Vert}}\over{k}} \right)^2
{{c_s^2 v_A^2}\over{(c_s^2 + v_A^2)^2}}
}
\right\}
=
{{c_s^2}\over{\beta_{pl}}}
( \beta_{pl}/2 +1 ) 
\left\{
1 +
\sqrt{
1 - 4 \left( {{k_{\Vert}}\over{k}} \right)^2
{{ \beta_{pl}/2 }\over{ ( 1 + \beta_{pl}/2 )^2 }}
}
\right\}
\buildrel {\beta_{pl} >> 1} \over
\longrightarrow
c_s^2
\label{vph}
\end{equation}

\noindent
and where the parameter beta of the plasma is defined
by $\beta_{pl} = 2 c_s^2 / v_A^2$.

Alfv\'en modes propagate
along or at an angle to the local magnetic field.
The Alfv\'en branch extends from low frequencies up
to the ion cyclotron frequency. In this frequency range
the Alfv\'enic dispersion relation is given by
$\omega = v_A |k_{\Vert}|$, where
$v_A = B_o/\sqrt{4 \pi n_{\rm th} m_i}$ is the Alfv\'en
velocity.

Alfv\'en and fast modes differ also for the direction of the
displacement vectors: the displacement of
Alfv\'en modes is always perpendicular to $B_o$,
while that of fast modes makes an angle
to the local magnetic field and in the case $\beta_{pl} \rightarrow 0$
it is perpendicular to $B_o$, while for $\beta_{pl} \rightarrow \infty$
it becomes radial (along $k$); a detailed discussion
on the decomposition of MHD modes can be found in
Cho \& Lazarian (2002), (2003) and in Kowal \& Lazarian (2006).

Slow modes has ``-'' before the square root in Eq.(\ref{vph})
and for $\beta_{pl} >> 1$ they have the dispersion relation of
Alfv\'en modes. We will not include slow modes in our calculations
in the in the particle acceleration process by large scale
modes (Sect.~5) as they have a phase velocity in the ICM much smaller 
than that of the fast modes and thus are less important.

At MHD--scales Alfv\'en and slow modes might be of some relevance
in discussing the particle acceleration process 
either because they can accelerate particles, or
because in principle 
they may provide a particle pitch--angle scattering
process\footnote{The Alfv\'enic mode as well as the slow mode
gets anisotropic for scales less
than $l_A$, which makes the scattering inefficient, however (Chandran
2000, Yan \& Lazarian 2002).} which is required by acceleration
processes driven by other modes (\S~4,5).

\subsubsection{Coupling between turbulent modes
in the ICM}

Although the complex dynamics
of galaxy clusters and the relatively large
value of the Reynolds number of the ICM are likely to make the ICM itself
a turbulent medium, it is somewhat difficult to have
a clear idea of the relative importance of the different turbulent
modes in the ICM.
Indeed this depends on the nature of the turbulent forcing and on the
mode coupling between different modes in the ICM.

We shall assume that a sizeable part of turbulence
at large scales (namely at scales where the magnetic tension does not
affect the turbulent motions) is in the form of compressible motions.
This is reasonable 
as these modes are expected to be easily generated 
in a high beta medium even in the unfavourable case of solenoidal 
turbulent forcing.
This is proved by closure calculations carried out
in the case of $\beta_{pl} >> 1$.
Indeed when motions are hydro-- in nature the coupling between solenoidal and
compressible motions is efficient and 
the excitation of compressible modes by the solenoidal modes
is driven by the {\it incompressible} pressure arising from
non--linear interaction between solenoidal modes themselves
(Bertoglio et al. 2001).
These studies have shown that the fraction of energy in the form of
compressible modes is found to scale with 
$\propto {\it M}_s^2 \times {\it R}e$
for ${\it M}_s^2 \times {\it R}e\,<10$ (${\it M}_s<1$ is the
turbulent Mach number),
while for ${\it M}_s^2 \times {\it R}e \,\geq10$
the scaling is expected to flatten
(Bertoglio et al. 2001; Zank \& Matthaeus 1993).
Obviously a solenoidal turbulent forcing, which limits the energy
of compressible modes to be smaller than that of 
solenoidal modes (even in the super--Alfv\'enic case),
is probably not appropriate for galaxy clusters 
where turbulence is likely to be excited by compressible forcings, 
and this might result in a larger ratio between compressive
and solenoidal modes (at least for super--Alfv\'enic motions).

\noindent
Situation may be radically different at smaller scales
where the magnetic field tension affects turbulent motions, i.e.
in the MHD-- regime, $l \leq l_A$.
In this case, MHD numerical simulations have shown that
a solenoidal turbulent forcing gets
the ratio between the amplitude of Alfv\'en and 
fast modes in the form (Cho \& Lazarian 2003) :

\begin{equation}
{{(\delta V)_c^2}\over{(\delta V)_s^2}}
\sim
{{(\delta V)_s v_A }\over{ c_s^2 + v_A^2 }}
\label{transfercl03}
\end{equation}

\noindent
which essentially means that coupling between these two
modes may be important only at $l \approx l_A$ (in the 
MHD-- regime it should be $(\delta V)_s \leq v_A$)
since the drain of energy
from Alfv\'enic cascade is marginal when the
amplitudes of perturbations become weaker.
Most importantly in galaxy clusters it is $c_s^2 >> v_A^2$
and thus the ratio between the amplitude of 
Alfv\'en and fast modes at scales $l < l_A$ 
is expected to be small, 
$(\delta V)_c^2/(\delta V)_s^2 \leq 
(v_A / c_s)^2 \sim 10^{-2}$ (this for solenoidal forcing at $l \approx l_A$).

\subsection{Dissipation of turbulence in the ICM}

\subsubsection{Collisional regime \& viscous dissipation}

Viscosity is important in the dissipation of
turbulent eddies in the collisional regime.
In this regime the cascade of hydro-- motions 
is maintained down to a scale
$l_{diss} \sim L_o ( {\it Re} )^{-{3\over 4}}$ at which
the viscous dissipation rate equals the wave energy
transfer rate.
The damping rate of hydro-- motions at scale $l$ due to the 
viscosity is :

\begin{equation}
\Gamma_k^{\nu} \sim {{\nu_K }\over{l^2}}
\label{dampvis}
\end{equation}

\noindent
here $\nu_K$ is a reference value of the
kinetic viscosity which gives the main uncertainty in the calculations.

As a simplified and conservative approach we can assume that 
${\bf B}_o$ is initially ordered and that the first super--Alfv\'enic
turbulent eddies, injected at large $L_o >> l_{mfp}$ scales, 
initiate a cascading and thus
that the bending of the field lines follows this cascading.
Turbulent motions along ${\bf B}_o$ experience the strongest viscous
dissipation 
which can be grossly estimated by using the classical formulation
of (unmagnetised) viscosity.
By taking physical conditions appropriate for
the central Mpc of hot galaxy clusters, the
dissipation scale of these parallel motions 
reads :

\begin{equation}
l_{diss} \simeq l_{mfp} \left( {{v_i}\over{3 V_L}} \right)^{3 \over 4}
\left( {{L_o}\over{l_{mfp}}} \right)^{1\over 4}
\approx  l_{mfp}
\left(  {{V_L }\over{10^3{\rm km/s}}} \right)^{-{3 \over 4}}
\left( ({{L_o}\over{300 \,{\rm kpc}}}) ( {{n_{th}}\over
{10^{-3}}}) ( {{8 \, {\rm keV} }\over{T}})^{1\over 2} 
( {{\ln \Lambda}\over{40}} ) \right)^{1 \over 4}
\label{ldiss_standard}
\end{equation}

\noindent
while turbulent motions transverse to ${\bf B}_o$ experience a much 
smaller viscosity and shall cascade at scales $<< l_{mfp}$.
The cascading of these transverse (quasi--perpendicular) motions
at a given scale takes a time of the order of the the bending time
scale of the magnetic field on the same scale and these motions become
the responsible for the bending of the field lines on scales
$<< l_{mfp}$, potentially down to scales $\approx l_A$.
We note that indeed 
recent {\it Bayesian} analysis of RM show that magnetic fields in 
galaxy clusters could be tangled at least on scales $\approx$ kpc
(Vogt \& Ensslin 2005), smaller scales being inaccessible to observations,
thus suggesting that the bending of the field lines happens on
scales $<< l_{mfp}$.

\noindent
As discussed in \S~3.2 the bending of the field
lines on scales $< l_{mfp}$ reduces the effective
particle mean free path yielding a reduction of the viscosity.
Viscosity indeed depends on the flux of the momentum which
is transported by particles and this is 
determined by the diffusion of the particles that
carry this momentum from the layers moving with different velocities. 
By limiting this diffusion the turbulent--bending of the field lines  
decreases the viscosity and thus the dissipation of turbulence
itself.

\noindent
Thus the turbulent eddies which cascade afterwards evolve 
in a very tangled magnetic field and experience an effective viscosity
which we shall adopt in the form $\nu_K \approx 1/3 v_i l_A$,
and the {\it effective} dissipation scale, $l^b_{diss}$, we would
encounter in the case of super--Alfv\'enic turbulent ICM becomes :

\begin{equation}
l^b_{diss} \approx l_{diss} \left({{l_A}\over{l_{mfp}}} \right)^{3 \over 4}
\approx
{1\over {45}} \, l_{mfp}
\left({{V_L}\over{10^3{\rm km/s}}}\right)^{-3}
({{L_o}\over{300 \,{\rm kpc}}}) 
\left( {{B}\over{\mu G}} \right)^{{9 \over 4}}
\left({{n_{th}}\over{10^{-3}{\rm cm}^{-3}}}\right)^{-{1\over 8}}
\left( {{T}\over{\rm 8 \, keV}} \right)^{-{{13} \over 8}}
( {{\ln \Lambda}\over{40}} )
\label{vcutmag}
\end{equation}

Also in this case the effect driven by plasma instabilities
in the ICM may affect our estimates.
In particular, the scattering of thermal ions
induced by these instabilities may 
additionally decrease the effective viscosity in the ICM, and 
this might reinforce our conclusions
that, even assuming collisional physics, 
the bulk of compressible turbulent motions 
in the ICM is expected to be dissipated 
only at small scales, $\leq$ kpc.

\subsubsection{Collisionless regime}

The viscous damping is not important in the collisionless
regime, i.e. when the scales of interest 
are smaller than the particle's mean free path or when 
the time--scales of interest are shorter than the 
particle's collision time.
When the diffusive--trajectory of particles is not
driven only by collisions (as indeed in the super--Alfv\'enic
turbulent--magnetized case, 
\S~3.2, 3.4.1) the most appropriate way to define the collisionless
regime is in terms of collision frequency, and we shall
use collisionless physics for the turbulent modes
when the frequency of these modes is larger than the 
ion--ion collision frequency $\nu_{ii}$ (e.g., Eilek 1979) : 

\begin{equation}
\nu_{ii}
\simeq
{4 \over 3} \sqrt{\pi}
{{e^4 n_{th} \ln(\Lambda) }\over{
m_p^{1/2} ( k_B T )^{3/2} }}
\label{nuii}
\end{equation}

\noindent
Magnetosonic modes dissipate energy in the collisionless
regime in accelerating
charged particles especially
via Transit--Time--Damping (e.g., Schlickeiser
\& Miller 1998) which is particularly severe
in high beta--plasma conditions like those in the ICM.
In terms of scales, from Eq.(\ref{nuii}) and $\omega = V_{ph} k$,
the collisionless regime
for magnetosonic waves in the ICM starts approximatively
at the scale of the ion mean free path $l_C \sim l_{mfp}$ (Eq.~\ref{lmfp}).
Thus from a general point of view, 
in order to understand the way compressible modes dissipate 
in the ICM it is necessary to compare the viscous dissipation
scale, $l^b_{diss}$, with the collisionless scale $l_C$: if 
$l^b_{diss} < l_C$ the cascading process of these turbulent modes
would reach collisionless scales before
being significantly affected by viscosity and energy will be
dissipated via collisionless dampings, while in the opposite case
turbulence will be dissipated by viscosity.

\noindent
From Eqs.(\ref{ldiss_standard}--\ref{vcutmag}) 
we immediately have that the bulk of compressive turbulence 
in the hot ICM is likely to be dissipated via collisionless dampings.
Indeed in hot (and massive) galaxy clusters 
it is found that viscosity is not efficient enough 
to dissipate the turbulent motions,
unless the large--scale
velocity of these motions is relatively small, $V_L < 300$ km/s,
namely in case of very low turbulence.
At the same time, however, an efficient dissipation of 
turbulent motions may happen 
in strongly magnetized ($B \geq 5 \mu$G), lower
temperature and high density regions which are 
conditions appropriate at the center of clusters with 
cooling flows ({\it cool cores}).
Here viscosity may potentially become an important source of dissipation
of the turbulent eddies.

\noindent
It should be mentioned that plasma instabilities might complicate the
picture.
On one hand, their straightforward 
effect is to decrease the effective viscosity in the
ICM, however, on the other hand they introduce a new relevant scattering
frequency of ions which could be larger than the ion--ion scattering
frequency (Eq.~\ref{nuii}) and 
the net result might be that the collisionless regime gets into play at
smaller scales.
As in the previous Sections we discard this possible
effect which would deserve detailed investigation.

The nature (collisional or collisionless) of the 
turbulent dissipation in astrophysical plasma is a crucial
point.
In Tab.~1 we report the case for several astrophysical situations
undertaking different physical conditions, processes and 
scales of interest.
A collisionless dissipation of compressive turbulence is believed to
be eventually operating in a few other astrophysical regions such as
in solar flare plasma and in the Galactic Halo.
It is important to note here that 
stochastic particle acceleration is indeed suggested to 
power the hard X--ray flares observed in the Sun
(e.g., Miller, La Rosa \& Moore 1996; Petrosian, Yan \& Lazarian 2006).
We note that the beta of plasma in these filaments is extremely small
and thus even in the case of strong turbulence the
collisionless dissipation of compressive
modes should happen at scales, $l \leq l_A$, at which turbulent
motions are MHD in nature.
On the other hand turbulence in the ICM is super--Alfv\'enic 
(essentially due to the high beta of plasma)
and the collisionless regime in the hot ICM starts at scale 
$l > l_A$ were compressive motions are still hydro-- in nature.

\begin{table*}
\begin{tabular}{|c|c|c|c|c|c|c|}
\hline 
 & GC & CC & galactic halo & HIM & WIM & Sun
\tabularnewline
\hline
T(K)& $10^8$ & $3 \times 10^7$ & $2\times 10^6$& $ 10^{6}$& $8 \times 10^3$ & $10^7$
\tabularnewline
\hline
$c_s$(km/s)& 1650 & 900 & 130 & 90 & 8 & 360
\tabularnewline
\hline
n$_{th}$(cm$^{-3}$)&
$10^{-3}$ & $5 \times 10^{-2}$ & $10^{-3}$ &
$4\times10^{-3}$& 0.1& $10^{10}$
\tabularnewline
\hline
{ $l_{mfp}${\it (cm)}}&
$5\times 10^{22}$ & $10^{20}$ & $4\times 10^{19}$& $2\times10^{18}$&
$6\times10^{12}$& $10^8$
\tabularnewline
\hline
L$_o$(pc)&
$1-5\times 10^5$ & $1-5\times 10^5$ & 100 & 100 & 50 & $3\times 10^{-10}$
\tabularnewline
\hline
B($\mu$G)&
1 & 10 & 5& 2& 5& $10^8$
\tabularnewline
\hline
$c_s^2/v_A^2$&
500 & 100 & 0.3 & 3.5& 0.1& 0.03
\tabularnewline
\hline
damping&
{\it collisionless}$^*$ & {\it collisionless} ? 
& {\it collisionless} & collisional &
collisional & {\it collisionless}$^{**}$
\tabularnewline
\hline
\end{tabular}
\caption{The reference parameters of astrophysical plasma 
and relevant damping. The dominant damping mechanism for turbulence is
given in the last line. GC= hot galaxy clusters
($^*$ $V_L > 300$ km/s), CC= cool--cluster cores,
HIM= hot ionized ISM, WIM= warm ionized ISM, SUN= Solar flare plasma
($^{**}$ Alfv\'enic turbulent--Mach number $M_A \geq 0.3$ is assumed).}
\end{table*}

\subsection{Conclusion I: Turbulent Scenario in the ICM}

Given the above discussions it is possible to 
set up a simplified and operative scenario of turbulence in the ICM
to adopt in this paper.

\noindent
Within a simplified picture of turbulence that we consider
here, super--Alfv\'enic turbulence is made by a mix of magnetosonic modes
(essentially similar to sound modes)
and incompressible--Kolmogorov turbulent eddies (which roughly
correspond to the Alfv\'en modes in the MHD regime).

\noindent
We shall assume that 
turbulence is injected at large scales $L_o \sim 300-500$ kpc
most likely by a complex mixture of compressive
and solenoidal forcing.
The typical velocity of the turbulent eddies at the injection scale
is expected to be
around $V_L \sim 500-1000$ km/s which makes turbulence sub--sonic,
with $M_s = V_L/c_s \approx 0.3-0.8$, but
strongly super--Alfv\'enic, with $M_A=V_L/v_A \geq 10$.
Turbulent motions at large scales are thus essentially hydrodynamics and
the cascading of compressive (magnetosonic) modes may couple with
that of solenoidal motions (Kolmogorov eddies).

\noindent
Assuming typical conditions in hot (and massive) galaxy clusters
we find that even in the unmagnetized case viscosity would still
allow hydro-- motions to cascade down to scales of the order of
$\approx l_{mfp}$.
In the magnetized case viscosity is believed to be partially suppressed. 
In addition when turbulence is super-Alfv\'enic 
hydro-- motions can easily bend the
magnetic field lines affecting the 
{\it effective} mean free path of ions 
which happens to become limited approximatively to the MHD scale, 
$l_A$.\footnote{This provided that turbulent
eddies may reach the MHD scale without being dissipated
(\S~5.1.3, \S~5)}
The value of the effective viscosity, even for motions along
the magnetic field lines, 
is thus expected to be
considerably reduced with respect to the classical
unmagnetized value and one may adopt a reasonable value of the
Reynolds number ${\it R}e \geq 10^3$.

\noindent
The important consequence of this picture is that 
both solenoidal and compressive modes in hot galaxy clusters
would not be strongly affected by viscosity at large scales
and an inertial range
is established, provided that the velocity of the eddies at large
scales exceeds $\approx$ 300 km/s.
We shall assume that a sizeable part of the large scale
turbulence is done by magnetosonic (essentially sound) modes.
At collisionless scales, $l < 10-50$ kpc, these
modes are affected by strong collisionless dampings with both 
thermal and relativistic particles (\S~4.3)
and thus they are expected to be the modes which dominate
the particle acceleration process.
Our claim about the existence of this well developed 
turbulent cascade which establishes an inertial range
from large scales to the collisionless scales would be even
reinforced when the possible effect of plasma instabilities
is considered.

\noindent
Although in this paper we focus on the particle
acceleration by hydro-- magnetosonic 
modes, it is worth mentioning that the mode composition at smaller
scales, $l<<l_A$, in the ICM should becomes much complex.
We shall assume that 
Alfv\'en modes are present at these MHD scales in the
ICM since in principle 
the cascading of solenoidal modes might reach very small scales, 
due to the lack of large--scale collisionless dampings
for these modes,  
and also because several mechanisms can convert a fraction
of the energy flux of large--scale turbulent cascade
in the injection of Alfv\'en modes at smaller scales
(e.g. Kato 1968; Eilek \& Henriksen 1984; Lazarian \& Beresnyak 2006).
At scales $l \leq l_A$, 
the coupling between Alfv\'en  
and compressible modes gets changed
and only slow modes are cascaded by Alfv\'enic modes (GS95,
Lithwick \& Goldreich 2001, Cho \& Lazarian 2002), 
while the cascading of fast modes is not particularly
sensitive to the presence of the other modes.
Given that, and since
magnetosonic modes are expected to be damped at 
scales larger than (or similar to) $l_A$ (\S~4,5), 
the spectrum of the ICM--turbulence at $l << l_A$
is expected to be populated only by Alfv\'en and slow modes.
These modes however would 
get anisotropic at these scales (unless injected at these scales
by some mechanism) 
and this should reduce their contribution to the scattering
and acceleration of fast particles via gyro--resonance.

\section{Compressible turbulence in the collisionless regime}

Compressible turbulence in galaxy clusters is thus made of
large scale hydro--motions with frequencies
essentially {\it infinitely} small with respect to $\Omega_i/\beta_{pl}$
($\Omega_i$ being the Larmor frequency of ions).
The basic physics of these low--frequency compressible modes 
in the collisionless regime can be derived by mean of the quasi--linear
theory and has been investigated in several seminal papers
(e.g., Melrose 1968; Barnes 1968;
Baldwin, Bernstein \& Weenink 1969;
Barnes \& Scargle 1973, hereafter BS73)\footnote{For hydromagnetic waves
with frequency of the order of $\Omega_i/\beta_{pl}$ see Foote \&
Kulsrud (1979).}.
This Section extends previous studies as we derive specific
and operative expressions for the physical properties 
of these modes which are of interest for the present paper
(e.g., energy decomposition of the mode, TTD--damping rate)
and discuss their dependence on the mode--propagation angle.
We focus on the case of long--wavelength modes in a magnetized plasma
dominated by thermal particles 
as it should be the case of the ICM.
Here we report the main formulae, while details and derivation of the
main equations are given in the Appendices.

\subsection{Geometry of the Mode and Dielectric Tensor}

\noindent
We define the turbulent fluctuations
associated with the electric and magnetic field as :

\begin{equation}
{\bf E} =
{\cal R}\left({\bf E}_k \exp\{ i({\bf k}\cdot {\bf r} - \omega t) \}
\right)
\label{e}
\end{equation}

\noindent
and

\begin{equation}
{\bf B} = {\cal R}\left({\bf B}_k
\exp\{ i({\bf k}\cdot {\bf r} - \omega t) \}
\right)
\label{b}
\end{equation}

\noindent
where ${\cal R}()$ stands for the Real part.
In the collisionless regime it is usual to start with 
fixing the geometry of the mode propagation and of the
electric field fluctuations.
Without loss of generality we may chose the particular
system where the {\it y}--component of the wavevector
of the modes vanishes, i.e. :

\begin{equation}
{\bf k} = \left(
k_{\perp}, 0, k_{\Vert}
\right)
\label{k}
\end{equation}

\noindent
For this choice the amplitude of the electric field
(and spatial Fourier transform of the electric field
of the mode) is given by (e.g., BS73) :

\begin{equation}
{\bf E}_k = \left(
0, E_{\perp}, E_{\Vert}
\right)
\label{E}
\end{equation}

\noindent
The amplitude of the magnetic field
of the mode comes from the Faraday low,
${\bf B}_k = {{c}\over{\omega}}
{\bf k} \times {\bf E}_k$ :

\begin{equation}
{\bf B}_k = {{c}\over{\omega}} 
\left( - k_{\Vert} E_{\perp},
- k_{\perp} E_{\Vert},
k_{\perp} E_{\perp}
\right)
\label{B}
\end{equation}
 
\noindent
As a starting point we
assume the presence of several, $\alpha$, species of
particles with particle momentum given by :
 
\begin{equation}
{\bf p}_{\alpha}= \left(
p_{\perp} \cos \phi, p_{\perp} \sin \phi, p_{\Vert} \right)=
m_{\alpha} \gamma
\left(v_{\perp} \cos \phi, v_{\perp} \sin \phi, v_{\Vert}
\right)
\label{p}
\end{equation}

\noindent
and indicate with $\hat{f}_{\alpha}$ the normalized particle
distribution in the momentum space of species $\alpha$
($f_{\alpha}(p)=N_{\alpha} \hat{f}_{\alpha}(p)$).

\noindent
The properties of a wave propagating in a magnetized
plasma in the collisionless regime depend on the dielectric
tensor of the plasma.
In the general case, the dielectric tensor of the
magnetized plasma is given by
(e.g., Melrose 1968; see also Tsytovich 1977 for the unmagnetized case) :

\begin{eqnarray}
K_{ij}= \delta_{ij}+
2\pi \sum_{\alpha} m_{\alpha}
\left( {{\omega_{p,\alpha} }\over{
\omega}} \right)^2 \int \int
dp_{\perp} p_{\perp} dp_{\Vert}
\Big[
{{v_{\Vert}}\over{v_{\perp}}}
\Big(
v_{\perp} {{\partial }
\over{\partial p_{\Vert}}}
- v_{\Vert} {{\partial }
\over{\partial p_{\perp}}}
\Big) \hat{f}_{\alpha}(p) b_i b_j +
\sum_{n=-\infty}^{\infty}
{{ \left( V_i V_j^* \right)_{\alpha}}\over{
\omega - n \Omega_{\alpha} - k_{\Vert} v_{\Vert} }}
\times \nonumber \\
\Big\{
{{ \omega - k_{\Vert} v_{\Vert}}\over{v_{\perp}}}
{{\partial }
\over{\partial p_{\perp}}}
+ k_{\Vert} {{\partial }
\over{\partial p_{\Vert}}}
\Big\}
\hat{f}_{\alpha}(p) \Big]
\label{dielectrictensor1}
\end{eqnarray}

\noindent
where $\omega_{p,\alpha}=\sqrt{ 4\pi N_{\alpha} e_{\alpha}^2/m_{\alpha}}$
is the plasma frequency for the species $\alpha$,
$b_i = \left( {{{\bf B}_o}\over{|B_o|}} \right)_i$
is the unit vector along the magnetic field,
$\Omega_{\alpha} 
= (e_{\alpha} B_o/m_{\alpha} c)/\gamma$ is the Larmor frequency
of particles $\alpha$,

\begin{equation}
\left( V_i V_j^* \right)_{\alpha}=
{\pmatrix{ \Big( {{v_{\perp} n }\over{z}} \Big)^2 J_n^2(z)
& {\it i} {{v_{\perp}^2 n }\over{z}} J_n(z) J_n^{\prime}(z)
&
{{v_{\perp} v_{\Vert} n }\over{z}}
J_n^2(z)
\cr
-{\it i} {{v_{\perp}^2 n }\over{z}} J_n(z) J_n^{\prime}(z)
&
v_{\perp}^2 \Big( J_n^{\prime}(z) \Big)^2
&
-{\it i} v_{\perp} v_{\Vert} J_n(z) J_n^{\prime}(z)
\cr
{{v_{\perp} v_{\Vert} n }\over{z}} J_n^2(z)
&
{\it i} v_{\perp} v_{\Vert} J_n(z) J_n^{\prime}(z)
&
v_{\Vert}^2 J_n^2(z)
\cr} }_{\alpha}
\label{VV}
\end{equation}

\noindent
and $z_{\alpha}=k_{\perp} p_{\perp}/m_{\alpha}
\Omega^{\alpha}_o$
($\Omega^{\alpha}_o=\Omega \gamma$ is the classical Larmor frequency)
is an adimensional parameter which scales with the ratio
between the frequency of the mode and the particle Larmor frequency, 
and with the ratio 
between the particle velocity and the phase velocity of the
mode, $z_{\alpha} \approx (\omega/\Omega^{\alpha}_o)( v_{\alpha}/V_{\rm ph})$.
We notice that magnetosonic modes with long wavelength, $l >>$pc, 
always have $z_{\alpha} << 1$.
In this case a more suitable expression for the dielectric tensor
can be obtained
by expanding the {\it Bessel--functions} in  
Eqs.~\ref{dielectrictensor1}--\ref{VV} in the limit
$z_{\alpha} << 1$ (Appendix A).

\subsection{Energy of the Mode}

\noindent
The energy of a mode in a magnetized plasma
is done by the sum of the energy associated with
the fluctuations
of the electric and magnetic fields, $W_E$ and $W_B$,
and by the energy contributed by
particles to the modes, $W_P$.
The total energy is then :

\begin{equation}
W = W_B + W_E + W_P
\label{energytot}
\end{equation}

\noindent
In the collisional regime (and adiabatic equation of state)
$W_P$ is given by the contributions from the kinetic
energy, $W_K$, and from a potential energy, $W_{\Phi}$,
associated with pressure fluctuations,
and a simple equipartition condition exists
(e.g., Denisse \& Delcroix 1963; Melrose 1968), namely:

\begin{equation}
W_B + W_{\Phi} \approx W_E + W_K
\label{equip}
\end{equation}

\noindent
In the collisionless regime the medium is described
in terms of the dielectric tensor
and it is not possible to
define $W_P$ in a meaningful way.
Thus one has to use Eq.(\ref{energytot}) as a definition
for $W_P$, with the total energy, $W$, defined independently.
The total energy of the mode is given by (e.g., Barnes 1968; see also
Melrose 1968 \& Tsytovich 1972 for equivalent
expressions):

\begin{equation}
W(k,\omega)=
{1\over{16 \pi}}
\Big[
{B^{*}_k}_i {B_k}_i +
{E_k}_i^*
{{\partial }\over{\partial \omega}}
\Big( \omega K_{ij}^h \Big)
{E_k}_j \Big]_{\omega_i =0}
\label{ww}
\end{equation}

\noindent
where $K_{ij}^h$ stands for the Hermitian part of
the dielectric tensor.
In this case the first term
in Eq.(\ref{ww}) accounts for magnetic field fluctuations,
while (from Eq.~\ref{energytot}) the term

\begin{equation}
{1\over{16 \pi}} \left({E_k}_i^*
{{\partial }\over{\partial \omega}}
\Big( \omega K_{ij}^h \Big)
{E_k}_j \right)=
{{|E_k|^2 }\over{16 \pi}}
+
\sum_{\alpha} ( W_P(k) )_{\alpha}
\label{wwterm}
\end{equation}

\noindent
accounts for the contribution to the mode energy
from the electric field fluctuations
and from particles (Barnes 1968).

\noindent
In the quasi--linear regime
the energy of the magnetic field fluctuations
is related to that of
electric field fluctuations
by (e.g., Melrose 1968):

\begin{equation}
W_B = \left( {{c }\over{V_{\rm ph} }} \right)^2
\left\{ 1 - \big| {{\bf k}\over{k}} \cdot {{\bf E_k}\over{|E_k|}}
\big|^2
\right\} W_E
\label{wb_we}
\end{equation}

\noindent
which can be taken $W_B \simeq (c/V_{\rm ph})^2 W_E$ since 
under the physical conditions of interest for this paper
magnetosonic modes have $E_{\perp}/E_{\Vert} <<1$ (Appendix B).
Thus combining Eq.(\ref{energytot}) with Eqs.(\ref{ww}--\ref{wb_we})
it is easy to get the ratio between the total energy in the
mode and that associated with the different components,
$W_B$, $W_E$, and $W_P$.

\begin{figure}
\resizebox{\hsize}{!}{\includegraphics{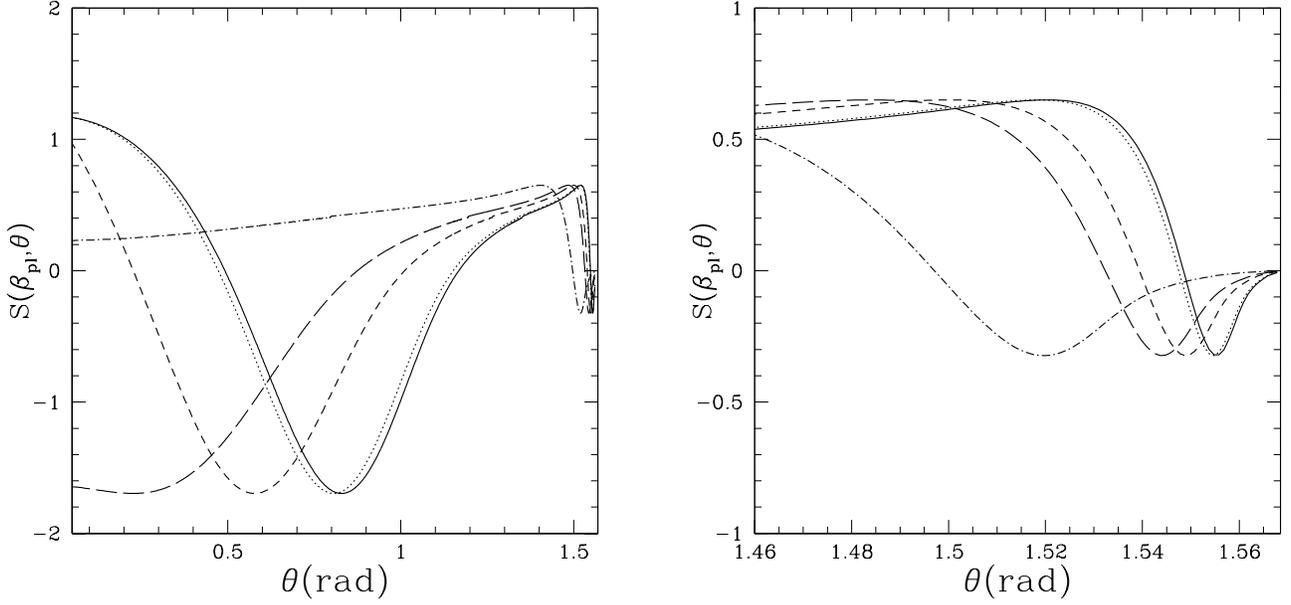}}
\caption[]{ Panel {\bf a)}
The expression
$S(\beta_{pl},\theta)$ is given
as a function of $\theta$.
The behavior around $\theta = \pi/2$
is highlighted in the right panel.
$k_B T=$8.6 keV is assumed.
Calculations are reported for : $c_s^2/v_A^2$=$\beta_{pl}/2$=
100 (solid line),
10 (dotted line), 1 (dashed line), 0.5 (long--dashed line),
and 0.1 (dash--dotted line).
}
\label{fig:s}
\end{figure}

\noindent
Thermal particles in the ICM should provide the
dominant contribution to the
total energy of turbulent modes.
Thus we make the
approximation that the dielectric tensor of the
ICM is described by
that of an electron--proton magnetized plasma in thermal equilibrium.
Assuming a Maxwellian distribution
for the thermal electrons and protons in the ICM :

\begin{equation}
f_{\alpha}(p)= N_{\alpha} \hat{f}_{\alpha}(p)=
{{ N_{\alpha} }\over{(2 \pi m_{\alpha} k_B T )^{3/2}}}
\exp \big\{ - {{p^2}\over{2 m_{\alpha} k_B T}} \big\}
\label{maxw}
\end{equation}

\noindent
in Appendix B, from Eqs.(\ref{dielectrictensor1}--\ref{VV}) and
Eq.(\ref{ww}) we show that 
the total energy of a fast mode is :

\begin{eqnarray}
W(k,\theta) =
{{ | B_k |^2 }\over{16 \pi}}
\Big\{ 1 +
{{\beta_{pl}}\over{2}}
\Big[
\left({{V_{\rm ph}}\over{c_s}}\right)^2 +
{3\over 5} \Big( {{k_{\perp}}\over{k}} \Big)^2
\Big( 2 - {\cal S}(\beta_{pl},\theta) \Big)+
{1\over{\beta_{pl}}}
\left( {{ V_{\rm ph} }\over{c}} \right)^2
\left( {3\over 5} \beta_{pl} +2 \right)
\Big]
\Big\}
\label{W_main}
\end{eqnarray}

\noindent
where the function ${\cal S}(\beta_{pl},\theta; \{ f_{rel}(p),T \})$
(Fig.~\ref{fig:s}) accounts for the terms of the 
dielectric tensor in the form :

\begin{equation}
\int dp_{\Vert} dp_{\perp} p_{\perp}^m
{{ \partial f_{\alpha} / \partial p_{\Vert} }\over
{k_{\Vert} v_{\Vert} - \omega}}
\end{equation}

which all come from the collisionless resonance between
particles and modes with $n=0$ in Eq.(\ref{dielectrictensor1})
(see \S~5.1 \& Appendix B).
Provided $f_{\alpha}$ is an even function of $p_{\Vert}$,
only particles with velocity larger than the phase velocity
of the mode can contribute to ${\cal S}$, since they should
satisfy the condition $k_{\Vert} v_{\Vert} \sim \omega=V_{\rm ph} k$.
The velocity of the selected resonant particles scales as
$v \sim V_{\rm ph}(\beta_{pl})/\cos(\theta)$, thus 
with increasing the angle between ${\bf k}$ and $B_o$, $\theta$, 
particles with increasing velocities may contribute to this term.
Formally particles with
$v \rightarrow \infty$ contribute to ${\cal S}$
for $\theta \rightarrow \pi/2$, and this
gets ${\cal S} \rightarrow 0$ in this limit (Fig.~\ref{fig:s}).
Two {\it wave--like behavior} of ${\cal S}$ can be recognized
in Fig.~\ref{fig:s}: the first one, for $\theta \leq 1$, marks 
the contribution from protons, while the second one, for larger 
$\theta$, marks that from electrons, which are faster than protons.
The resonance condition, $k_{\Vert} v_{\Vert} \sim \omega$, also 
drives the shift of these {\it wave--like behavior} 
toward smaller values of $\theta$ with decreasing $\beta_{pl}$ :
when $\beta_{pl}$ decreases the resonance between
the mode and a fixed portion of the particle distribution comes
up at smaller values of $\theta$.

\begin{figure}
\resizebox{\hsize}{!}{\includegraphics{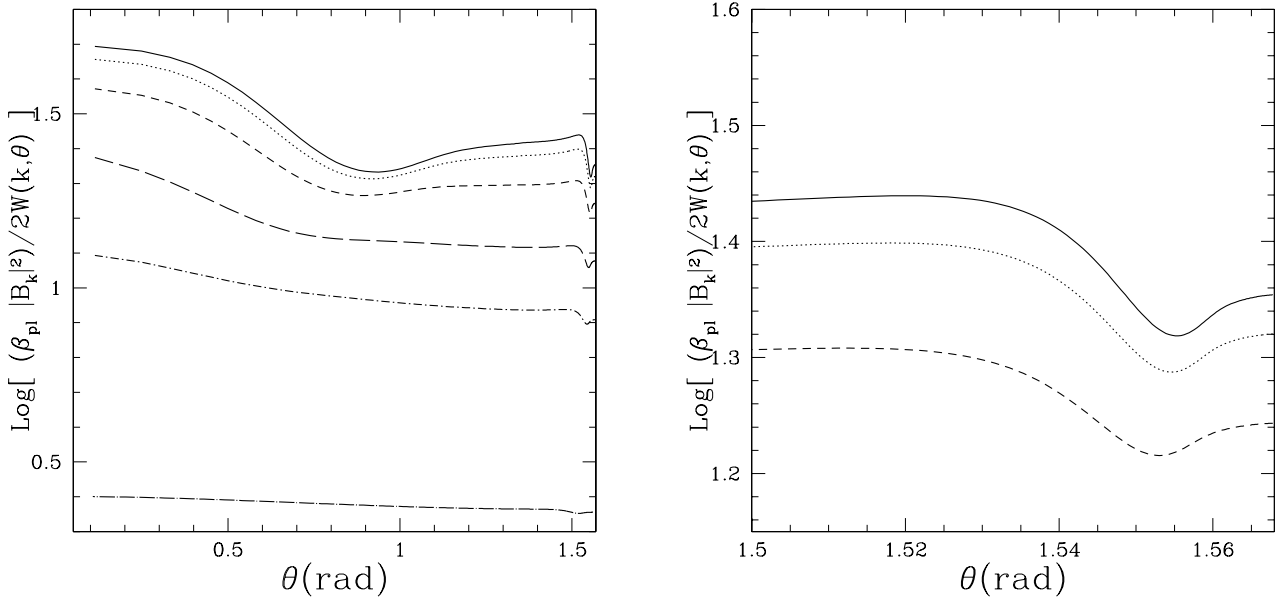}}
\caption[]{
The ratio between the magnetic energy density
of the mode and the total energy density of the mode is given
as a function of $\theta$ (for a better comparison
the quantity is multiplied by $16 \pi \beta_{pl}/2$).
The behavior around $\theta = \pi/2$
is highlighted in the right panel.
$k_B T=$8.6 keV is assumed.
Calculations are reported for: $c_s^2/v_A^2$=$\beta_{pl}/2$=
100 (solid line),
10 (dotted line), 3 (dashed line), 1 (long--dashed line),
0.5 (short--dashed--dotted line), and 0.1 (long--dashed--dotted line).
}
\label{fig:bkwk}
\end{figure}

In Fig.~\ref{fig:bkwk} we report the ratio between magnetic and
total energy of a mode propagating at an angle $\theta$ 
as a function of $\beta_{pl}$; this ratio is independent of the
wavenumber k of the mode.
Two {\it wave--like behaviors} (due to the
contribution from ${\cal S}$--terms) are visible:
the first one, for $\theta \leq 1$, marks the contribution
from protons, and the second one, for larger $\theta$, from electrons.
For small values of $\beta_{pl}$ it is $W_B \approx W/2$ and thus
the quantity $\beta_{pl} |B_k|^2/ 2W \propto \beta_{pl}$, on the
other hand, for large values of $\beta_{pl}$ it is
$W_B \propto W/\beta_{pl}$ and $\beta_{pl} |B_k|^2/ 2W$ becomes
independent of $\beta_{pl}$.

\begin{figure}
\resizebox{\hsize}{!}{\includegraphics{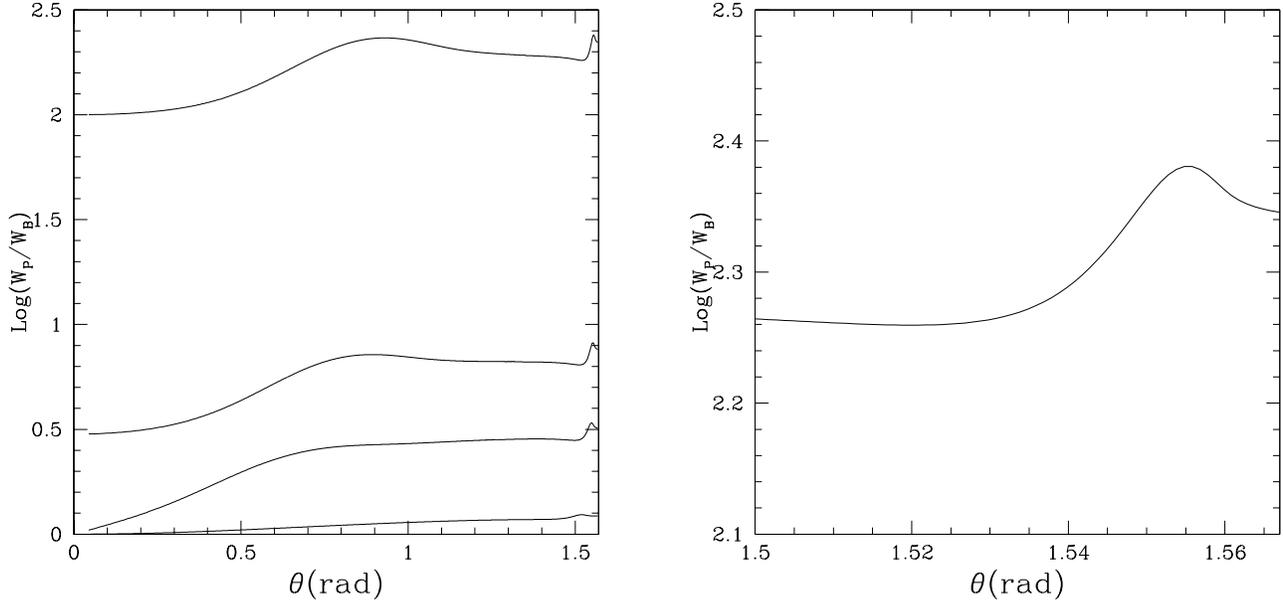}}
\caption[]{
The ratio between particle energy and magnetic energy
in the mode is reported 
as a function of $\theta$; $k_B T$=8.6 keV is assumed.
Calculations are reported for: $c_s^2/v_A^2$=0.1,
1, 3 and 100 (from the bottom to the top of the diagram).
The behavior at $\theta \sim \pi/2$
is highlighted in the right panel for $c_s^2/v_A^2 =100$.
}
\label{fig:energies}
\end{figure}

\noindent
Finally, in Fig.~\ref{fig:energies} we report the ratio between
particle energy and magnetic energy in the mode for different
values of $\beta_{pl}$ (see caption).
For $\beta_{pl} << 1$, $W_P$ reaches equipartition with $W_B$ similarly
to the case of collisional and low $\beta_{pl}$ plasmas.


\subsection{Turbulence Damping: TTD--resonance (n=0)}

A compressible mode in the collisionless regime
experiences strong collisionless damping with thermal and relativistic
particles and gets modified.
In this Section we report relevant formulae for the
collisionless damping rate via TTD--resonance of magnetosonic
waves which will be used in the present paper (\S~5).

\noindent
The damping coefficient of the mode
can be obtained by
the standard formula for the linear growth rate of the
mode in the quasi--linear theory
(e.g., BS73)\footnote{With this
formula it is $\partial W/\partial t = - \Gamma W$}:

\begin{equation}
\Gamma = - {\it i} \Big(
{{ E_i^* K^a_{ij} E_j }\over{ 16 \pi W }}
\Big)_{\omega_i =0}\omega_r
\label{damping_start_noA}
\end{equation}

\noindent
where $K^a_{ij}$ stands for the anti--Hermitian part of
the dielectric tensor, and $\omega_r$ is the 
real part of $\omega$.
The general formula for
the collisionless damping rate $(n=0, \pm 1, ..)$
is (Appendix C, and BS73):

\begin{equation}
\Gamma (k,\theta)=
-{{ \pi}\over{16 \omega_r W(k,\theta)}}
{{k_{\Vert}}\over{|k_{\Vert}|}}
\sum_{\alpha , n}
\omega_{p,\alpha}^2
\int_0^{\infty}
dp_{\perp}
\int_{-\infty}^{\infty}
dp_{\Vert} p_{\perp}^2
\Psi_n^{\alpha}
\Big\{
\left(
{{\omega}\over{k_{\Vert}}}
- v_{\Vert} \right)
{{\partial \hat{f}_{\alpha}(p)
}\over{\partial p_{\perp}}}
+
v_{\perp}
{{\partial \hat{f}_{\alpha}(p)
}\over{\partial p_{\Vert}}}
\Big\} \,\,\,
\delta \left({{p_{\Vert} }\over{m_{\alpha}}}
+
{{ n \Omega_{o,\alpha} - \omega_r \gamma }\over
{k_{\Vert}}}
\right)
\label{damping_1_noA}
\end{equation}

\noindent
where

\begin{equation}
\Psi_n^{\alpha}=
2 \Big|
{\it i} J^{\prime}_n(z_{\alpha}) E_{\perp}
+ {{ p_{\Vert} }\over{p_{\perp}}} J_n(z_{\alpha}) E_{\Vert}
\Big|^2
\label{theta_n_noA}
\end{equation}

\noindent
In this paper we focus on the case $n=0$ (Transit Time Damping, 
discussed in \S~5)
which is the most important collisionless
resonance between magnetosonic waves and particles in the ICM.
In the case of long--wavelength magnetosonic waves
the damping rate due to TTD--resonance with thermal electrons
and protons with number density $N_{e/p}$,
is given by (Appendix C):

\begin{equation}
\Gamma_{e/p}(k,\theta) =
\sqrt{ {{\pi}\over 8} }
{{ |B_k |^2}\over{W(k,\theta)}}
{\cal H}
\Big(1- {{V_{\rm ph}}\over{c}} {{k}\over{ |k_{\Vert}| }} \Big)
{{ V_{\rm ph}^2}\over{B_o^2}}
\left( {{k}\over{|k_{\Vert}|}} \right)
\left( {{k_{\perp}}\over{k}} \right)^2
{{ \left( m_{e/p} k_B T \right)^{1/2}}\over{
1 - ( {{V_{\rm ph} k}\over{c k_{\Vert} }} )^2 }} N_{e/p}
\exp \Big\{ - {{ m_{e/p} V_{\rm ph}^2 }\over
{2 k_B T}} {{\left( {{k / k_{\Vert}}} \right)^2 }\over{
1 - ( {{V_{\rm ph} k}\over{c k_{\Vert} }} )^2 }}
\Big\}
k
\label{damping_th_LFM_noA}
\end{equation}

\noindent
where ${\cal H}(x)$ is the Heaviside step function (1 for $x>0$,
and 0 otherwise), and
the ratio $|B_k|^2/W$ is given by Eq.(\ref{W_main}).

\noindent
Actually for a fixed value of $\beta_{pl}$, the damping rate scales
with $\sqrt{T}$ and this makes the damping strong in the
case of galaxy clusters ($T \sim {10^{7-8}}$K).
For $\beta_{pl} >>1$ the TTD damping rate from thermal electrons
is $\Gamma_e/\omega_r \approx \sqrt{3 \pi x /20} \exp(-5 x/3) \sin^2 \theta$, 
where $x=(m_e/m_p)/\cos^2\theta$, which is sufficiently 
small\footnote{$\Gamma_e/\omega_r$ has a maximum 
value $\approx 0.2$} to make the linear--theory approach
adopted here still reasonable.
Eq.(\ref{damping_th_LFM_noA}) is a general expression of the damping 
rate due to TTD resonance with thermal particles from which well known
formulae can be readily re--obtained. For instance in the case of
low $\beta_{pl}$ it is $V_{\rm ph} \rightarrow v_A$ and
$( |B_k|^2 / 16 \pi W ) \rightarrow 1/2$ and one gets the
usual TTD--damping rate of fast modes with thermal electrons 
(e.g., Akhiezer et al.~1975; Achterberg 1981; Miller 1991):

\begin{equation}
\Gamma_{e/p}(k,\theta) \rightarrow
\sqrt{ {{\pi}\over 2} }
{{m_e }\over{m_p}}
{{v_{te} }\over{v_A}}
{{\sin^2\theta}\over{|\cos\theta|}}
\exp \Big\{ - {{ v_A^2 }\over
{2 v_{te}^2 \cos^2\theta }} \Big\}
v_A k
\label{damping_th_LFM_noA_lowbeta}
\end{equation}

\noindent
where we define $v_{te} = \sqrt{ k_B T/m_e} $.
A formula equal to Eq.~\ref{damping_th_LFM_noA_lowbeta} is
also given for $\beta_{pl}<<1$ in Ginzburg (1961) and Shafranov (1967)
without adopting the simplified quasi--linear approach.
These authors also report a non--quasi--linear formula for the damping 
rate of thermal electrons and protons under the {\it particular} condition 
of $c_s << V_{ph} << v_{te}$, in which case
the plasma dielectric tensor can be largely simplified by expanding 
the {\it Z--function} (Appendix B, Eqs.~\ref{F}--\ref{plasmaz})
of electrons and protons for large (protons) and small
(electrons) arguments. In this case the normalization of the 
formula for the damping rate of protons is 5 times larger than that
in Eq.~\ref{damping_th_LFM_noA_lowbeta}, while the formula
for the damping of electrons is equal to Eq.~\ref{damping_th_LFM_noA_lowbeta}
(this {\it asymmetry} in the electron--proton contribution comes from the 
expansions of the {\it Z--function} in the two
opposite regimes for electrons and protons).
Still since it is $c_s << V_{ph}$ the damping from
protons is negligible and Eq.~\ref{damping_th_LFM_noA_lowbeta} is
equivalent to the result reported by these authors.

\noindent
The damping rate
due to ultra--relativistic electrons and protons is given by
(Appendix C, and BS73):

\begin{equation}
\Gamma_{e/p}(k,\theta)=
- {{\pi^2}\over{8}}
{{
| B_k |^2 }\over{W(k,\theta)}}
\left(
{{k_{\perp}}\over{k}}
\right)^2
\left(
{{k}\over{|k_{\Vert}|}}
\right)
{\cal H}
\Big(1- {{V_{\rm ph}}\over{c}} {{k}\over{|k_{\Vert}|}} \Big)
{{ N_{e^{\pm}/p} \,\, V_{\rm ph}^2}\over
{B_o^2}} \,\, k
\big(
1 - ( {{V_{\rm ph} k}\over{c k_{\Vert} }} )^2 \big)^2
\int^{\infty}
p^4 dp
\left( {{\partial \hat{f}_{\alpha}(p)
}\over{\partial p}}\right)_{e/p}
\label{damping_ur_noA}
\end{equation}

\noindent
while the damping rate due to generic non--ultra relativistic and
non--thermal particles is given in Appendix C 
(Eq.\ref{damping_2_trans}).

In Fig.~\ref{fig:damping} we report the damping
rate from both thermal and relativistic particles
under conditions typical of massive and hot galaxy clusters.
The most important damping for a mode propagating at
small angles ($\theta \leq 1$) is that with thermal protons,
on the other hand, a mode propagating at larger angles
is damped by thermal electrons.
We find that under viable physical conditions the damping due to
relativistic particles is formally relevant only in a narrow range 
of the values of $\theta$ (close to $\theta \sim \pi/2$), and 
that it accounts for only a few percent of the total damping rate.

For a given temperature of the plasma, $T$, 
the strength of the damping rate decreases with 
decreasing $\beta_{pl}$ as the phase velocity of the modes increases
with respect to the thermal velocity and this makes the
particle--mode resonance more difficult.

We notice that the overall damping rate is anisotropic with a relatively
narrow peak
at $k/k_{\Vert} \sim 30$ (for high $\beta_{pl}$)
where the bulk of thermal electrons
resonates with the modes.
On the other hand, as discussed in \S~3 ,
the ICM turbulence is super--Alfv\'enic
and thus the turbulent modes
can easily bend the magnetic field lines.
The time scale of the bending of lines
from hydro--motions
on a scale $l$ is expected to be
$\approx$ a fraction of $l/v_l$, where $v_l$ is the
rms velocity of the turbulent eddies at the scale $l$.
The bending of the lines by hydro--motions
on the shortest scales is thus the most efficient so that
we can grossly estimate this bending time--scale,
$\tau_{bb}$, as $\approx$ a fraction of $l_A/v_A$; 
eddies on scales below $l_A$
cannot significantly bend the field lines.\footnote{The
wandering of the magnetic field at scales $l \leq l_A$ is
discussed in Yan \& Lazarian (2004).}
This value of $\tau_{bb}$ should be compared with
that of the damping time at collisionless
scales which is grossly (from Eqs.\ref{damping_th_LFM_noA}
and \ref{W_main}) $\Gamma(k)^{-1}(l_{mfp}) \approx  
\sqrt{m_p/m_e} l_{mfp}/c_s$.
The relevant time--scale for 
isotropization of the pitch angle
$\theta$ due to line--bending is thus
faster than the damping process, i.e.
$\tau_{bb} < 1/\Gamma(l_{mfp})$, in the case \footnote{Here 
we assume that turbulent eddies reach scales $\leq l_A$
(\S~5.1.3, Fig.\ref{fig:tacc}a), in case
$l_{cut} \geq l_A$ the bending time--scale gets grossly
of the order of a fraction of 
the damping time--scale at the cut--off scale, 
which would still
be sufficient to have some isotropization} :

\begin{equation}
\beta_{pl} >> \left( {{L_o}\over{l_{mfp}}} \right)
\left( {{V_L}\over{c_s}} \right)^{-3}
\left( {{m_e}\over{m_p}} \right)^{1/2}
\label{bending}
\end{equation}

\noindent
The condition in Eq.(\ref{bending}) is
always satisfied in the ICM, at least
under the hypothesis of this paper, and thus we shall use
an effective damping rate {\it for the bulk of the spectrum} 
of magnetosonic modes
which comes from the contribution from different $\theta$s
and is defined by :

\begin{equation}
\langle \Gamma_{e/p}(k) \rangle=
\int_0^{\pi/2}
\Gamma_{e/p}(k,\theta) \sin\theta \, d\theta
\label{average}
\end{equation}

\begin{figure}
\resizebox{\hsize}{!}{\includegraphics{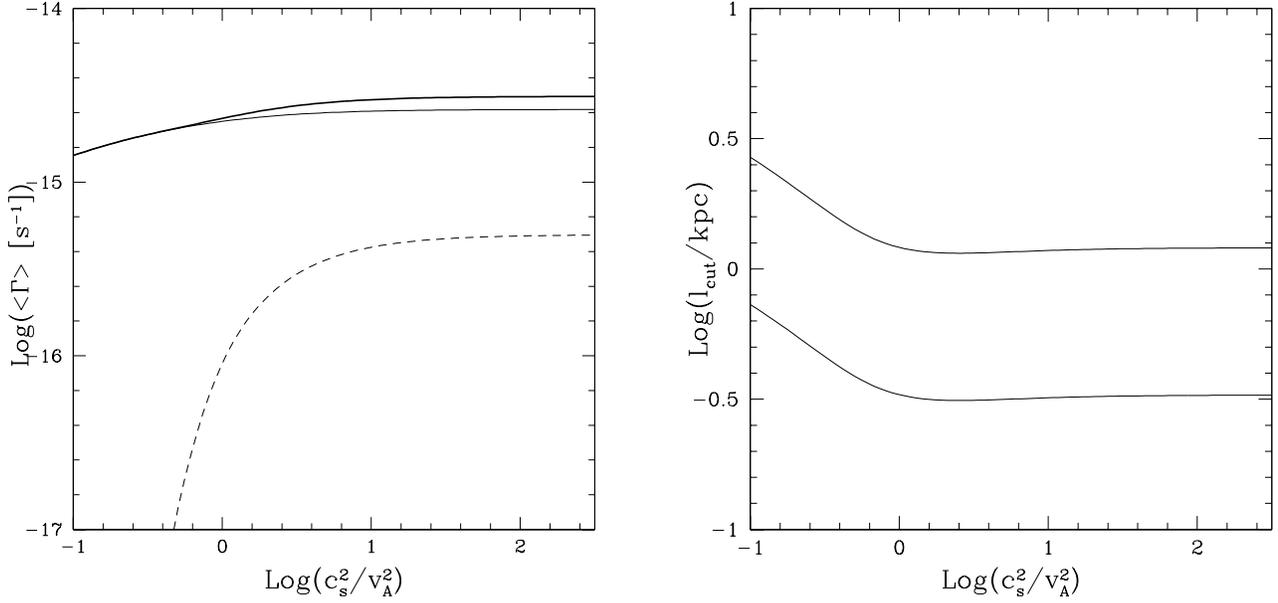}}
\caption[]{
{\bf Panel a)}
The average damping rates of magnetosonic modes (Eq.~\ref{average}),
due to TTD resonance with thermal electrons (solid line) and 
thermal protons (dashed lines), and total damping rate
(thick solid line) are reported as a function of
$c_s^2/v_A^2 (=\beta_{pl}/2)$; $k = 1$ kpc$^{-1}$ is taken.
{\bf Panel b)}
The turbulent cut--off scale (Eq.~\ref{kc}) is reported
as a function of $c_s^2/v_A^2 (=\beta_{pl}/2)$. 
Calculations are obtained assuming
$L_o$=300 kpc, 
and $(V_L/c_s)^2 =0.15$ (upper curve) and
$(V_L/c_s)^2 =0.3$ (lower curve).
In both panels we assume $k_B T = 8.6$ keV.
}
\label{fig:dampingcut}
\end{figure}

\noindent
This is reported in Fig.~\ref{fig:dampingcut}a as a function of
$c_s^2/v_A^2 (=\beta_{pl}/2)$ (for a given temperature
of the ICM, see caption). Damping of magnetosonic 
modes is found to be always dominated by thermal electrons because
they are faster than the phase velocity of these modes.
The contribution from thermal protons drops for $\beta_{pl} \leq
20$ since for smaller beta the phase velocity of the modes (Eq.\ref{vph})
increases with respect to the proton velocity  
and it is even more difficult for protons to satisfy
the resonant condition.

\begin{figure}
\resizebox{\hsize}{!}{\includegraphics{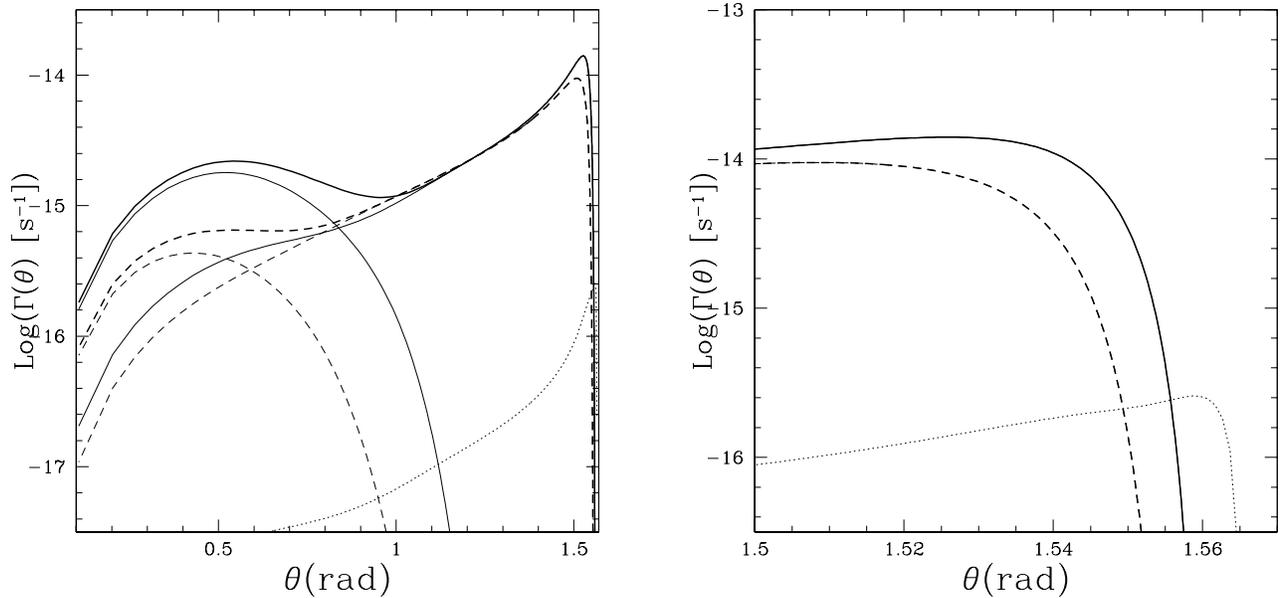}}
\caption[]{
The damping rates of magnetosonic modes due to TTD resonance
with thermal electrons (upper curves in the right
end of the panel) and with thermal protons (upper curves in the
left end of the panel), and the total damping rate
(thick curves) are reported as a function of $\theta$.
The behavior at about $\theta = \pi/2$
is highlighted in the right panel.
Calculations are reported for : $c_s^2/v_A^2$=100 (solid lines),
and 1 (dashed lines), and taking $k=1$ kpc$^{-1}$ and $k_B T=$8.6 keV.
The damping rate with relativistic protons is also 
reported in both panels (dotted lines): in this case 
we assume an energy distribution in the form
$f(p) \propto p^{-4.2}$ and an 
energy density $\sim 5$\% of the thermal one.
}
\label{fig:damping}
\end{figure}

Finally, let us comment that it is $ \langle \Gamma \rangle/\omega_r << 1$ and
this further motivate the practical use of the quasi--linear theory
in this paper.

\section{Stochastic particle acceleration in Galaxy Clusters}

In this Section we discuss the particle acceleration 
process in the ICM via resonant and non--resonant mechanisms
with compressible modes.

\subsection{Resonant {\it Transit Time Damping} Acceleration}

\subsubsection{Introduction}

Compressible (and incompressible) 
low--frequency MHD waves can strongly affect particle
motion through the action of the
mode--electric field via gyroresonant interaction (e.g.,
Melrose 1968), the condition for which is :

\begin{equation}
\omega
- k_{\Vert}v_{\Vert}
-n {{\Omega}\over{\gamma}} =0
\label{resonance}
\end{equation}

\noindent
where $n=\pm 1$, $\pm 2$, .. gives the
first (fundamental), second, .. harmonics
of the resonance, while
$v_{\Vert}=\mu v$ and $k_{\Vert}=\eta k$
are the parallel (projected along $B_o$)
speed of the
particles and the wave--number, respectively.
In general gyroresonance is a process important only 
for modes at very small scales, $l << l_A$.
However, as anticipated in \S~3.5, 
at these scales fast modes are probably absent in the ICM due to
strong resonant dampings (\S~5.1.3, Figs.~\ref{fig:damping},
\ref{fig:dampingcut}, \ref{fig:tacc}) 
and because they do not
couple with the Alfv\'enic cascade (\S~3.3.2, 3.5).

Interestingly enough, the compressible component of
the magnetic field of compressible 
modes (i.e. the component along ${\bf B}_o$
in the case of oblique propagation)
can interact with particles through the $n=0$ resonance.
This interaction is called {\it transit--time damping}
(e.g., Fisk 1976; Eilek 1979;
Miller, Larosa \& Moore 1996;
Schlickeiser \& Miller 1998).
An important aspect of this interaction is the need of
isotropization of particle momenta during acceleration
(e.g., Schlickeiser \& Miller 1998).
This is because the $n=0$ resonance changes only the
component of the particle momentum parallel to the
seed magnetic field.
This would cause an increasing degree of anisotropy
of the particle distribution 
and thus the deriving acceleration
would become less and less efficient with time.
Under our working picture,
particle--pitch angle scattering in the ICM 
can be provided by several processes discussed
in the literature. 
Those include electron firehose instability which is
indeed driven by pressure--anisotropies in high beta plasma
(Pilipp \& V\"olk 1971; Paesold \& Benz 1999), and 
gyro-resonance by Alfv\'en (and slow) modes at small scales, provided
that these modes are not too much anisotropic (cf. Yan \& Lazarian
2004). The latter condition means that the Alfv\'enic modes are
considered for scales not much less than $l_A$, provided that the
turbulence injection is isotropic. In addition, gyro-resonance was
discussed for the 
electrostatic lower hybrid modes generated by
anomalous Doppler resonance instability due to
pitch angle anisotropies
(e.g., Liu \& Mok 1977; Moghaddam--Taaheri et al.
1985) and by gyroresonant interaction
with whistlers (e.g., Steinacker \& Miller 1992). The latter
process, however, is somewhat more problematic than the
Alfv\'enic mode scattering, as
whistler turbulence is even more anisotropic than the Alfv\'enic one
(Cho \& Lazarian 2004).
Finally, 
instabilities within cosmic ray fluid look as a safer bet for isotropizing
cosmic rays. For instance, 
Lazarian \& Beresnyak (2006) proposed isotropization of cosmic rays 
due to gyroresonance 
instability that arises as the distribution of cosmic rays gets
anisotropic in phase space. This instability that is customary discussed
for plasma rather than for cosmic rays (see Gary et al. 1994, Kulsrud 2004)
would guarantee that in
the environments of galaxy clusters the TTD will 
not be quenched.

\subsubsection{Diffusion Coefficient}

The momentum--diffusion coefficient, $D_{pp}$, of particles
can be calculated by deriving the first--order corrections 
due to small amplitude plasma turbulence to
the orbits of particles in a uniform magnetic field, and
ensemble averaging over the statistical properties of the
turbulence (e.g., Jokipii 1966).
The resulting analytic expressions for the pitch--angle 
and momentum-- diffusion 
coefficients due to TTD resonance with fast modes in a low beta plasma 
can be found in Schlickeiser \& Miller (1998).

An additional and self--consistent way to derive the 
momentum--diffusion coefficient 
from the quasi--linear theory is to use an argument of detailed
balancing. The diffusion coefficient of a $\alpha$--species
is indeed related to the damping rate of the modes themselves
with the same particles, and one has (e.g., Eilek 1979; 
Achterberg 1981) :

\begin{equation}
\int d^3p E_{\alpha} \left( {{ \partial f_{\alpha}(p)}
\over{\partial t}} \right)
=
\int d{\bf k}
\Gamma^{\alpha}(k,\theta) {\cal W}({\bf k})
\label{dpp_1}
\end{equation}

\noindent
where $E_{\alpha}$ is the energy of a particle of
species $\alpha$, and ${\cal W}({\bf k})$ is the total
energy of the
modes in the elemental range $d{\bf k}$.
This is given by :

\begin{equation}
{\cal W}({\bf k}) =
{\cal W}_E({\bf k}) \left( {{W}\over{W_E}} \right)_{\bf k}
=
{{ {\cal W}_E(k) }\over{4 \pi k^2 }} \left( {{W}\over{W_E}}
\right)_{\bf k}
\label{w_1}
\end{equation}

\noindent
where $(W/W_E)_{\bf k}$ is the ratio between
the total and the electric energy
in a single mode propagating at ${\bf k}$ (\S~4.2),
and ${\cal W}_E(k)$ is the electric--field energy of the modes
in the elemental range $dk$.
In Eq.(\ref{w_1}) we have assumed an isotropic spectrum
of the electric field fluctuations which is an appropriate
assumption for super--Alfv\'enic turbulence and fast modes
(e.g., Cho \& Lazarian 2003).

\noindent
If isotropy of the particle momenta is {\it maintained}, 
the time evolution of the particle distribution function
is related to the diffusion coefficient by :

\begin{equation}
{{ \partial f_{\alpha}(p)}
\over{\partial t}}
=
{1\over{p^2}}
{{ \partial }
\over{\partial p}}
\left(
p^2 D_{pp} {{ \partial f_{\alpha}(p)}
\over{\partial p}}
\right)
\label{dpp_2}
\end{equation}

\noindent
and thus Eq.(\ref{dpp_1}) reads:

\begin{equation}
\int d^3p {{E_{\alpha} }\over{p^2}}
{{ \partial }\over{\partial p}}
\left(
p^2 D_{pp} {{ \partial f_{\alpha}(p)}
\over{\partial p}}
\right)
=
{1\over 2} \int dk \int d\theta \sin(\theta)
\Gamma^{\alpha}(k,\theta)
{\cal W}_E(k)
\left( {{W}\over{W_E}}
\right)_{(\theta,k)}
\label{dpp_3}
\end{equation}

\noindent
Here we are interested in
deriving the diffusion coefficient in the case of
relativistic species in the ICM (\S~5.1.3, 5.5).
The damping with these particles
(Eq.~\ref{damping_ur_noA}) can be
expressed in the form :

\begin{equation}
\Gamma^{\alpha}(k,\theta)=
- \Gamma^{\alpha}(\theta) \, k \int p^4 dp {{ \partial f_{\alpha}(p)}
\over{\partial p}}
\label{dpp_gamma}
\end{equation}

\noindent
and from partial integration of Eq.(\ref{dpp_3}) and from
Eqs.(\ref{dpp_gamma}) and (\ref{damping_ur_noA}) taking
${\cal W}_B(k) = (c/V_{\rm ph})^2 {\cal W}_E(k)$, 
one gets :

\begin{equation}
D_{pp}(p)={{\pi^2}\over{2\, c}}
p^2
{{ 1 }\over{B_o^2}}
\int_0^{\pi/2} d\theta V_{\rm ph}^2
{{ \sin^3(\theta) }\over{ |\cos(\theta) | }}
{\cal H}\left(1 - {{V_{\rm ph}/c}\over{\cos \theta}}
\right)
\left(
1 - ( {{V_{\rm ph}/c}\over{\cos \theta}} )^2 \right)^2
\int dk
{\cal W}_B(k) k
\label{dpp1}
\end{equation}

\noindent
This represents a self--consistent average 
(in terms of particle pitch--angle) momentum--diffusion coefficient
of isotropic particles with momentum $p$ which couple with fast 
magnetosonic modes via TTD resonance.
Eq.(\ref{dpp1}) in its low beta plasma limit (essentially 
$V_{\rm ph} \rightarrow v_A$ and $V_{ph}<<c$) is consistent with 
the expression (Eq.29) given in Schlickeiser \& Miller (1998) 
in its $z=k_{\perp} v_{\perp}/\Omega <<1$ limit
and averaged over the particle pitch--angle
\footnote{It is sufficient to integrate (average)
Eq.(29) in Schlickeiser \& Miller (1998) over the particle pitch--angle
using the properties of the delta--function, to solve this
integration and to expand the Bessel function in Eq.(29) for small
arguments.}.

\subsubsection{Acceleration efficiency in the ICM}

As summarized in \S~3.5 we focus on a picture in which
compressible turbulence is injected at large scales by the action
of cluster mergers and accretion of matter.
Provided that large scale turbulence in the
ICM is not significantly affected by the ion--viscosity
(\S~3.4), an inertial range is established 
due to the combination of turbulence injection
and cascading.
For isotropic turbulence the diffusion equation in the
k--space is given by :

\begin{equation}
{{\partial {\cal W}(k,t) }\over{\partial t}}
=
{{\partial}\over{\partial k}}
\left(
k^2 D_{kk}
{{\partial}\over{\partial k}}
( {{ {\cal W}(k,t) }\over{k^2}} )
\right)
- \sum_i \Gamma_i (k,t) {\cal W}(k,t)
+ I(k,t)
\label{modes_kinetic}
\end{equation}

\noindent
where $D_{kk}$ is the diffusion coefficient in the k--space,
$\Gamma_i(k,t)$ are the different damping terms
(\S~4.3),
and $I(k,t)$ accounts for the turbulence injection term.
The wave--wave diffusion coefficient of magnetosonic modes
(Kraichnan treatment; see also Zhou \& Matthaeus 1990;
Miller, La Rosa, \& Moore 1996 for low beta plasma) 
is given by \footnote{Here $\langle V_{\rm ph} \rangle$ is essentially 
a representative, 
averaged (with respect to $\theta$) phase velocity.}:

\begin{equation}
D_{kk}
\approx \langle V_{\rm ph} \rangle k^4 
\left(
{{ {\cal W}(k,t) }\over{\rho \langle V_{\rm ph} \rangle^2 }}
\right)
\label{Dkk}
\end{equation}

\noindent
We assume a constant (in time) injection spectrum of the modes in the simple
form $I(k)= I_o \delta (k - k_o)$ so that the stationary spectrum
of turbulence at the scales not significantly affected by dampings
($\Gamma_i \sim 0$) can be readily obtained from
Eq.(\ref{modes_kinetic}) : 

\begin{equation}
{\cal W}(k) =
\left(
{2 \over {7}} I_o \rho \langle V_{\rm ph}\rangle \right)^{ {1\over 2} }
k^{- {3 \over 2} }
\label{cascade_spectrum}
\end{equation}

\noindent
and the cascading time at a the scale $l = 2\pi/k$,
is given by :

\begin{equation}
\tau_{kk} \approx {{ k^3}\over{
{{\partial}\over{\partial k}}
( k^2 D_{kk} ) }}
= 
{2 \over {9}}
\left( {7 \over {2}} 
{{ \langle V_{\rm ph} \rangle \rho }\over{I_o}}
\right)^{ {1\over 2} }
k^{- {1\over 2}}
\label{cascading_time}
\end{equation}

\noindent
Provided that the dissipation of
compressible turbulence in the ICM
is collisionless (\S~3.4), the turbulence cascading
gets suppressed at a scale at which the resonant damping
time--scale, $\Gamma^{-1}$,
approach the cascading time.
This scale is given by Eq.(\ref{cascading_time}):

\begin{equation}
k_{cut} \approx {{81} \over {14}} 
{{I_o}\over{\rho \langle V_{\rm ph} \rangle}}
\left( {{\langle \Gamma(k) \rangle}\over{k}} \right)^{-2}
\label{kc}
\end{equation}

\noindent
where $\langle \Gamma \rangle$ is the average collisionless TTD damping
term given by Eqs.(\ref{damping_th_LFM_noA}),
(\ref{damping_ur_noA}) and (\ref{average}).
The value of the cut--off scale
is reported in Fig.~\ref{fig:dampingcut}b as a function
of the beta of the plasma for physical conditions
in the ICM (see caption): we find that if turbulence is energetic
enough (actually for the values used in the \S~5.3--5.5)
compressible modes are dissipated at $\approx$ sub--kpc scales.
The cut--off scale slightly increases in the case of small
$\beta_{pl}$ as the cascading of magnetosonic modes becomes 
less efficient (the cascading
time--scale goes as $\tau_{kk} \propto \langle V_{\rm ph} \rangle$ 
and, fixed $c_s$, increases for small $\beta_{pl}$).
Actually the cascading of compressible motions is likely to
reach MHD scales before being dissipated, $l_{cut} \leq l_A$ 
(Fig.~\ref{fig:tacc}a), and in this case it is also worth to mention 
that an Alfv\'enic turbulence can be activated by the cascading
of these compressible motions.

Eq.(\ref{modes_kinetic}) is appropriate to
describe the time evolution of the
total spectrum of isotropic turbulent modes.
On the other hand, formally
in the collisionless regime 
the ratio between the energy of the fields ($E$ and $B$)
and that associated with particles
changes with the mode--propagation angle 
(Figs.~\ref{fig:s}--\ref{fig:bkwk},
Appendix B). 
However the induced anisotropy is 
within a factor of 2--3 for a stationary $B_o$, and 
it should be efficiently smoothed out by the 
effect of the bending of the field lines (\S~4.3).
Thus we shall adopt isotropy as a viable approximation,
and define the
energy associated with the magnetic field fluctuations as:

\begin{equation}
{\cal W}_B(k,t) \sim {1\over{\beta_{pl}}}
\langle {{ \beta_{pl} |B_k|^2 }\over{ 16 \pi W(k) }} \rangle
{\cal W}(k,t)
\label{wb_w}
\end{equation}

\noindent
where for consistency $|B_k|^2/W$ is
taken from Eq.(\ref{W_main}) and $\langle \rangle$ indicates
the average over the propagation angle of the modes.

\noindent
The TTD diffusion coefficient in the particle--momentum
space is then obtained from Eqs.(\ref{dpp1}), (\ref{cascade_spectrum}),
and (\ref{wb_w}) in the form :

\begin{equation}
D_{pp}(p,t)=
{{\pi}\over{8}} \, {{p^2}\over{c}}
\langle {{ \beta_{pl} |B_k|^2 }\over{16 \pi W}} \rangle
{{ 1 }\over{c_s^2}}
\left( {{ 2 I_o \langle V_{\rm ph} \rangle}\over{7 \rho}} \right)^{ {1\over2} }
{ k_{cut}(t) }^{ {1\over2} } \int_0^{\pi/2}
d\theta V_{\rm ph}^2
{{ \sin^3(\theta) }\over{ |\cos(\theta) | }}
{\cal H}\left(1 - {{V_{\rm ph}/c}\over{\cos \theta}}
\right)
\left(
1 - ( {{V_{\rm ph}/c}\over{\cos \theta}} )^2 \right)^2
\label{dpp_ICM}
\end{equation}

\noindent
Eq.(\ref{dpp_ICM}) allows a prompt estimate of the acceleration
efficiency via TTD resonance, once the injection rate per unit mass
of the compressible turbulence ($I_o/\rho$) and the injection scale,
$k_o$ (or $L_o$), are fixed :

\begin{equation}
{{I_o}\over{\rho}} \approx
{\cal C} V_L^3 \, k_o
\left( {{V_L}\over{\langle V_{\rm ph} \rangle}} \right)
\label{iosurho}
\end{equation}

\noindent
where ${\cal C} \approx 5-6$ is a numerical factor which can be readily
obtained by taking $I_o/\rho \approx V_L^2 / \tau_{LL}$ and 
Eq.(\ref{cascading_time}).
The resulting {\it systematic}
acceleration rate, $\tau_{acc}$, is given by :

\begin{equation}
\tau_{acc}
= p^3 \big\{ 
{{\partial }\over{\partial p}}
\left( p^2 D_{pp}(p) \right)
\big\}^{-1}
\label{tauacc}
\end{equation}

\noindent
The {\it systematic} acceleration time from TTD resonance
does not depend on particle momentum (see also Fig.\ref{fig:times_loss})
and is reported in Fig.(\ref{fig:tacc}b) as a function of
$c_s^2/v_A^2(= 2\beta_{pl})$: for a given temperature
(and $\beta_{pl} > 1$) the acceleration efficiency
scales approximatively with $\sqrt{T}$ and 
is found to be almost independent from the value of $\beta_{pl}$.
The important point here is that the strength of the
TTD--acceleration efficiency, powered by compressible
turbulence with large--scale rms velocity $V_L^2/c_s^2 \approx 0.3$, is
found to give a {\it systematic} acceleration time of the order of
$\sim 10^8$yrs which is sufficient to 
accelerate electrons up to energies of several GeV, and
this may produce diffuse synchortron radio emission in 
$\mu$G--magnetized media (\S~6).

\subsection{Nonresonant acceleration}

\subsubsection{Introduction}

Resonant TTD acceleration is not the only process by which 
compressible turbulence may accelerate cosmic rays in the ICM.
For instance, fast particles can
be accelerated also by large scale compressible motions 
(e.g., Ptuskin 1988; Chandran 2003; Chandran \& Maron 2004;
Cho \& Lazarian 2006). 
Compression changes the particle momentum according to :

\begin{equation}
{{\partial p}\over{\partial t}}
=
- {1\over 3} p \,\, \nabla \cdot {\bf V}_l
\label{compression}
\end{equation}

\noindent
If the medium is neither expanding nor contracting 
it is $\langle \nabla \cdot {\bf V}_l \rangle=0$ and thus particles will
not experience regular changes in energy.
On the other hand if ${\bf V}_l$ is a turbulent field a 
statistical
acceleration effect (analogous to a classical 
second order Fermi process)
may exist.
This is essentially because particles would statistically
experience more compression than expansion.

\subsubsection{Diffusion Coefficient}

Limiting to the case $V_l^2 << c_s^2$ and provided that  
the turbulent velocity of the medium has correlation scales
much longer than the effective particle mean free path, 
the diffusion coefficient in the particle momentum space, $D_{pp}$,  
and the {\it total} (turbulent advection and diffusion) spatial
diffusion coefficient, $D_*$,  
can be obtained by standard procedure in plasma physics in the
quasi--linear approximation.
These are (Ptuskin 1988):

\begin{equation}
D_{pp} =
{2 \over 9}
p^2 D \int_{k}
{{ dy y^2 {\cal V}(y) }\over
{c_s^2 +y^2 D^2 }}
\label{dpp_compression_1}
\end{equation}

\noindent
and 

\begin{equation}
D_* = D \left( 1 + {4 \over 3} \int_{k}
{{ dy  {\cal V}(y) }\over
{c_s^2 +y^2 D^2 }}
\right)
\label{dxx*}
\end{equation}

\noindent
where $D$ is the spatial particle--diffusion coefficient
(without considering the effects induced by the 
nonresonant compressible coupling itself, 
Eq.~\ref{dpp_compression_1}), and ${\cal V}(y)$ is defined as :

\begin{equation}
\int {\cal V}(y) dy = V_L^2
\label{vo2_compression}
\end{equation}

\noindent
In this regime slow and fast diffusion limit exist. 
In the slow limit the rate of particle diffusion out of compressible
eddies is slower than the wave period, $\tau_w \sim l/c_s$, 
i.e. $\tau_{diff} \sim l^2/D  >> \tau_w$ and $c_s^2 >> k^2 D^2$.
Here the process is mainly contributed by the action
of the smaller eddies in the spectrum of the modes 
and it becomes faster as this minimum scale gets smaller
(e.g., Cho \& Lazarian 2006). From Eq.(\ref{dpp_compression_1}) we
find :

\begin{equation}
D_{pp} \sim 
{1 \over 9}
p^2 \left( {{V_L}\over{c_s}} \right)^2
D \left(
{1\over{L_o l^2}} \right)^{2/3}_{l \sim l_{min}}
\label{dpp_compression_1_SL}
\end{equation}

\noindent
For small minimum--turbulent scales this process formally 
becomes extremely efficient, however the minimum scale of 
the bulk of compressible turbulent
eddies in the ICM cannot be very small as these modes
are strongly damped (\S~5.1.3, Fig.~\ref{fig:dampingcut}).

\noindent
In the opposite case, 
in the fast diffusion limit, 
particles leave the eddies before they turnover,
i.e. $\tau_{diff} << \tau_w$ and $c_s^2 << k^2 D^2$.
Here the process is mainly contributed by the action
of the largest eddies which contain the bulk of
the turbulent energy, 
and from Eqs.(\ref{dpp_compression_1}) \& (\ref{vo2_compression})
we find :

\begin{equation}
D_{pp} \sim
{2 \over 9}
p^2 {{ V_L^2 }\over {D}}
\label{dpp_compression_1_FL}
\end{equation}

\noindent
An important point discussed in \S~3.2 is that 
particle--spatial diffusion itself is likely to be affected by the
turbulent bending of the 
magnetic field lines which gets the effective ion mean free path
$\sim l_A$.
Compared to the Coulomb or gyroresonance scattering
the diffusion with the characteristic scale $l_A$ does not involve any
changes of the particle energy via scattering. Therefore the particle may
diffuse slowly, but the only change in energy
will be due to large scale compressions
(cf. Cho \& Lazarian 2006).
We thus shall adopt a very simplified form of the spatial
diffusion coefficient in
Eqs.(\ref{dpp_compression_1}--\ref{dpp_compression_1_FL}) :

\begin{equation}
D \sim {c \over 3} \beta \max \Big\{ l_{cut}\, , \,
\min\{l_A \, , \, l_{mfp} \} \Big\}
\label{dxx}
\end{equation}

\noindent
The combination between Eqs.(\ref{dpp_compression_1}) and 
a turbulent--driven spatial diffusion coefficient
(e.g., Eq.~\ref{dxx}) 
provides an important refinement of the evaluation of  
the cosmic--ray acceleration  
via compressible long wave turbulence, and 
may have important consequences in the case of the particle acceleration in
the ICM (\S~5.3).

\noindent
Finally,
we want to remind that Eq.(\ref{dpp_compression_1})
is obtained 
by neglecting the effect of possible 
additional scattering processes due to resonant particle--wave
interactions.
The presence of instabilities in cosmic rays 
may create an additional slab-type
Alfv\'enic component that would produce additional gyroresonance
acceleration and reduce the effective mean
free path (Lazarian \& Beresnyak 2006). 
Conservatively we do not discuss this possibility in the
present paper. 

\begin{figure}
\resizebox{\hsize}{!}{\includegraphics{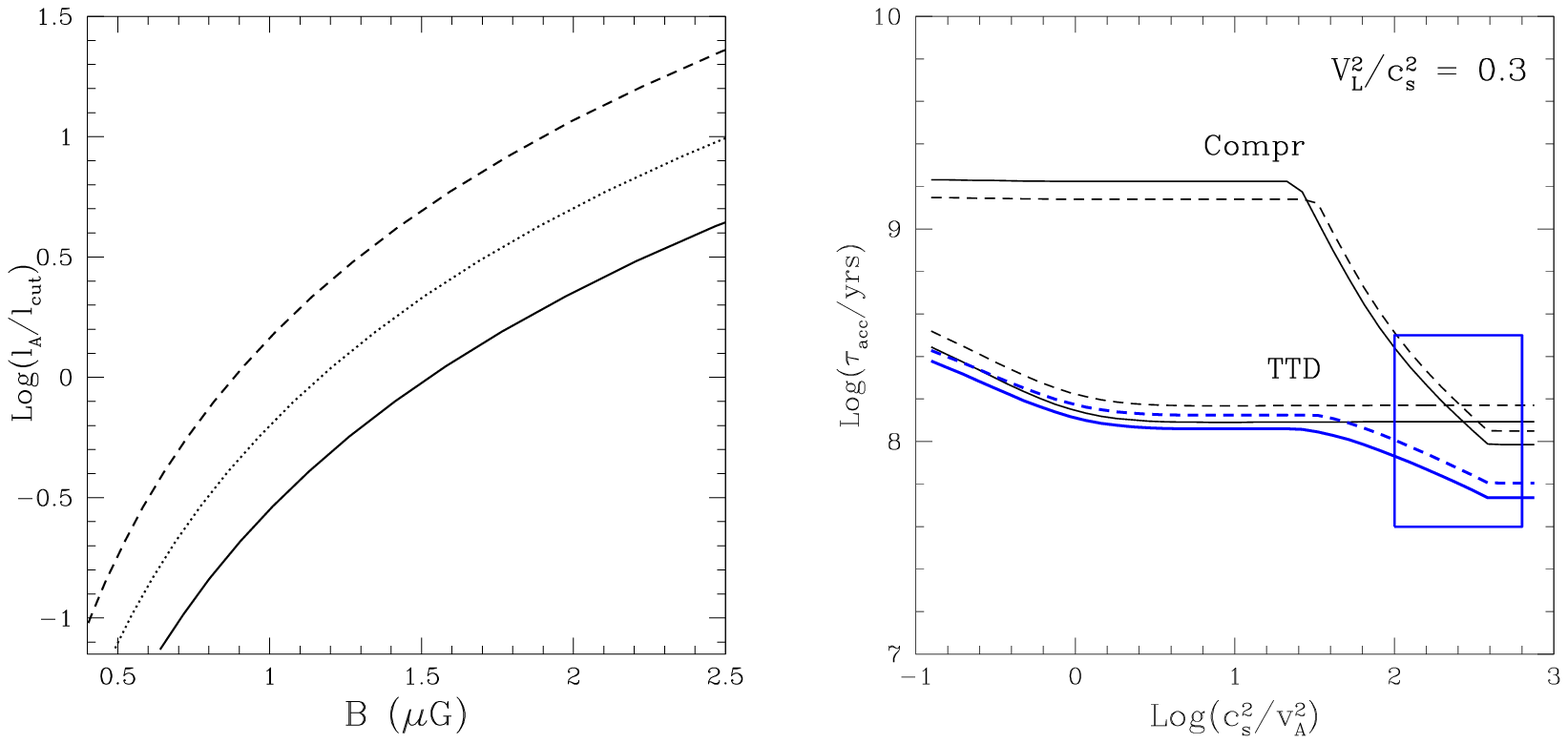}}
\caption[]{
{\bf Panel a):}
The ratio $l_A/l_{cut}$ is reported as a function of $B$.
Calculations are reported for $k_B T=$6 keV (dashed lines),
9 keV (dotted lines), and 12 keV (solid lines); 
$(V_L/c_s)^2=0.3$, $n_{th}=10^{-3}$cm$^{-3}$,
and $L_o=$ 300 kpc were assumed in the calculations.

\noindent
{\bf Panel b):}
The acceleration time (Eq.~\ref{tauacc}) is reported 
as a function of $c_s^2/v_A^2$ in the case of 
nonresonant compressive
acceleration (Eqs.~\ref{dpp_compression_1}) 
and resonant TTD
acceleration (Eq.~\ref{dpp_ICM}).
Calculations are reported for $k_B T=$7 keV (dashed lines) and
$k_B T=$11 keV (solid lines); 
$(V_L/c_s)^2=0.3$, $n_{th}=10^{-3}$cm$^{-3}$ , and
$L_o=$ 300 kpc were assumed in the calculations.
The acceleration time from the combined effect of the two mechanisms
is also shown (tick lines).

\noindent
The box marks the relevant range of the values of 
$c_s^2/v_A^2 (=\beta_{pl}/2)$ 
in the hot ICM, and the acceleration time
necessary to boost relativistic electrons at several GeV
(this accounts for both synchrotron and inverse Compton losses
with redshift $\sim$0--0.3).
}
\label{fig:tacc}
\end{figure}

\subsubsection{Acceleration efficiency in the ICM}

In this Section we calculate the 
efficiency of the particle
acceleration from large--scale nonresonant compression
in the ICM.

\noindent
Taking a Kraichnan scaling for the super--Alfv\'enic 
compressible turbulence, 
${\cal V}(k) \approx V_L^2 k^{-3/2}/L_o^{1/2}$,
from Eq.(\ref{dpp_compression_1}) we have :

\begin{equation}
D_{pp} \simeq
{2 \over 9} D \, p^2 {{ V_L^2}\over{L_o^{1/2}}}
\int_{1/L_o}^{1/l_{cut}}
{{ dy \, y^{1/2} }\over{c_s^2 + D^2 y^2 }}
\label{dpp_compression_KR}
\end{equation}

\noindent
where the spatial--diffusion coefficient is
given by Eq.(\ref{dxx}).

\noindent
The resulting {\it systematic} acceleration time is independent
of particle momentum (at least in the ultra--relativistic
case, see also Fig.~\ref{fig:times_loss})
and is reported in Fig.~\ref{fig:tacc}b.
For a given temperature of the plasma, $T$, in the case
of small $\beta_{pl}$ 
the nonresonant compression is formally very
inefficient because for large values of the
magnetic field 
the particle spatial--diffusion coefficient is large 
(essentially $D \approx 1/3 \, \beta \, c \, l_{mfp}$, $l_{mfp}$ 
from Eq.~\ref{lmfp}).
On the other hand, in the case of large $\beta_{pl}$ the
acceleration efficiency increases because turbulence
bends the magnetic field lines at scales smaller than $l_{mfp}$
and the effective particle mean free path is 
$\approx l_A$ (which scales as $\beta_{pl}^{-3/2}$);
saturation for large $\beta_{pl}$ is reached when
$l_{cut} \geq l_A$ (Fig.~\ref{fig:tacc}).

\noindent
The reference value of $\beta_{pl}$ in the ICM is
in the range $\beta_{pl} \sim 200-1000$ (i.e. $B \sim 0.5 - 3 \mu$G, with
$n_{th} \sim 10^{-4} - 10^{-3}$cm$^{-3}$ and $k_B T\sim 7-10$ keV),
and formally under these conditions 
we find that the 
acceleration efficiency from nonresonant compression driven by
relatively energetic turbulence (caption) is similar
to that due to the TTD--resonance.

\noindent
As already pointed out in \S~5.2.2, in the derivation of
Eq.(\ref{dpp_compression_1}) (or Eq.\ref{dpp_compression_KR}) 
it was assumed that the effective particle mean free path is
much smaller than the scale of the turbulent eddies.
This condition is formally violated in the case of small
$\beta_{pl}$ in Fig.~\ref{fig:tacc}b where the smaller 
turbulent eddies are $< l_{mfp}$ (mean free path 
$l_{mfp} \approx 10-50$ kpc).
On the other hand, this does not happen for larger $\beta_{pl}$, 
since in this case the particle {\it effective} mean free path, $\approx l_A
\approx l_{cut}$, 
is actually comparable to (or smaller than) the smallest  
turbulent eddies.

\subsection{Overall effect of compressible turbulence}

As discussed in \S~5.1.1 the TTD resonance is 
expected to be an efficient mechanism in the ICM, provided that particle
isotropy is preserved. Yet the TTD alone might not be efficient enough
in maintaining such isotropy 
because both $D_{pp}(\mu,p)$ and $D_{\mu \mu}(\mu,p)$
are strongly maximized for particles moving at small angles
with the direction of the seed magnetic field.
However additional resonant processes acting on small scales 
might easily maintain particle isotropy.
If these mechanisms are really at work in the ICM they should
also affect the spatial diffusion, $D$, of the particles and thus
the efficiency of the nonresonant compression mechanism.
Formally with decreasing $D$ the nonresonant coupling 
with eddies in the fast diffusion limit becomes more efficient,
and, at the same time, 
a larger range of scales of the eddies couples with particles
in the slow diffusion regime which is very efficient; 
actually this is what happens with increasing the beta of the
plasma in Fig.~\ref{fig:tacc}b.
However, if the spatial diffusion is strongly suppressed,
namely when $D < c_s l_{cut}$ in Eqs.(\ref{dpp_compression_1})
and (\ref{dpp_compression_KR}),
one gets into the slow diffusion limit at any turbulent scale, and
a decrease of $D$ yields a corresponding decrease in 
the efficiency of the nonresonant compression 
(Eq.~\ref{dpp_compression_1_SL}).
Thus future studies using self--consistent spatial diffusion
coefficients will be of great importance.

The turbulent bending of the field lines which happens in
the super--Alfv\'enic case cannot change the pitch angle of particles
which would preserve the adiabatic invariant, however in the high beta
ICM turbulent bending is associated with turbulent compressions which
indeed power the nonresonant acceleration mechanism and might provide
a source of particle--pitch angle isotropization.
The spatial diffusion coefficient
is related to that in the pitch angle
as (order of magnitude) $D \approx c^2/D_{\mu \mu}$, and
the resulting time--scale of the pitch
angle scattering, $\approx D_{\mu \mu}^{-1}$, is indeed
much shorter than the acceleration time of fast particles
(which is $\approx 10^7-10^8$yrs).

\noindent
This is important since it implies that the action of large scale
compressible turbulence in the ICM is twofold. 
On one hand particles diffusing
through the compressible turbulent eddies experience
substantial nonresonant stochastic acceleration. On the other hand, 
even without requiring additional processes at small scales, 
this might contribute to help 
in maintaining particle--momentum isotropization, so that 
the compressive
component of the turbulent magnetic field (that along ${\bf B}_o$)
may also couple efficiently 
with particles via TTD resonance without greatly change the 
particle spatial diffusion.

\noindent
These two mechanisms, TTD resonance and nonresonant compression, 
are driven by the same turbulent modes and 
involve independent particle--mode couplings and thus,
as a first approximation, the
acceleration process may be thought as the combination of the
two effects; the deriving particle acceleration time 
is also reported in Fig.(\ref{fig:tacc}b).

\section{Compressive turbulence and particle re--acceleration model
in galaxy clusters}

As already anticipated in the Introduction 
direct evidence for relativistic electrons diffused on Mpc
scales in the ICM
comes from radio halos and relics (e.g., Feretti 2005), while
the hard X--ray tails detected in a few
clusters may result from inverse Compton scattering of
the Cosmic Microwave Background photons by the same
electrons (e.g., Fusco--Femiano et al.~2004; Rephaeli, Gruber \&
Arieli 2006).

The particle re--acceleration model
is a promising possibility to explain the properties of the 
giant radio halos and possibly also the strength of the
hard X--ray tails.
This scenario assumes that turbulence is injected in a 
substantial fraction, Mpc$^3$, of the cluster volume during
cluster--cluster mergers, and that relativistic electrons
already present in the ICM and accumulated at $\gamma \approx 100$
are re--accelerated for a typical time--scale of $\leq$Gyr
(e.g., Brunetti et al.~2001,04; Petrosian 2001; Fujita et al.~2003).
Alternatively these seeds electrons to be re--accelerated
could be secondary
products of hadronic interactions (Brunetti \& Blasi 2005).

In this Section, after a brief review of the injection 
processes of cosmic rays in galaxy clusters and of the most relevant
channels of energy losses (\S~6.1), we provide calculations in the context
of the particle re--acceleration model which include the effect
of TTD--resonance and nonresonant acceleration due to compressible
turbulent modes injected at large scales.

\subsection{Cosmic Ray injection in the ICM}

There is a general consensus on the fact that several
mechanisms of injection of cosmic rays may be at work
in the ICM, and that once injected the bulk of these cosmic rays
does not escape the cluster 
(e.g., Berezinsky, Blasi \& Ptuskin 1997;
Ensslin et al. 1998; Voelk \& Atoyan 1999).

Collisionless shocks are generally recognized as efficient particle
accelerators through the so-called diffusive shock acceleration
(DSA) process (Drury, 1983; Blandford \& Eichler 1987).
This mechanism has been invoked several times as an
efficient acceleration process in clusters of galaxies
(Takizawa \& Naito 2000; Blasi 2001; Miniati et al. 2001;
Fujita \& Sarazin 2001; Ryu et al. 2003).
Present simulations confirm the analytical claim that 
shocks with Mach number larger than 2--3 are rare (Gabici \&
Blasi 2003), and claim 
that the energy content
in the form of cosmic rays in massive clusters
may be of the order of a few percent
of the thermal energy (Pfrommer et al. 2006; Jubelgas et al. 2006).
The bulk of the energy of these cosmic rays is injected in the
cluster outskirts by
shocks with a Mach number of the order of $\sim 3$, the real
efficiency of these shocks is however uncertain and it is
generally computed in the {\it test particle limit} and according to
the so-called  {\it thermal leakage} model (e.g., Kang \& Jones, 1995).

A contribution to the injection of cosmic rays in clusters of
galaxies may come from Active Galactic Nuclei which indeed
might fill the ICM with relativistic particles and
magnetic fields, {\it extracted} from the accretion power of their central
black hole (Ensslin et al., 1997).
Similarly to Active Galactic Nuclei,
powerful Galactic Winds may also inject relativistic particles and
magnetic fields in the ICM (V\"olk \& Atoyan 1999). Although the present day
level of starburst activity is low, it is expected that these winds were more
powerful during starburst activity in early galaxies, as also suggested
by the iron abundances in galaxy clusters (V\"olk et al. 1996).

\subsection{Energy Losses}

\subsubsection{Electrons and Positrons}

In the conditions typical of the ICM, ultra-relativistic electrons rapidly
cool down through inverse Compton 
and synchrotron emission, and accumulate at 
Lorentz factors $\gamma \sim 100-500$ where they may survive for a few
billion years before cooling further down in energy through Coulomb
scattering and eventually thermalize.
Energy losses and relevant time--scales of relativistic electrons 
in the ICM are discussed in several papers (e.g., Sarazin 1999; 
Petrosian 2001; Brunetti et al.~2004; Pfrommer \& Ensslin 2004).
In Fig.~\ref{fig:times_loss}a we report the particle life--time 
as a function of the Lorentz factor: the life--time has a peak
at $\gamma \approx 10^2-10^3$ where the cooling of electrons 
is slower and where particles may accumulate providing
a seed populations to be re--accelerated in the context of the
re--acceleration model.
More specifically, 
Fig.~\ref{fig:times_loss}a is obtained for typical physical
conditions in cluster cores and in the cluster outskirts:
in the external regions of clusters electrons survive
since Coulomb losses are less severe and in principle
these particles can
be accumulated for cosmological time--scales
at energies $\gamma \sim 100-1000$.
On the other hand, in cluster cores the higher thermal density 
limits the maximum life--time of electrons at less than 1 Gyr.

\subsubsection{Protons}

Once injected the 
relativistic cosmic--ray protons do not suffer catastrophic 
radiative--energy losses.
The only relevant channel of energy losses for these particles 
in the ICM is given by hadronic collisions which however get a 
typical particle life--time which is larger than a Hubble time
for $\sim$ GeV particles.
This, together with the long time necessary to the bulk of
these cosmic rays to diffuse out of clusters, 
makes clusters themselves reservoir in which cosmic ray
protons are confined and may accumulate over cosmological epochs  
(e.g., V\"olk et al.~1996; Berezinsky, Blasi \& Ptuskin 1997).

\noindent
On the other hand, mildly and sub- relativistic protons may be significantly
affected by Coulomb energy losses, which in turn change the particle spectrum
with respect to the injection spectrum.
The rate of Coulomb losses is (e.g., Schlickeiser 2002) :

\begin{equation}
{{dp}\over{dt}}_i (\beta_p) \approx - {{6}\over{\sqrt{\pi}}} \times
10^{-29} n_{th} 
\left[ \int_0^{\beta_p/\beta_e}
dy \exp\{-y^2\} -
{{\beta_p}\over{\beta_e}}
\left( 1 + {{m_e}\over{m_p}} \right)
\exp \big\{ - ({{\beta_p}\over{\beta_e}})^2 \big\}
\right]
\label{coulomb_i}
\end{equation}

\noindent
where $\beta_p$ and $\beta_e \sim 0.18 (T/10^8 K)^{1/2}$ are the 
velocity in units of the light speed of thermal electrons 
in the ICM and of the cosmic ray protons, respectively.

\begin{figure}
\resizebox{\hsize}{!}{\includegraphics{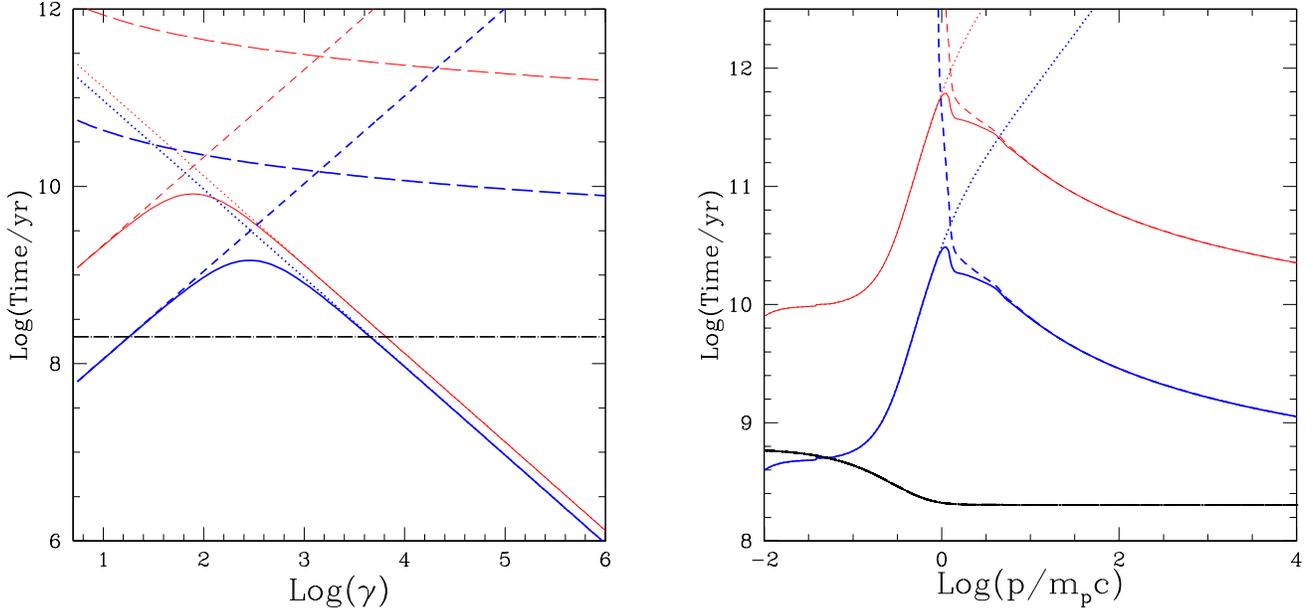}}
\caption[]{
{\bf Left panel}: The life--time of relativistic electrons in the
ICM at $z=0.2$ as a function of the Lorentz factor.
Thick (blue) lines are for cluster cores and thin (red) lines
are for cluster periphery.
We report the total life--time (solid lines) and the life--times
due to single processes: Coulomb losses (dashed lines),
synchrotron and IC losses (dotted lines), and bremsstrahlung
losses (long dashed lines).

\noindent
{\bf Right panel}: The life--time of cosmic--ray protons
in the ICM at $z=0.2$ as a function of the particle momentum.
Thick (blue) lines are for cluster cores and thin (red) lines
are for cluster periphery.
We report the total life--time (solid lines) and the life--times
due to single processes: Coulomb losses (dotted lines) and
pp--collisions (dashed lines).

\noindent
In both panels calculations in the cores (thick--blue)
are obtained for $B= 3 \mu$G and $n_{th}= 2 \times 10^{-3}$cm$^{-3}$,
and in the periphery (thin--red) for $B= 0.5 \mu$G 
and $n_{th}= 10^{-4}$cm$^{-3}$.
For comparison, the dash--dotted lines in both panels give the acceleration 
time--scale which is used in Fig.\ref{fig:fp}; note that
the increase of this
time--scale in the case of sub--relativistic protons is due to the decrease
of the efficiency of the non--resonant compression at sub--relativistic
energies.
}
\label{fig:times_loss}
\end{figure}

\noindent
As in the case of leptons, the details of the mechanisms of energy losses
of cosmic ray hadrons in the ICM 
can be found in several papers (e.g., Blasi \& Colafrancesco 1999;
Pfrommer \& Ensslin 2004; Brunetti \& Blasi 2005).
In Fig.~\ref{fig:times_loss}b we report the particle life--time 
as a function of the particle momentum.
Fig.~\ref{fig:times_loss}b is obtained for typical physical
conditions in cluster cores and in the cluster outskirts:
it is clear that even in the cluster cores where losses are
much severe, the bulk of relativistic protons has a life--time
of the order of an Hubble time.
Only protons with kinetic energy larger than about 200 GeV and
smaller than about 30 MeV in the cluster cores
have life--times smaller than a couple of Gyrs,
while just out of the core regions the life--time 
of these particles grows (time $\propto n_{th}^{-1}$) 
and all these particles are expected to survive for 
cosmological time--scales.

\begin{figure}
\resizebox{\hsize}{!}{\includegraphics{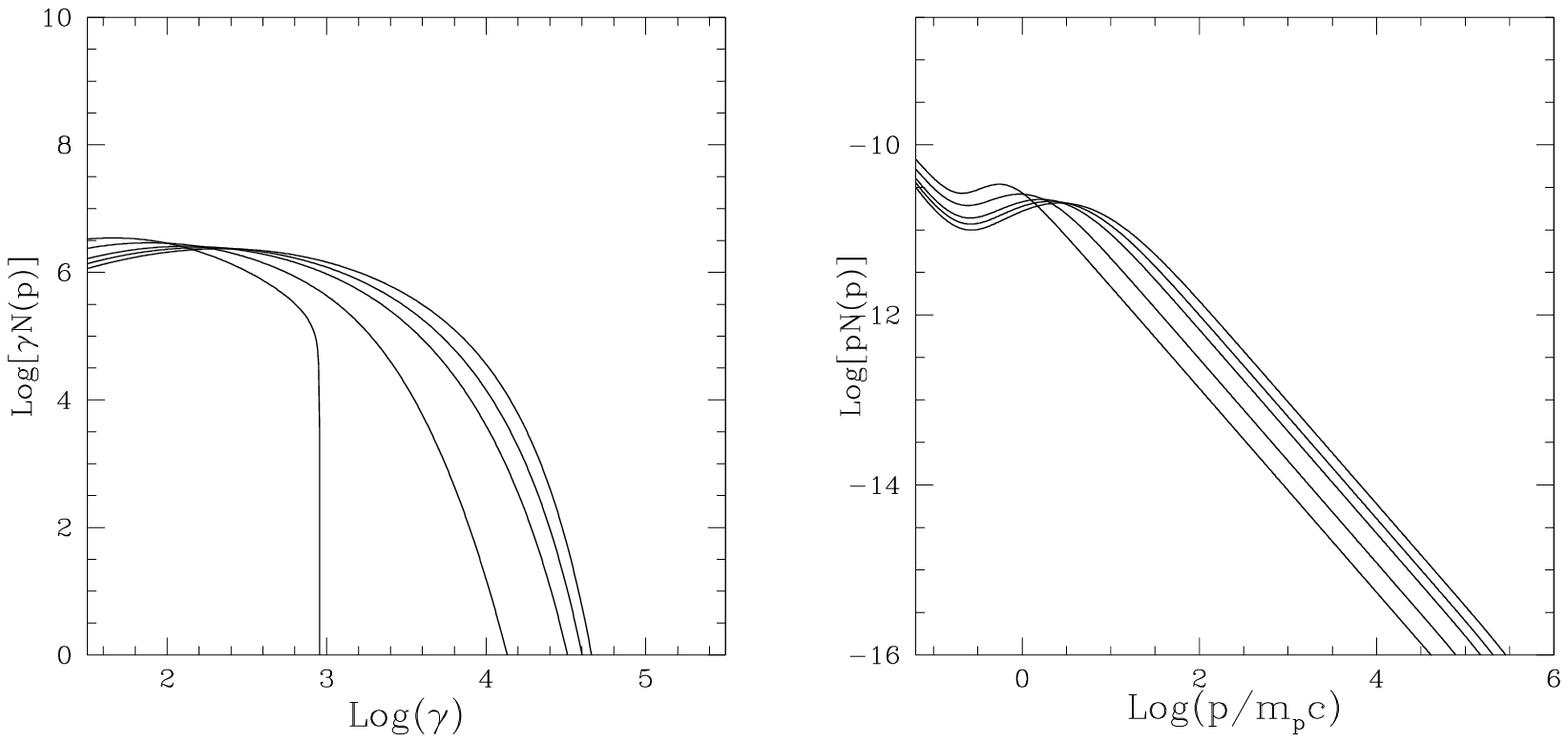}}
\caption[]{
{\bf Left Panel}: Time--evolution of the spectrum of relativistic
electrons as a function of the Lorentz factor.

\noindent
{\bf Right Panel}: Time--evolution of the spectrum of cosmic
ray protons as a function of the particle momentum.

\noindent
In both panels calculations are reported for: $t=0$, $4 \times 10^{15}$,
$8 \times 10^{15}$, $10^{16}$, $1.2 \times 10^{16}$sec from the start
of the re--acceleration phase.
Calculations are performed assuming $(V_L/c_s)^2=0.18$, $L_o = 300$ kpc,
$n_{th}=10^{-3}$, $k_B T$=9 keV, $B= 1 \mu$G, and redshift z=0.1 (for IC
losses).}
\label{fig:fp}
\end{figure}

\subsection{Numerical Calculations}

In this Section we calculate the time--evolution of the
spectrum of the relativistic particles 
stochastically re--accelerated by turbulent modes
in the ICM.

\subsubsection{Formalism}

As discussed in \S~5 we shall
assume isotropy of the particle momenta and of the modes, 
and in this case the time evolution of the
spectrum of the turbulent modes
and of the particles 
can be formally derived by a set of coupled kinetic equations.
The time evolution of the spectrum of the
leptonic component is given by :

\begin{equation}
{{\partial N_{\pm}(p,t)}\over{\partial t}}=
{{\partial }\over{\partial p}}
\left[
N_{\pm}(p,t)\left(
{{dp}\over{dt}}_{\rm rad} + {{dp}\over{dt}}_{\rm i}
-{2\over{p}} \big\{ D^{r}_{\rm pp} + D^{c}_{\rm pp}
\big\} \right)\right] +
{{\partial }\over{\partial p}}
\left[
\big\{ D^r_{\rm pp} + D^{c}_{\rm pp} \big\}
{{\partial N_{\pm}(p,t)}\over{\partial p}}
\right] +
Q_{e^{\pm}}(N_p(p,t); p,t) \, ,
\label{elettroni}
\end{equation}

\noindent
where $N_{-}$ and $N_{+}$ stands for electrons and positrons
($N = 4 \pi p^2 f$, $f$ is used in \S~4),
respectively, and where 
the terms $dp/dt_{rad}$ and $dp/dt_{i}$ account for
radiative (synchrotron and IC) and Coulomb losses,
while $D^r_{pp}$ and $D^c_{pp}$ are the resonant (TTD, Eq.\ref{dpp_ICM})
and the non--resonant (from turbulent--compression,
Eq.~\ref{dpp_compression_KR}) particle momentum--diffusion
coefficients.
In Eq.~\ref{elettroni} we also formally include an injection term,
$Q_{e^{\pm}}(N_p(p,t);p,t)$, which depends on the spectrum of
cosmic ray protons, and is necessary in the case that the
re--accelerated electrons and positrons are injected by hadronic collisions
in the ICM (see Brunetti \& Blasi 2005).

\noindent
The time evolution of the spectrum of 
cosmic ray protons is given by :

\begin{equation}
{{\partial N_p(p,t)}\over{\partial t}}=
{{\partial }\over{\partial p}}
\left[
N_p(p,t)\left( {{dp}\over{dt}}_{\rm i}
-{2\over{p}} \big\{ D^{r}_{\rm pp} + D^{c}_{\rm pp}
\big\}
\right)\right] +
{{\partial }\over{\partial p}}
\left[
\big\{ D^r_{\rm pp} + D^{c}_{\rm pp} \big\}
{{\partial N_p(p,t)}\over{\partial p}}
\right] 
,
\label{protoni}
\end{equation}

\noindent
where the term $dp/dt_{i}$ accounts for Coulomb losses (Eq.~\ref{coulomb_i}),
while the particle depletion due to
proton--proton collisions can be neglected. Because the population
of cosmic ray protons in the ICM essentially comes from the accumulation
of these particles during cosmological time--scales,  
we do not consider the source term in Eq.(\ref{protoni}) which would
account for the contribution from freshly injected protons.

The evolution of the spectrum of the
compressible turbulent modes is given by Eq.(\ref{modes_kinetic})
described in \S~5.1.3, where all the dampings are formally
derived in combination with Eqs.(\ref{elettroni}--\ref{protoni}).
The effect of the non--resonant damping on the spectrum
of the modes can be 
neglected since turbulent--compression acts efficiently only
on relativistic particles in the ICM (\S~5.2), and this gets a
net damping rate which is much smaller than that via TTD--resonance 
with the thermal ICM.

\subsubsection{Assumptions}

In this paper we adopt the particle re--acceleration model
assuming that a seed population of relativistic
electrons and protons in the ICM is re--accelerated by turbulence
injected at large scales during a merger event.
For simplicity we do not study the more complex issue of 
the re--acceleration of secondary electrons and positrons injected 
by proton collisions in the ICM (Brunetti \& Blasi 2005).

\noindent
The {\it new of this paper} is that compressible turbulence is used
as the driving of particle re--acceleration, and accordingly
the detailed 
diffusion coefficients obtained in \S~5 and the scenario and properties
of turbulence discussed in \S~3-4 
are used in the calculations.
For seek of clarity 
the main assumptions and physical parameters  
used in the calculations are listed below :

\begin{itemize}
\item[]{\it i)}
We consider physical parameters appropriate for massive galaxy clusters:
$T \approx 10^8$K,
$n_{th} \approx 10^{-3}$cm$^{-3}$, $B \approx 0.5 - 3 \, \mu$G.

\item[]{\it ii)}
Turbulence is assumed to be sub--sonic, with 
$V_L^2 << c_s^2$, and injected at large scales $L_0 \approx 300-500$ kpc
for a typical cluster--cluster crossing time.

\item[]{\it iii)}
The initial spectrum of electrons, $N_e(p,t=0)$, 
is derived by assuming that electrons are injected in the ICM
in a single event and then evolve passively for $\approx 1-3$ Gyr 
before being re--accelerated (see Brunetti et al.~2004).

\item[]{\it iv)}
The initial spectrum of protons, $N_p(p,t=0)$, is derived by assuming
that protons are continuously injected in the ICM for a long 
period, $\approx 3-5$ Gyr,  
with a constant injection spectral rate $Q \propto p^{-2.2}$
before being re--accelerated 
(see Brunetti et al.~2004).

\item[]{\it v)}
Damping terms from relativistic species in 
Eq.(\ref{modes_kinetic}) are neglected in the calculations as
they become important
only if the relativistic component gets a relevant fraction of the
thermal energy of the ICM (\S~4.3).

\item[]{\it vi)}
The damping term due to thermal particles is taken stationary 
because, under the assumption ({\it ii)}, 
the thermal properties of the ICM are not 
significantly modified with time.
\end{itemize}

Under conditions {\it ii)}, {\it v)} and {\it vi)} the spectrum of 
the modes is also stationary and this is given by
Eq.(\ref{cascade_spectrum}).

\subsubsection{Main Results}

Once large scale turbulence is injected in the ICM, magnetosonic modes
take a relatively long time to cascade at collisionless scales :

\begin{equation}
\tau_{kk} ({\rm Gyr})
\approx 0.6 \left( {{L_o}\over{300 \, {\rm kpc} }} \right)
\left( {{ V_L }\over{10^3 {\rm km/s} }} \right)^{-1}
\left( {{M_s}\over{0.5}} \right)^{-1}
\label{taucascade}
\end{equation}

\noindent
In the re--acceleration scenario
this is an unavoidable temporal gap, of a fraction of a Gyr,
between the injection of the first turbulent eddies and the beginning 
of the particle re--acceleration process.
When turbulence reaches collisionless scales the
acceleration process starts and particles take a time, of the
order of the re--acceleration time, to be significantly 
boosted in energy. 
A relevant example of the time evolution of the electron and proton
spectrum during the re--acceleration period is reported 
in Fig.~\ref{fig:fp} assuming $V_{L}^2 \sim 0.18 \, c_s^2$
(see caption): 
the seed electrons initially accumulated at $\gamma \sim 10^2-10^3$
are efficiently re--accelerated up to $\gamma \approx 10^4-10^5$.

\noindent
Radio (and hard X--rays) observations can be well explained
in terms of a high energy {\it tail} of emitting
relativistic electrons at energies of several GeV
(e.g. Schlickeiser et al. 1987; Brunetti et al. 2001;
Petrosian 2001).
We calculate the evolution of the re--accelerated particles
under the conditions given in \S~6.3.2 and find that,
quite independently from the initial electron and
proton spectrum, an appreciable high energy tail
of relativistic electrons at these energies
is produced approximatively
for $(V_L/c_s)^2 \geq 0.15 (1+z/0.1)^2$\footnote{The term
$(1+z/0.1)$ comes from IC losses (as for $B \approx \mu$G
synchrotron losses are sub-dominant).}.
In this case compressible turbulence injected 
at large scales in galaxy clusters may actually 
trigger efficient particle 
re--acceleration, and potentially explain the 
diffuse Mpc radio sources observed in massive galaxy clusters
and the hard X--rays in excess to the thermal X--ray emission.

\noindent
A spectral break, at $\approx$GHz frequencies, 
is observed in the synchrotron spectrum of a few
radio halos and this is interpreted in favour of the re--acceleration
scenario (e.g., Brunetti 2004; Feretti 2005).
Under the conditions {\it i)} in Sect.~6.3.2, 
such a break requires the presence of a corresponding break
in the spectrum of the emitting electrons at energies $\approx$5--10 GeV,
and we find that this is reproduced by our re--acceleration model
in the case of {\it moderate} turbulence, typically 
$(V_L/c_s)^2 \sim 0.15-0.25$, while in the case of more energetic turbulence
electrons can be re--accelerated at larger energies and the corresponding
synchrotron break is shifted at several GHz.

\noindent
Also protons are efficiently re--accelerated. 
Relativistic protons are not subject to radiative losses 
and since the re--acceleration efficiency scales with the energy
of the particles (\S~5.1--5.2) the spectrum is simply shifted at
higher energies and the slope of the injection spectrum
is essentially
preserved during the re--acceleration (Fig.~\ref{fig:fp}).

Under our assumptions
({\it v)}, \S~6.3.2), 
it is $\Gamma_{th} >> \Gamma_{rel}$ and the spectrum of the
turbulent fluctuations in terms of magnetic field, ${\cal W}_B$,
does not depend on the presence of cosmic rays,
thus protons cannot significantly 
affect the acceleration process of relativistic
electrons. This marks an important difference with
Alfv\'enic re--acceleration, 
in which case the dominant damping of the modes comes from the resonance
with relativistic protons and thus these protons affect the electron
acceleration ({\it Wave--proton Boiler}, Brunetti et al.~2004).

\noindent
An additional point to stress here is that, because 
$\Gamma_{th} >> \Gamma_{rel}$, 
the fraction of the turbulent energy which goes into
the cosmic rays via TTD--resonance 
is simply $\approx \Gamma_{rel}/\Gamma_{th}$ and is fixed by
the fraction of the energy in the ICM which is in the
form of cosmic rays. An additional
contribution to the energy of the re--accelerated particles 
comes from the non--resonant compression.
In our calculations (assuming that cosmic rays store 
a few percent of the thermal energy in the ICM) 
the total fraction of the turbulent energy which goes into
non--thermal particles is of the order of $\approx 2-5 \%$.

In case of long re--acceleration periods, actually $\geq 3-4$ times
the re--acceleration time (Eq.~\ref{tauacc}),
a non negligible fraction of the electron number is boosted
toward the maximum energy. Here an equilibrium between acceleration
and losses is reached and most of the energy flux from
the damping of the turbulence with these particles 
is radiated away via synchrotron and IC
by the same re--accelerated particles.
Thus in principle for very long re--acceleration periods
the total energy of the electron population should saturate
and the spectrum is expected to slowly approach stationary conditions.
On the other hand since cosmic ray protons are free from
energy losses, the energy flux from the damping of
the turbulence is totally stored in the form
of particle energy and this gives an {\it unbalance} between electron and
proton acceleration.
In the re--acceleration scenario this {\it unbalance} is not
expected to be large, indeed turbulence is injected during cluster mergers
and 
the duration of a re--acceleration period is constrained by the
cluster--cluster crossing time and by the turbulence cascading
time, and these cannot significantly exceed about 1 Gyr
(see Eq.~\ref{taucascade}).
In addition, present studies of the number counts of giant radio halos 
in galaxy clusters limit the life--time of these sources
at about $\approx 1$ Gyr 
(e.g., Hwang 2004) and this additionally constraints 
the duration of stochastic 
particle re--acceleration periods in galaxy clusters.
Actually, given these limits, we find that assuming 
$(V_{L}/c_s)^2 \sim$ 0.15--0.25 (which is required to
provide the necessary electron re--acceleration up to $\approx 5-10$ GeV)
and a duration of the re--acceleration {\it phase}
in the range 0.4--1 Gyr, 
the total energy of the cosmic ray protons and that of the relativistic
electrons are both boosted by a factor 1.5--4, and the
{\it unbalance} is not substantial.

\section{Discussion}

\subsection{Major Results}

The problem of proton and electron stochastic 
re--acceleration by compressible motions is a complex one. 
The efficiency of acceleration 
depends on the spectrum of compressible turbulent motions. The extend and the
shape of this spectrum, in its turn depends on the processes of plasma damping.
In the case of the hot 
ICM the corresponding issues have not yet been clarified 
sufficiently in the literature.

As a result, we had to address those issues one by one. Namely, we started
with the problem of describing turbulence in ICM in \S~3. 
First of all, we provided arguments suggesting that turbulence 
is expected to be present in the medium in hot (and massive)
galaxy clusters. This is also because the ICM is magnetized and this
implies a partial suppression of the plasma viscosity.
The suppression of the viscosity in a magnetized medium is a well known
effect and has been addressed at least for laminar flows 
(e.g., Simon 1955).
In the case of the strongly super--Alfv\'enic turbulence in the
ICM an additional effect comes due to the bending of the field
lines. Field lines are bended on scales $< l_{mfp}$ and this affects the 
ion diffusion process and thus viscosity.
An additional suppression of the viscosity might come from
the effect of plasma instabilities which affect the ion--ion mean free path,
but that are not considered in this paper.
Then, as we are interested in the compressible motions we
discussed their
generation in super--Alfv\'enic and MHD turbulence along with providing 
the estimates for collisional and
collisionless damping of such motions. 
The outcome of \S~3 is a validation of a basic
features of scenario according to which the energy can be injected due to
cluster
mergers on large scales and energize the particles in the ICM.

The quantitative treatment of the particle re--acceleration requires a much
more rigorous treatment of mode spectrum and damping which was non trivial.
In \S~4 we make use of collisionless physics and 
quasi--linear theory and derive general formulae for the spectrum of the 
compressible modes,
basically the ratio between the energy in magnetic fluctuations
and the total energy in the mode, and for the damping rate 
in magnetized plasma,
and re--obtained expressions known in the literature, as a particular cases
of our approach. 
The importance of the derived formulae goes beyond our particular case
of study, as a rigorous description of damping is important for many other
astrophysical important situations, e.g. in galactic environments 
(see Yan \& Lazarian 2004).
 
Having at hand a description of compressible super--Alfv\'enic and
MHD turbulence with specified injection 
and damping scales we studied in \S~5 proton and electron stochastic
re--acceleration by compressible modes. 
We focus on particle acceleration from magnetosonic modes and neglect
the contribution from slow modes and Alfv\'en modes. Slow modes are 
sub--dominant for particle re--acceleration as they have a phase
velocity $<<$ than that of fast modes and sound waves, in addition
both slow modes and Alfv\'en modes get anisotropic at small scales
(if injected at large scales) and this reduces the efficiency of
gyro--resonance acceleration.
We showed that because of
efficient damping of fast modes at small scales the acceleration
by gyroresonance is suppressed, i.e. only extremely
high energy protons with large gyroradius
can find magnetic perturbations to resonate with. 
Thus we study stochastic re--acceleration by both non--resonant large scale
compressions and resonant TTD, and 
clarified the regimes when the non--resonant large scale compressions 
is important.
The acceleration picture that is drawn from this paper is complex.
In the case of super--Alfv\'enic turbulence in the ICM the
turbulent bending of magnetic field lines limits particle spatial diffusion.
Because line bending is associated with turbulent compression
fast particles diffusing through the compressible turbulent eddies
may experience efficient stochastic acceleration via {\it Fermi II}
nonresonant--turbulent compression.
The same particles can also experience coupling with these
compressible eddies via TTD resonance which is found to be efficient in the
ICM provided that particle pitch--angle isotropization is maintained.

Finally, in \S~7, we apply our results to the case of the particle re--acceleration
scenario which is proposed to explain radio halos (and hard X--ray tails)
in galaxy clusters.
Our calculations showed that the acceleration of energetic particles in
galaxy clusters may be efficient. 
Relativistic electrons in the ICM can be re--accelerated against radiative 
and Coulomb losses up to energies of
several GeV (or more) assuming that compressible turbulence at
large scales stores a {\it non--negligible} fraction of the thermal
energy, namely $(V_L/c_s)^2 \geq 0.15$.
These electrons would emit Mpc--scale synchrotron radiation 
up to GHz frequencies
(or more) provided that the magnetic
field in the ICM is at $\approx \mu$G level on these scales.
In addition it also comes out that the re--acceleration of these electrons
happens without transferring too much energy to protons, which
might alleviate possible problems of earlier re--acceleration models 
that appealed to Alfv\'en modes.

\subsection{Simplifying assumptions}

In other words, the proposed re--acceleration scenario, which makes use
of compressible modes, is a plausible one and
deserves further studies. 
At the moment it includes several simplifications. In particular,
plasma instabilities can decrease further mean free path of protons, which
would decrease damping of turbulence. As a result, compressible modes could
cascade to smaller scales, making, for instance, gyroresonance 
acceleration by fast modes more efficient.
An effect related to the gyrokinetic instability in accelerated particles 
(Lazarian \& Beresnyak 2006) may act in a different direction suppressing
compressible motions at small scales, however on the other hand
the Alfv\'enic component
generated by this instability may also accelerate particles.
Plasma instabilities might also affect the diffusion of fast 
particles and this might be important in the calculation of the 
efficiency of the acceleration from nonresonant compression.

\noindent
In addition, reconnection processes taking place in the
magnetized plasma should be able to accelerate particles on their own.
Within small volume current sheets, the percentage of accelerated 
particles is small.
However, stochastic reconnection model in Lazarian \& Vishniac (1999) allows
acceleration of a substantial part of particles at the expense of 
the magnetic energy in the turbulent plasmas 
(Gouveia Dal Pino \& Lazarian 2005).
Therefore we believe that our treatment would underestimate the 
actual acceleration; 
further research should clarify the actual picture.

Our derivations of the damping rates are valid when the ratio of the
imaginary to the real part of the mode--frequency is much less than unity. 
This is generally true in the ICM and is
a natural assumption for dealing with turbulence cascade, 
as in the opposite regime, no cascading is 
possible and the energy dissipates at the injection scale.

\subsection{Relation to Earlier Works}

Stochastic particle acceleration in galaxy clusters
has been addressed by several papers (e.g., Schlickeiser et al.~1987;
Brunetti et al.~2001; Petrosian 2001; Fujita et al.~2003).
This work appeals to compressible motions to re--accelerate particles in the
ICM. 
Earlier detailed time--dependent 
calculations of the problem of re--acceleration 
was addressed in Brunetti et al. (2004) and Brunetti \& Blasi (2005), 
where Alfv\'en modes were used for the purpose. 
Such an approach is adequate if, for
instance, Alfv\'en modes are injected by some mechanism
at small scales. One possibility is that 
this might happen in the gyrokinetic instability scenario 
(Lazarian \& Beresnyak 2006), and this provides
an interesting possibility that we consider elsewhere. 
If, however, Alfv\'en modes are injected at large scales
the Alfv\'enic component at the scale of energetic particle gyroradius 
gets very anisotropic and interacts very inefficiently
with the particles (Chandran 2000; Yan \& Lazarian 2002). 
Thus fast compressible modes should be considered. 
These modes are isotropic 
and may scatter particles efficiently as we have
demonstrated above. 

In some aspects the scenario suggested in the present paper
for the ICM is similar to that adopted to calculate 
the {\it scattering} of galactic cosmic rays in Yan \& Lazarian (2004),
and to that adopted in the ICM 
by Cassano \& Brunetti (2005).

\noindent
The present paper is a {\it theoretical extension} of the work of
Cassano \& Brunetti (2005) where a more simplified treatment
of the resonant TTD re--acceleration of
electrons by fast modes was used to derive the statistical 
properties of non--thermal emission in galaxy clusters.

\noindent
Yan \& Lazarian (2004) discuss cosmic rays propagation
in Milky Way thus focusing on MHD turbulence at scales $l<l_A$ and mostly
low beta plasma. On the other
hand, here we concentrated on the
acceleration by motions at scales larger than $l_A$ and high beta plasma,
 the conditions which are relevant to the clusters of galaxies.
This made our calculations of the particle--mode damping rates
and of the particle--diffusion coefficients different from
those in Yan \& Lazarian (2004). 
In particular, we had to 
re--adopt many of the plasma results for high beta plasma and 
to treat differently magnetic field wandering.

Finally, an additional new of this paper is that 
we not only considered acceleration of electrons 
by the TTD resonance, but
acceleration of fast protons 
and electrons subjected to both the TTD resonance 
and the large scale compressions.

\section{Short Summary}

The paper above explains the non-thermal emission observed in galaxy 
clusters as a consequence of electron re--acceleration by compressible 
turbulence. In this scenario turbulence 
is injected at the scale of galaxy mergers and cascades to
small scales where the bulk of energetic particle acceleration happens.
The turbulence is described by using the recent advances in understanding 
of MHD turbulence. The paper incorporates:\\
I. A model of compressible turbulence in galaxy clusters. In this 
model the energy is injected at the scale of galaxy mergers and cascades to 
small scales where the bulk of energetic particle acceleration happens.\\
II. Calculations of the plasma damping and energy of the mode
for an arbitrary angle of
wave propagation to magnetic field and a rather general model of plasma.\\
III. Calculations of acceleration of protons and electrons by compressible 
motions in ICM plasma and a detailed application to the particle
re--acceleration scenario to explain radio halos and possibly hard X--ray
tails.

Our results 
show that electrons obtain a substantial part of the energy
transfered
to the energetic particles, which fits well to the existing observational
constraints.

\section{Acknowledgments}
We thank the referee V. Petrosian for very useful comments
and discussions, and R. Cassano and V. Dogiel for comments on the
manuscript.
We acknowledge partial support from MIUR through grant Cofin2004.
GB thanks partial support from MIUR through grant PRIN2005.
AL acknowledges NASA NNG05GGF57G and the NSF Center for Magnetic
Self--Organization in Laboratory and Astrophysical Plasmas.

\appendix

\section{Dielectric Tensor in the long wavelength limit}

\noindent
A relevant case for many astrophysical situations is that
of long--wavelength modes for which it is 
$|z_{\alpha}| \sim k p/m_{\alpha} \Omega_o^{\alpha} <<1$.
This is also the case of the turbulent modes in the 
ICM of interest in the present paper.
The dielectric tensor (Eq.\ref{dielectrictensor1}) is in the
form :

\begin{equation}
K_{ij}= \delta_{ij} -\omega^{-2} \sum_{\alpha} R^{\alpha}_{ij}
\label{dielectrictensor}
\end{equation}

\noindent
Barnes \& Scargle (1973) calculated the
tensor $R_{ij}$ from Eqs.\ref{dielectrictensor1}--\ref{VV}
and under the conditions $z_{\alpha}=
k_{\perp} p_{\perp}/m_{\alpha} 
\Omega^{\alpha}_o << 1$
and $X^{\alpha}_1 \equiv (k_{\Vert}p_{\Vert}/m_{\alpha} - 
\omega \gamma)/\Omega^{\alpha}_o << 1$.
In this case, by taking into account that
$J_n (z_{\alpha})= (z_{\alpha}/2)^n 
\sum_m (-z_{\alpha}^2/4)^m / m! \Gamma(n+m+1)$, one finds
(BS73) :

\begin{equation}
R_{ij} \simeq
-2 \pi \sum_{\alpha}
\omega_{p,\alpha}^2
\int_o^{\infty} dp_{\perp}
\int_{-\infty}^{\infty}
dp_{\Vert} \Big\{
p_{\perp}^2 \Lambda^{\alpha}_{ij} \Big[ (\omega -k_{\Vert}
v_{\Vert} )
\cdot {{\partial \hat{f}_{\alpha}(p) }
\over{\partial p_{\perp}}}
+ k_{\Vert} v_{\perp} {{\partial \hat{f}_{\alpha}(p)
}\over{\partial p_{\Vert}}}
\Big] +
m_{\alpha} p_{\Vert} \Big(v_{\perp} {{\partial
\hat{f}_{\alpha}(p)
}\over{\partial p_{\Vert}}}
-v_{\Vert } {{\partial \hat{f}_{\alpha}(p)
}\over{\partial p_{\perp}}}
\Big) \delta_{i3} \delta_{j3} \Big\}
\label{rij}
\end{equation}

\noindent
where

\begin{equation}
\Lambda^{\alpha}_{ij}=
{1 \over{2 \Omega_o^{\alpha}}}
\pmatrix{ X_1^{\alpha}  & -{\it i} &
- \Big({{p_{\Vert} }\over{p_{\perp} }} \Big)
z_{\alpha}
\cr
{\it i} & X_1^{\alpha} - {{z_{\alpha}^2}\over
{2 X_1^{\alpha} }} &
-{\it i} \Big( {{ p_{\Vert} }\over{ p_{\perp} }} \Big)
{{ z_{\alpha} }\over{ X_1^{\alpha} }} \cr
- \Big( {{ p_{\Vert} }\over{ p_{\perp} }} \Big)
z_{\alpha} &
{\it i} \Big( {{p_{\Vert} }\over{ p_{\perp} }} \Big)
{{ z_{\alpha} }\over{ X^{\alpha}_1}} &
{{ z_{\alpha}^2 -2 }\over{X_1^{\alpha}}}
\Big( {{ p_{\Vert} }\over{ p_{\perp} }} \Big)^2
\cr} +
{\cal O}(z_{\alpha}^2)
\label{lambdaij}
\end{equation}

\noindent
and where the term
${\cal O}(z_{\alpha}^2)$ comes from the contribution
from the $n\geq 2$ resonances in Eq.(\ref{dielectrictensor1}).

\noindent
In an isotropic plasma, Eqs.(\ref{rij}--\ref{lambdaij}) can be
further simplified by introducing the total energy of species
$\alpha$ :

\begin{equation}
{\cal E}_{\alpha}=
2 \pi N_{\alpha} \int \int
dp_{\perp} p_{\perp}
dp_{\Vert} \hat{f}_{\alpha}(p)
m_{\alpha} c^2 \gamma \, ,
\label{energy_bs}
\end{equation}

\noindent
and the pressure of species $\alpha$ :

\begin{equation}
{P}_{\alpha} =
{P}_{\perp}^{\alpha}=
\pi N_{\alpha}
\int \int
dp_{\perp} p_{\perp}
dp_{\Vert}
\hat{f}_{\alpha}(p)
p_{\perp} v_{\perp}=
{P}_{\Vert}^{\alpha}=
2 \pi N_{\alpha}
\int \int
dp_{\perp} p_{\perp}
dp_{\Vert}
\hat{f}_{\alpha}(p)
p_{\Vert} v_{\Vert} \, .
\label{p_bs}
\end{equation}

\noindent
which is $P_{\alpha}= N_{\alpha} k_B T$ for a Maxwellian
distribution of $\alpha$ particles.

By integrating Eqs.~\ref{rij}--\ref{lambdaij}, 
introducing Eqs.(\ref{energy_bs}--\ref{p_bs}), 
and requiring no net charge in the
plasma (i.e., $\sum_{\alpha} N_{\alpha} e_{\alpha} =0$),
the components of the tensor are given by :

\begin{equation}
R_{11}= - {{4 \pi k_{\Vert}^2 c^2}\over{B_o^2}} 
\left( {{\omega}\over{k_{\Vert}c}} \right)^2
\sum_{\alpha}
\left( {\cal E}_{\alpha} + {P}_{\alpha} \right)
\label{r11}
\end{equation}

\begin{equation}
R_{12}= R_{21}= 0
\label{r12}
\end{equation}

\begin{equation}
R_{13}= R_{31}=0
\label{r13}
\end{equation}

\begin{equation}
R_{22}=R_{11}+
{{8 \pi k_{\perp}^2 c^2 }\over{B_o^2}} \sum_{\alpha} {P}_{\alpha}
- \sum_{\alpha} {{ \pi  k_{\perp}^2 N_{\alpha} }\over{ B_o^2 }}
\langle p_{\perp} v_{\perp}^3 \rangle_{\alpha}
- {{ 2 \pi^2 k_{\Vert} k_{\perp}^2 c^2 }\over{B_o^2}}
\sum_{\alpha} N_{\alpha} \int \int p_{\perp} dp_{\perp}
dp_{\Vert} {{ p_{\perp}^2 v_{\perp}^2}\over{\omega -k_{\Vert}
v_{\Vert}}}
{{\partial \hat{f}_{\alpha}(p)
}\over{\partial p_{\Vert}}}
\label{r22}
\end{equation}

\begin{equation}
R_{23}=-R_{32}=
-{{4 \pi^2 {\it i} k_{\perp} c }\over{B_o}}
\sum_{\alpha} N_{\alpha} e_{\alpha} \omega
\int \int p_{\perp} dp_{\perp}
dp_{\Vert} {{ p_{\perp} v_{\perp}}\over{\omega -k_{\Vert}
v_{\Vert}}}
{{\partial \hat{f}_{\alpha}(p)
}\over{\partial p_{\Vert}}}
\label{r23}
\end{equation}

\begin{equation}
R_{33}=
- {{8 \pi^2 \omega^2 }\over{k_{\Vert}}}
\sum_{\alpha} N_{\alpha} e_{\alpha}^2
\int \int p_{\perp} dp_{\perp}
dp_{\Vert} {{ 1 }\over{\omega -k_{\Vert}
v_{\Vert}}}
{{\partial \hat{f}_{\alpha}(p)
}\over{\partial p_{\Vert}}}
\label{r33}
\end{equation}

\noindent
which give all the components of the dielectric tensor
when inserted in Eq.(\ref{dielectrictensor}).

\noindent
As usual, the integrals in Eqs.(\ref{r22}--\ref{r33})
can be calculated using the {\it Sokhotskii--Plamelj} formula
by taking into account the causal condition
(e.g., Melrose 1968) :

\begin{equation}
{1 \over{\omega - k_{\Vert} v_{\Vert} + {\it i}0}}
= {\cal P} {1 \over{\omega - k_{\Vert} v_{\Vert} }}
- {\it i} \pi \delta\Big(\omega - k_{\Vert} v_{\Vert} \Big)
\label{plamelj}
\end{equation}

\noindent
where ${\cal P}$ is the {\it Cauchy principal value}, and
${\it i}0$ is an infinitesimal imaginary term.

\section{Energy of the Mode}

\noindent
The energy of the mode in a collisionless plasma
is given by (e.g., Melrose 1968; Barnes 1968): 

\begin{equation}
W(k,\omega)=
{1\over{16 \pi}}
\Big[
{B^{*}_k}_i {B_k}_i +
{E_k}_i^* 
{{\partial }\over{\partial \omega}}
\Big( \omega K_{ij}^h \Big) 
{E_k}_j \Big]_{\omega_i =0}
\label{ww_app}
\end{equation}

\noindent
where $K_{ij}^h$ stands for the Hermitian part of
the dielectric tensor (\ref{dielectrictensor}).
In the weak damping limit 
(i.e. ${\it Im}(\omega) << 1$) the Hermitian part of the
dielectric tensor (Eq.~\ref{dielectrictensor})
can be expressed as :

\begin{equation}
K^h_{ij} \simeq \delta_{ij}
-\sum_{\alpha} \Big(
{\cal M}_{ij,\alpha}^h (\omega_r) + {\it Im}(\omega) 
{{ \partial {\cal M}_{ij,\alpha}^h (\omega_r) }\over
{\partial \omega }} \Big)
\label{dielectrictensor_l}
\end{equation}

\begin{figure}
\resizebox{\hsize}{!}{\includegraphics{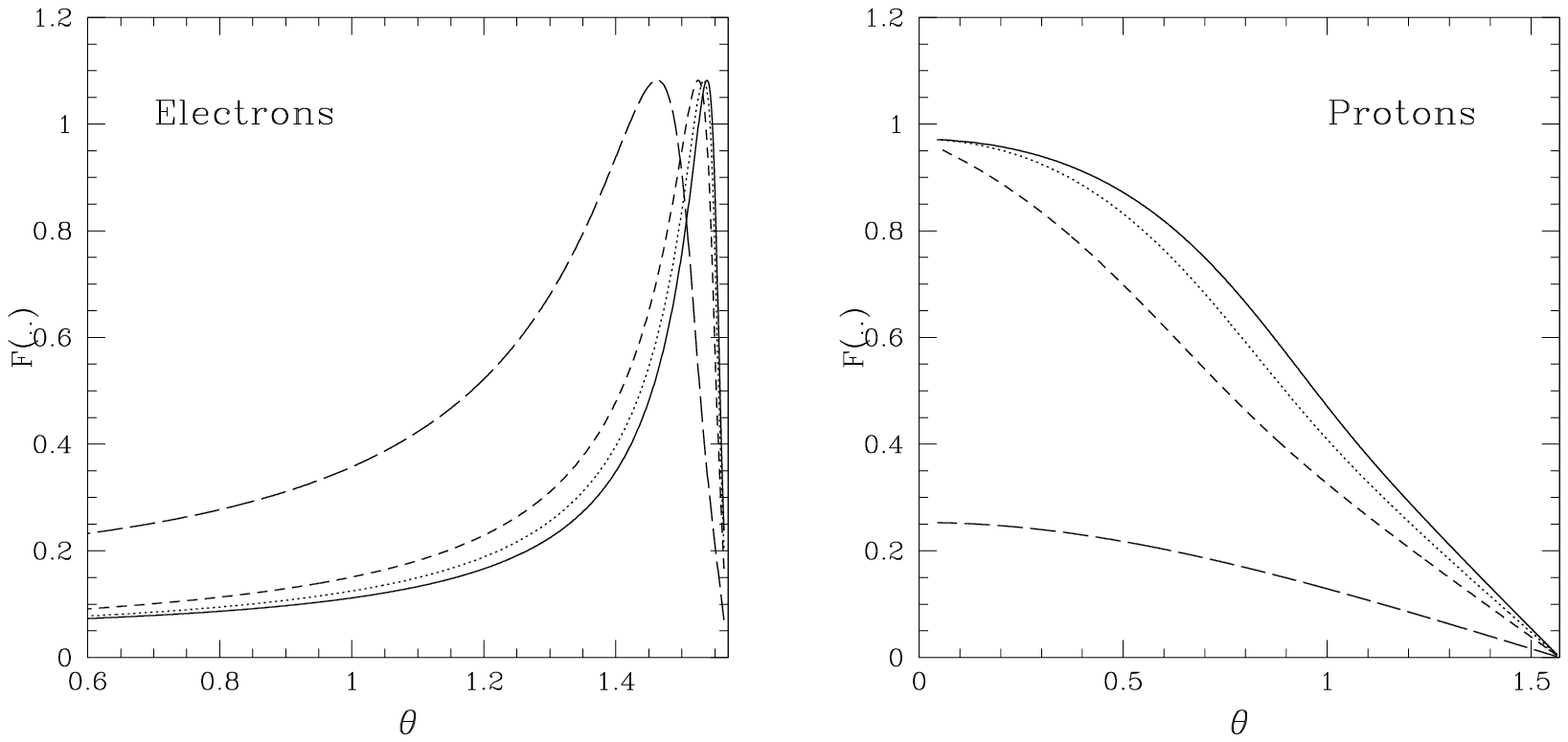}}
\caption[]{
The function $F$ is reported, as a function of the mode--propagation 
angle $\theta$, for electrons (left panel) and
for protons (right panel).
Results are shown for different values of 
$c_s^2/v_A^2 (=\beta_{pl}/2)$: 100 (solid lines), 3 (dotted lines),
1 (dashed lines), 0.1 (long--dashed lines).
In the calculations $k_B T = 8.6$ keV is assumed.
}
\label{fig:app_b1}
\end{figure}

\begin{figure}
\resizebox{\hsize}{!}{\includegraphics{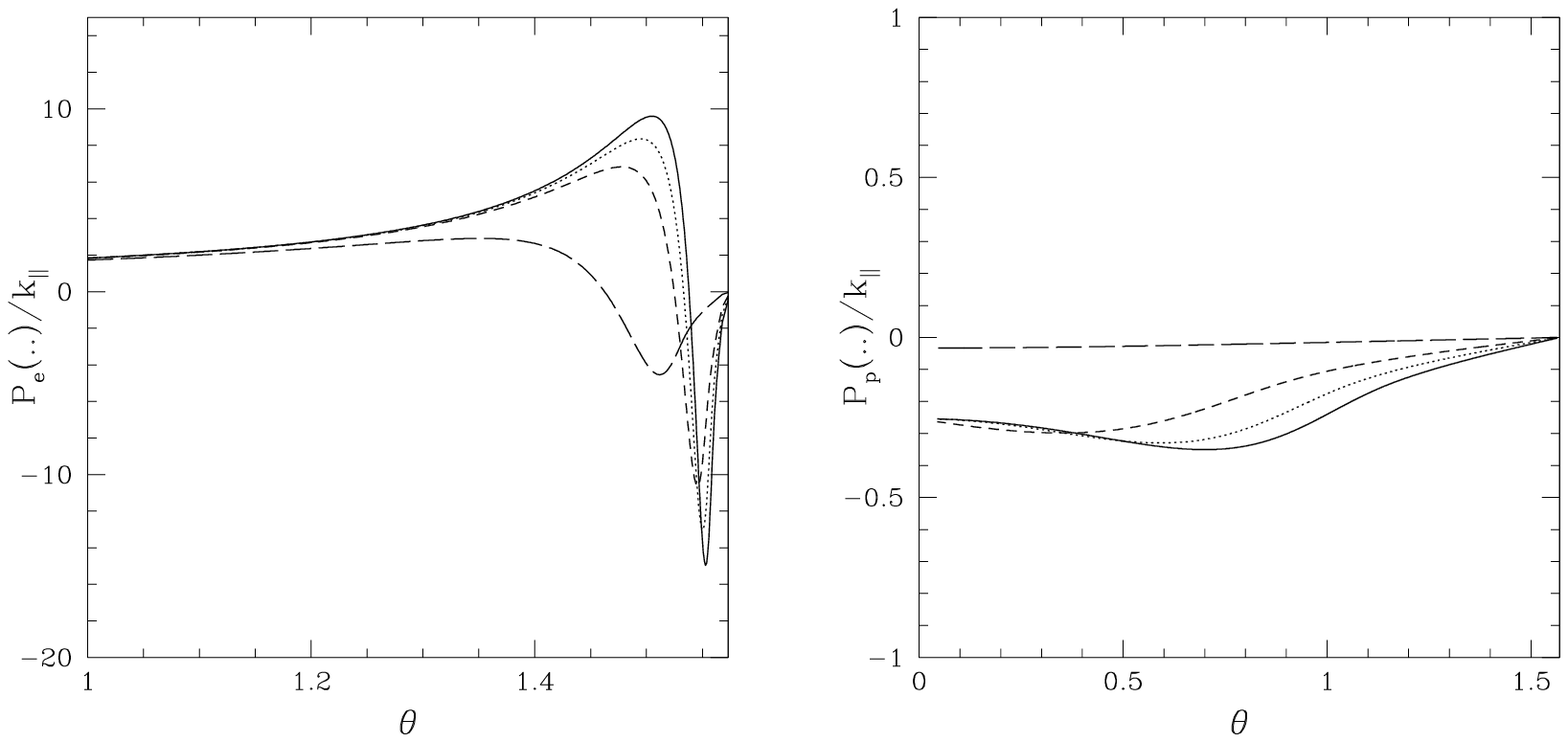}}
\caption[]{
The function ${\cal P}({\ldots})/k_{\Vert}$ 
is reported, as a function of the mode--propagation 
angle $\theta$, for electrons (left panel) and
protons (right panel).
Assumptions and line--styles are the same as 
in Fig.\ref{fig:app_b1}. 
}
\label{fig:app_b2}
\end{figure}

\noindent
where the components of the tensor 
${\cal M}_{ij,\alpha}= (R_{ij}^{\alpha} + R_{ji}^{\alpha})/2\omega^2$ 
are given by Eqs.(\ref{r11}--\ref{r33}).
We note that the x--component 
of the electric field associated with the mode
(and its spatial Fourier transform)
is =0 (Eq.\ref{E}), 
and thus only the components $K^h_{22}$, $K^h_{23}$, $K^h_{32}$, 
and $K^h_{33}$ contribute to Eq.(\ref{ww_app}).
The Hermitian part of the relevant components
of the tensor ${\cal M}_{ij}$ are
obtained from Eqs.(\ref{r22}--\ref{r33}, \& \ref{plamelj})
and Eq.(\ref{dielectrictensor_l}) :

\begin{equation}
{\cal M}_{22,\alpha}^h =
- {{4 \pi }\over{B_o^2}} \Big({\cal E}_{\alpha} + P_{\alpha} \Big)
+ \Big( {{k_{\perp} c }\over{\omega_r}} \Big)^2 
\Big(
{{8 \pi P_{\alpha} }\over{B_o^2}} 
- {{\pi N_{\alpha} \langle p_{\perp} v_{\perp}^3 \rangle }\over
{B_o^2 c^2 }} 
\Big)
- 2 \pi^2 {{ k_{\Vert} k_{\perp}^2 N_{\alpha} c^2 }\over
{B_o^2 \omega_r^2 }}
\int_0^{\infty} dp_{\perp} p_{\perp}^3 v_{\perp}^2
\left(
{\cal P}\int {{ dp_{\Vert}}\over{\omega_r - k_{\Vert} v_{\Vert}}}
{{ \partial \hat{f}(p) }\over{\partial p_{\Vert}}} \right)_{\alpha} \, ,
\label{m22}
\end{equation}

\begin{equation}
{\cal M}_{23,\alpha}^h = - {\cal M}_{32,\alpha}^h =
- 4 \pi^2 {{ k_{\perp} N_{\alpha} c e_{\alpha} }\over
{B_o \omega_r }}
{\it i} \int_0^{\infty} dp_{\perp} p_{\perp}^2 v_{\perp} 
\left(
{\cal P}\int {{ dp_{\Vert}}\over{\omega_r - k_{\Vert} v_{\Vert}}}
{{ \partial \hat{f}(p) }\over{\partial p_{\Vert}}} \right)_{\alpha} \, ,
\label{m23}
\end{equation}

\begin{equation}
{\cal M}_{33,\alpha}^h =
- {{8 \pi^2 N_{\alpha} e_{\alpha}^2}\over{k_{\Vert}}}
\int_0^{\infty} dp_{\perp} p_{\perp}
\left(
{\cal P}
\int {{ dp_{\Vert}}\over{\omega_r - k_{\Vert} v_{\Vert}}}
{{ \partial \hat{f}(p) }\over{\partial p_{\Vert}}} \right)_{\alpha}
\label{m33}
\end{equation}

\noindent
The energy spectrum of the mode (Eq.\ref{ww_app}) is thus given
by :

\begin{equation}
W(k,\omega)=
{1\over{16 \pi}}
\Big\{
|B_k|^2 + |E_k|^2
-|E_{\perp}|^2 {{\partial}\over{\partial \omega}}
\Big(\omega \sum_{\alpha}{\cal M}_{22,\alpha}^h \Big)
-|E_{\Vert}|^2 {{\partial}\over{\partial \omega}}
\Big(\omega \sum_{\alpha}{\cal M}_{33,\alpha}^h \Big)
- \Big[ E_{\perp}^* E_{\Vert} - E_{\perp} E_{\Vert}^* \Big]
{{\partial}\over{\partial \omega}} \Big(\omega \sum_{\alpha}
{\cal M}_{23,\alpha}^h \Big)
\Big\}
\label{ww2}
\end{equation}

\noindent
Here it is necessary to evaluate the ratio between the parallel
and perpendicular fluctuations of the electric field in
Eq.(\ref{ww2}).
In the low amplitude regime the Fourier--Laplace transform
of the mode electric field satisfies ${\bf \Psi} \cdot
{\bf E}({\bf k},\omega)=0$, where $\Psi_{ij}=R_{ij}+
c^2 k_i k_j - (k^2 c^2 -\omega^2)\delta_{ij}$ is the 
{\it Maxwell operator} (e.g., Melrose 1968; BS73;
Schlickeiser 2002).
Thus since it is ${E_k}_1=0$, one has :

\begin{equation}
{{E_{\Vert}}\over{{E_{\perp}}}}=
-
{{
\Psi_{12} + \Psi_{22} + \Psi_{32}}\over{
\Psi_{13} + \Psi_{23} + \Psi_{33}}}
\label{e/e_1}
\end{equation}

\noindent
A dimensional analysis of {\it Maxwell operator} 
in case of the long--wavelength modes gets
$\Psi_{32} >> \Psi_{22} >> \Psi_{12}$ and
$\Psi_{33} >> \Psi_{23} >> \Psi_{13}$ (BS73), 
and thus the ratio between the perpendicular and
parallel component of the mode electric field is :

\begin{equation}
E_{\perp} \simeq - E_{\Vert} 
{{ R_{33} }\over
{ R_{32} }}
\label{e/e}
\end{equation}

\noindent
which can be used in Eq.(\ref{ww2}) in combination
with Eqs.(\ref{r23}) and (\ref{r33}).
Here it should be noticed that it is
$|E_{\perp}|^2/|E_{\Vert}|^2 \approx |R_{33}|^2/
|R_{32}|^2 >> 1$ and thus that the fluctuations
of the electric field are perpendicular to $B_o$, 
and to the fluctuations of the magnetic field 
(see Eqs.\ref{k}--\ref{B}).
On the other hand, this does not immediately imply 
that the contribution from $|E_{\perp}|^2$ in
Eq.(\ref{ww2}) dominate on that from $|E_{\Vert}|^2$: 
a dimensional analysis of the elements of the
tensor $R_{ij}$ indeed shows that
$|E_{\perp}|^2 {\cal M}_{22}$ is of the same order of
$|E_{\Vert}|^2 {\cal M}_{33}$.

The most important contribution of particles
to the spectrum of the mode 
in the ICM (as in many other astrophysical cases)
is provided by thermal electrons and protons
which dominates the energy budget of the plasma.
In this case the particle distribution function of
electrons and protons is Maxwellian :

\begin{equation}
f_{\alpha}(p)= N_{\alpha} \hat{f}_{\alpha}(p)=
{{ N_{\alpha} }\over{(2 \pi)^{3/2}}} 
{{ \exp\{-p^2/(2m_{\alpha} k_B T)\}}\over{ (m_{\alpha} k_B T)^{3/2} }}
\label{maxw_app}
\end{equation}

\noindent
and, from Eqs.\ref{m22}--\ref{m33},
the components of the Hermitian part
of the dielectric tensor become :

\begin{equation}
{\cal M}_{22,\alpha}^h =
- {{4 \pi }\over{B_o^2}} \Big({\cal E} + P \Big)_{\alpha}
+ \Big( {{k_{\perp} c }\over{\omega_r}} \Big)^2
\Big(
{{8 \pi P_{\alpha} }\over{B_o^2}}
- {{\pi N_{\alpha} \langle p_{\perp} v_{\perp}^3 \rangle }\over
{B_o^2 c^2 }}
\Big)
+{{A^{\alpha}_{22}}\over{\omega^2}}
{\cal I}_{\alpha}
 M^{\alpha}_{22} +
{{A^{\alpha}_{22}}\over{\omega_r^2}}
{\cal I}_{\alpha}
\label{m22_m}
\end{equation}

\begin{equation}
{\cal M}_{23,\alpha}^h =
{{A_{23}^{\alpha} }\over{\omega}} {\it i} 
{\cal I}_{\alpha}
\label{m23_m}
\end{equation}

\begin{equation}
{\cal M}_{33,\alpha}^h =
A_{33}^{\alpha}
{\cal I}_{\alpha}
\label{m33_m}
\end{equation}

\noindent
where ${\cal I}_{\alpha}$ stands for the Cauchy principal value
in Eqs.(\ref{m22}--\ref{m33}), and 
we put :

\begin{equation}
A^{\alpha}_{22}=
{{16 \pi^2}\over{(2 \pi)^{3/2}}}
{{ k_{\Vert} k_{\perp}^2 c^2 N_{\alpha} }\over
{B_o^2 m_{\alpha}^2 }}
\sqrt{m_{\alpha} k_B T} \, ,
\label{a22}
\end{equation}

\begin{equation}
A^{\alpha}_{23}= {{8 \pi^2}\over{(2\pi)^{3/2}}}
{{ N_{\alpha} e_{\alpha} k_{\perp} c }\over
{B_o m_{\alpha}^{3/2} k_B^{1/2} T^{1/2} }} \, ,
\label{a23}
\end{equation}

\noindent
and

\begin{equation}
A^{\alpha}_{33}= {{8 \pi^2}\over{(2\pi)^{3/2}}}
{{ N_{\alpha} e_{\alpha}^2 }\over
{(m_{\alpha} k_B T)^{3/2} k_{\Vert} }} \, .
\label{a33}
\end{equation}

\noindent
In the case of 
Maxwellian distributions 
the Cauchy principal value in Eqs.(\ref{m22_m}--\ref{m33_m})
reads :

\begin{figure}
\resizebox{\hsize}{!}{\includegraphics{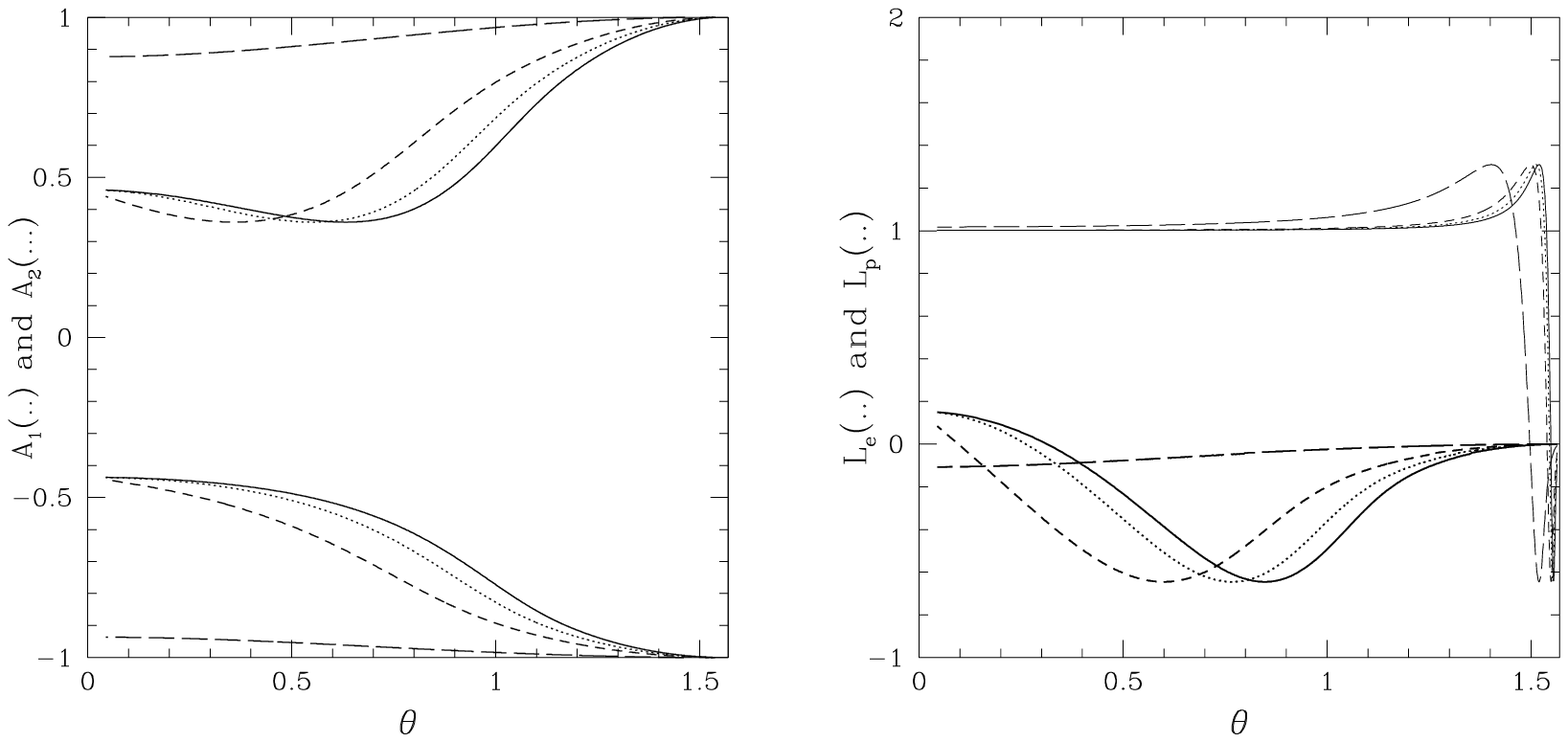}}
\caption[]{
{\bf Left Panel}: 
The expressions ${\cal A}_1/$ (bottom) and
${\cal A}_2$ (top) are reported  
as a function of the mode--propagation
angle $\theta$.

\noindent
{\bf Right Panel}: The function ${\cal L}$ is
reported, as a function of the mode--propagation
angle $\theta$, for electrons (thin lines) and
for protons (thick lines).

\noindent
In both panels, assumptions and line--styles are the same as
in Fig.\ref{fig:app_b1}.
}
\label{fig:app_b3}
\end{figure}

\begin{equation}
{\cal I}_{\alpha}=
\Big(
{\cal P} \int_{-\infty}^{\infty}
{{ dp_{\Vert} p_{\Vert} }\over{\omega_r - k_{\Vert} v_{\Vert} }}
\exp \{ -{{ p_{\Vert}^2}\over{2 m_{\alpha} k_B T}} \}
\Big)_{\alpha}
= 
- \sqrt{2\pi} 
\left( m_{\alpha} k_B T \right)^{1\over 2} {{m_{\alpha} }\over
{k_{\Vert}}} \Big\{ 1  - 
\tilde{\omega}_{\alpha} F(\tilde{\omega}_{\alpha}) 
\Big\}
\label{calP}
\end{equation}

\noindent
where we define 

\begin{equation}
F(\tilde{\omega})
\equiv 2 \exp\{-\tilde{\omega}_{\alpha}^2\}
\int_0^{\tilde{\omega}_{\alpha}}
dx \exp\{x^2\}
\rightarrow
\cases{
2 \tilde{\omega}_{\alpha} - {4\over 3}\tilde{\omega}_{\alpha}^3 
+ {8 \over {15}}
\tilde{\omega}_{\alpha}^5 \,\,\,\,\, {\rm for} \,\,\, \tilde{\omega} >> 1 \cr
     \cr
{1\over{\tilde{\omega}_{\alpha}}} +
{1\over{ 2 \tilde{\omega}_{\alpha}^3 }} +
{3 \over{4 \tilde{\omega}_{\alpha}^5}} \,\,\,\,\,\,\,\,\,\,\, {\rm for} 
\,\,\, \tilde{\omega} << 1
\cr}
\label{F}
\end{equation}

\noindent
which for real argument is $F(x)= - {\cal R}Z(x)$, and 

\begin{equation}
Z(x)= {1\over{\sqrt{\pi}}}
\int_{-\infty}^{\infty}
{{dt \exp\{-{t}^2\} }\over{t-x}}
\label{plasmaz}
\end{equation}

\noindent
is the well known 
{\it plasma dispersion function} (Fried \& Conte 1961;
Melrose 1968; Percival \& Robinson 1998).
The adimensional frequency, $\tilde{\omega}$, in 
Eq.~\ref{F} is defined (from Eq\ref{vph}) as :

\begin{equation}
\tilde{\omega}_{\alpha}(\beta_{pl},\theta)= 
\omega {{m_{\alpha} }\over{ k_{\Vert} }}
{1 \over {
\sqrt{2 m_{\alpha} K_B T}}} =
\sqrt{5\over3}
\left( {{k}\over{k_{\Vert}}} \right)
\left({{ m_{\alpha}}\over{m_p}} \right)^{1/2}
\left( {{ \beta_{pl}/2 +1}\over{ \beta_{pl} }} \right)^{1/2}
\left\{
1 +
\sqrt{ 1 - 4 \left( {{k}\over{k_{\Vert}}} \right)^2
{{ \beta_{pl}/2 }\over{ (1 + \beta_{pl}/2 )^2 }}}
\right\}^{1/2} \, .
\label{tildeomega}
\end{equation}

\begin{figure}
\resizebox{\hsize}{!}{\includegraphics{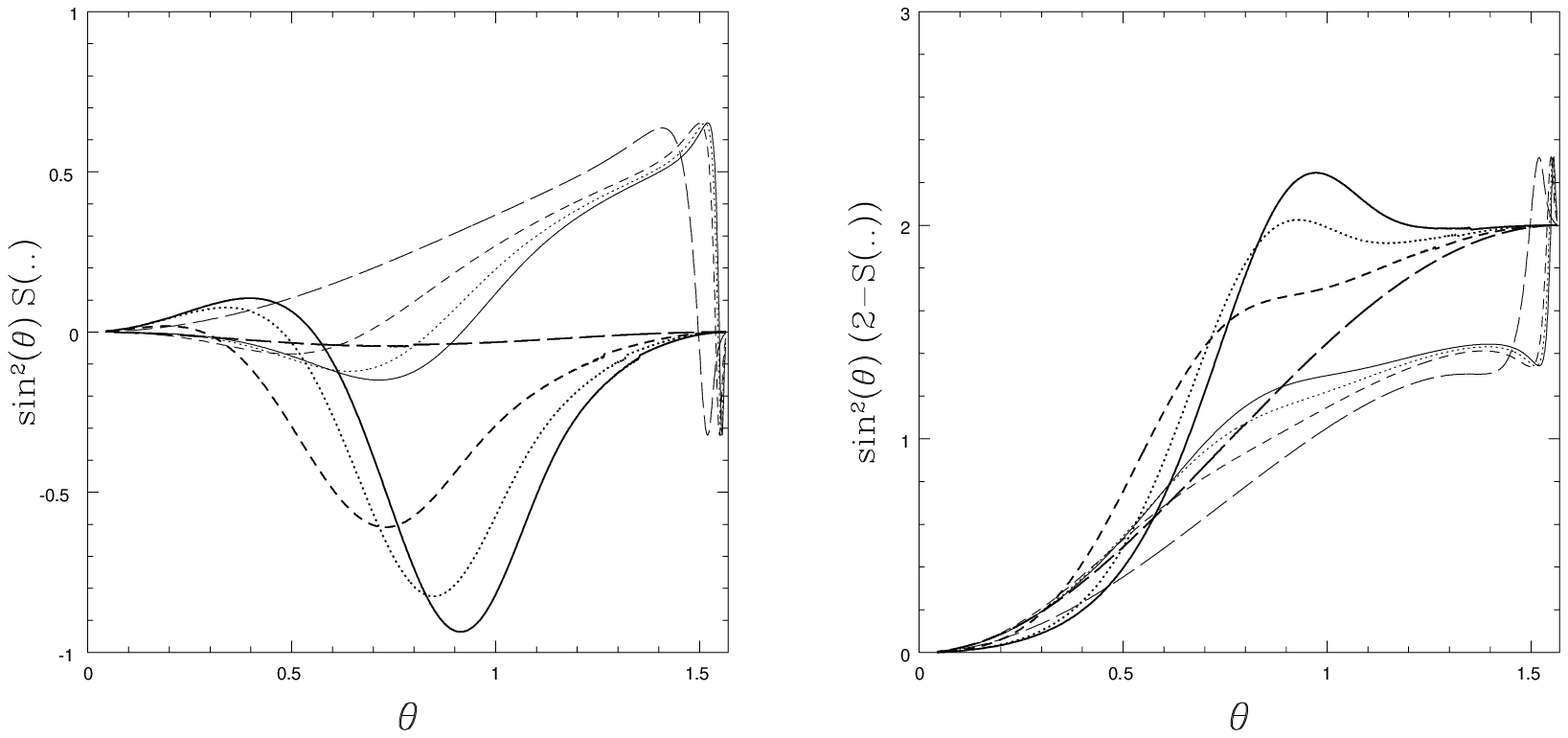}}
\caption[]{
{\bf Left Panel}:
The expression $k_{\perp}^2 {\cal S}(..)$
is reported 
as a function of the mode--propagation
angle $\theta$, for  electrons (thin lines) and
for protons (thick lines).

\noindent
{\bf Right Panel}:
The expression $k_{\perp}^2 (2-{\cal S})$
is reported
as a function of the mode--propagation
angle $\theta$, for  electrons (thin lines) and
for protons (thick lines).

\noindent
In both panels, assumptions and line--styles are the same as
in Fig.\ref{fig:app_b1}.

}
\label{fig:app_b4}
\end{figure}

\noindent
For $\tilde{\omega}_{\alpha} \sim 1$ the bulk of thermal
particles of species $\alpha$ undergoes $n=0$--resonance with
the mode.
The value of $\tilde{\omega}_{\alpha}$ increases with
increasing $\theta$ and goes to infinity with $\theta \rightarrow \pi/2$.
For a given ($\theta$, $\beta_{pl}$) the value of the adimensional 
frequency of electrons 
is about 40 times smaller than that of protons, thus electrons
experience $n=0$--resonance with the mode at larger angles than
protons.
The value of the adimensional frequency also depends on 
the $\beta_{pl}$: $\tilde{\omega}_{\alpha}$ increases with
decreasing $\beta_{pl}$, while $\tilde{\omega}_{\alpha} \rightarrow
\sqrt{5/3} \sqrt{m_{\alpha}/m_p}/\cos(\theta)$ for large $\beta_{pl}$.

\noindent
The behavior of the Cauchy principal value (Eq.~\ref{calP}) 
and of $F(\tilde{\omega}_{\alpha})$ is driven by the value
of $\tilde{\omega}_{\alpha}$.
In Fig.~\ref{fig:app_b1} we report $F(\tilde{\omega})$ for electrons
and protons for different values of $\beta_{pl}$. 
$F$ peaks at $\tilde{\omega} \sim 1$ and for electrons
this happens at larger $\theta$ than for protons.
With decreasing $\beta_{pl}$ the phase velocity of the mode 
increases and becomes significantly larger than 
the sound speed. This causes 
a shift of the peak of $F$ toward smaller $\theta$
in Fig.~\ref{fig:app_b1}, and also prevents
the $n=0$--resonance of the bulk of the protons in the case
of small $\beta_{pl}$.
In Fig.~\ref{fig:app_b2} we report the Cauchy principal value (Eq.~\ref{calP})
of thermal electrons and protons. 
The principal value goes to zero for $\theta \rightarrow \pi/2$
which essentially means that in this limit there is no particle--mode 
coupling via the $n=0$--resonance, and this is because formally
infinite particle's velocity is requested to resonate at these angles.
Also in this case the features of the curves are shifted at 
smaller angles with decreasing $\beta_{pl}$.

Given these results, 
the spectrum of the mode (Eq.~\ref{ww2})
in a magnetized--collisionless plasma
can thus be obtained in explicit form.
After some tedious algebra, from Eq.(\ref{ww2}),
Eqs.(\ref{m22_m}--\ref{F}), and Eq.(\ref{e/e}) we
find:

\begin{eqnarray}
W(k,\omega)=
{{ |B_k |^2 }\over{16 \pi}}
\Big\{
1 +
{{ \left( \omega / kc \right)^2 }\over{
1 - \left( {{k_{\Vert} }\over k} \right)^2
{1\over{1+ \omega^2 \langle {{ A_{33}^2 }\over{ A_{23}^2}} \rangle}} }}
\Big[
1 + {{ \omega^2 \langle {{ A_{33}^2 }\over{ A_{23}^2}} \rangle }\over{
1+ \omega^2 \langle {{ A_{33}^2 }\over{ A_{23}^2}} \rangle }}
\sum_{\alpha}
\Big\{
{{ N_{\alpha} m_{\alpha} c^2 + P_{\alpha} }\over{
B_o^2/4\pi}}
+{{P_{\alpha} }\over{
B_o^2/8 \pi}} \Big( {{ k_{\perp} c}\over{\omega}} \Big)^2
\times \nonumber\\
\Big[
1 - {\cal L}_{\alpha}
\left( 1 - {{ \Delta_{\alpha}^{{\cal L}}}\over{
2 {\cal A}_2(\beta_{pl},\theta) }}
\right)
-
{{N_{\alpha} \langle p_{\perp} v_{\perp}^3 \rangle_{\alpha} }\over{
8 P_{\alpha} c^2 }} \Big]
\Big\} \Big] \Big\}
\label{w_max}
\end{eqnarray}

\noindent
where we put 

\begin{equation}
{\cal A}_2(\beta_{pl},\theta)=
{{ (1 - \tilde{\omega}_p F_{\tilde{\omega}_p} )^2
\left( 1 +
{{1 - \tilde{\omega}_e F_{\tilde{\omega}_e} }\over{
1 - \tilde{\omega}_p F_{\tilde{\omega}_p} }}
\right)^2
+
{{ 5 \pi}\over{ 3}}
{{ V_{\rm ph}^2}\over{c_s^2}}
\big( {{k}\over{k_{\Vert}}} \big)^2
\exp\{ -2 \tilde{\omega}_p^2 \}
\left( 1 - ({{m_e}\over{m_p}})
\exp\{2 \Delta\tilde{\omega}_{p-e}^2 \}
\right) }\over{
(1 - \tilde{\omega}_p F_{\tilde{\omega}_p} )^2
\left[
1 -{{1 - \tilde{\omega}_e F_{\tilde{\omega}_e} }\over{
1 - \tilde{\omega}_p F_{\tilde{\omega}_p} }}
\right]^2+
{{ 5 \pi}\over{ 3}}
{{ V_{\rm ph}^2}\over{c_s^2}}
\big( {{k}\over{k_{\Vert}}} \big)^2
\exp\{ -2 \tilde{\omega}_p^2 \}
\left( 1 - ({{m_e}\over{m_p}})^{1\over 2}
\exp\{\Delta\tilde{\omega}_{p-e}^2 \} \right)^2 }} 
\label{<a33/a23>_ICM1}
\end{equation}

\noindent
where $\Delta\tilde{\omega}_{p-e}^2 = \tilde{\omega}_p^2
-\tilde{\omega}_e^2$, and 

\begin{equation}
\langle {{ A_{33}^2}\over{A_{23}^2}} \rangle=
{{ e_p^2 B_o^2}\over{( k_{\perp} c)^2
(k_B T)^2 k_{\Vert}^2}}
{\cal A}_2(\beta_{pl},\theta) \, ,
\end{equation}

\noindent
and where 

\begin{equation}
\Delta_{\alpha}^{{\cal L}}(\beta_{pl},\theta)
=
{{ 1 - \tilde{\omega}_{\alpha} F_{\tilde{\omega}_{\alpha}}
( 1- a_{\alpha} )
+ 2 a_{\alpha} \tilde{\omega}_{\alpha}^2
(1- \tilde{\omega}_{\alpha} F_{\tilde{\omega}_{\alpha}} )
}\over{{\cal L}_{\alpha}(\beta_{pl},\theta)
}} \, ,
\label{delta_l}
\end{equation}

\noindent
with $a_{\alpha}$ defined as :

\begin{equation}
a_e = -1 - 2 {\cal A}_1 
\label{ae}
\end{equation}

\noindent
and

\begin{equation}
a_p = -1 + 2 {\cal A}_1 
\label{ap}
\end{equation}

\noindent
in the case of electrons and protons, respectively, 

\begin{equation}
{\cal A}_1(\beta_{pl},\theta) =
{{  (1 - \tilde{\omega}_p F_{\tilde{\omega}_p} )^2
\left[ 1 - \left(
{{1 - \tilde{\omega}_e F_{\tilde{\omega}_e} }\over{
1 - \tilde{\omega}_p F_{\tilde{\omega}_p} }}
\right)^2 \right]
+
{{ 5 \pi}\over{ 3}}
{{ V_{\rm ph}^2}\over{c_s^2}}
\big( {{k}\over{k_{\Vert}}} \big)^2
\exp\{ -2 \tilde{\omega}_p^2 \}
\left( 1 - ({{m_e}\over{m_p}})
\exp\{2 \Delta\tilde{\omega}_{p-e}^2 \}
\right) }\over{
(1 - \tilde{\omega}_p F_{\tilde{\omega}_p} )^2
\left[
1 -{{1 - \tilde{\omega}_e F_{\tilde{\omega}_e} }\over{
1 - \tilde{\omega}_p F_{\tilde{\omega}_p} }}
\right]^2+
{{ 5 \pi}\over{ 3}}
{{ V_{\rm ph}^2}\over{c_s^2}}
\big( {{k}\over{k_{\Vert}}} \big)^2
\exp\{ -2 \tilde{\omega}_p^2 \}
\left( 1 - ({{m_e}\over{m_p}})^{1\over 2}
\exp\{\Delta\tilde{\omega}_{p-e}^2 \} \right)^2 }} 
\label{a1}
\end{equation}

\noindent
and where

\begin{equation}
{\cal L}_{\alpha}(\beta_{pl},\theta)=
1+
2 \tilde{\omega}_{\alpha}^2
\Big( 1 - \tilde{\omega}_{\alpha} F(\tilde{\omega}_{\alpha})
\Big) \, .
\label{l}
\end{equation}

\noindent
The terms ${\cal A}_1$, ${\cal A}_2$, and ${\cal L}_{\alpha}$
are reported in Fig.~\ref{fig:app_b3} for different values
of $\beta_{pl}$.

From Eq.(\ref{w_max}) with $(\omega^2 \langle A_{33}^2/A_{22}^2 \rangle) >>1$,
and $\omega = V_{\rm ph} k$, one gets the expression for the  
ratio between energy density in the magnetic field fluctuations
and total energy density of the mode which is used in this paper :

\begin{equation}
{{ | B_k |^2 }\over{W(k,\theta)}}
\simeq
16 \pi
\Big\{
1 +
{{\beta_{pl}}\over{2}}
\big[
\left( {{V_{\rm ph}}\over{c_s}} \right)^2 +
{3\over 5} \Big( {{k_{\perp}}\over{k}} \Big)^2
\Big( 2 - {\cal S}(\beta_{pl},\theta) \Big)+
{1\over{\beta_{pl}}}
\left( {{ V_{\rm ph} }\over{c}} \right)^2
\left( {3\over 5} \beta_{pl} +2 \right) 
\big]
\Big\}^{-1}
\label{B/W}
\end{equation}

\noindent
where 

\begin{equation}
{\cal S}(\beta_{pl},\theta; \{ f_{rel}(p),T \})=
\sum_{\alpha = e,p}
{\cal L}_{\alpha}(\beta_{pl},\theta)
\left( 1 - 
{{\Delta_{\alpha}^{{\cal L}}}\over{
2 {\cal A}_2(\beta_{pl},\theta)}}
\right)-
{{N_{\alpha} \langle p_{\perp} v_{\perp}^3 \rangle_{\alpha} }\over{
8 P_{\alpha} c^2 }}
\label{s}
\end{equation}

\noindent
is reported in Fig.~\ref{fig:app_b4} for different values of
$\beta_{pl}$; the expression for $|B_k|^2/W$ (Eqs.~\ref{B/W}--\ref{s})
is important in our calculations since it allows to obtain 
the value of the TTD--damping rate and of 
the TTD--acceleration efficiency (\S.~4.3--5.1).

\noindent
Both ${\cal L}$ and ${\cal S}$ goes to zero for $\theta \rightarrow
\pi/2$ (Figs.~\ref{fig:app_b3} \& \ref{fig:app_b4})
which means that there is no contribution to the
energy of the mode via $n=0$--resonance.
Also at small $\theta$ the contribution to the energy of
the mode due to particle--mode coupling via $n=0$--resonance goes to zero
(Fig.\ref{fig:app_b4}), and this is because in case of 
parallel propagation of the mode the compressible 
part (parallel) of the magnetic field fluctuations 
which drives the $n=0$--resonance goes to zero,  
$B_3 = c k_{\perp} E_{\perp}/\omega$ (Eq.~\ref{B}).
Finally, the same arguments used to comment Figs.~\ref{fig:app_b1}
\& \ref{fig:app_b2} can be used here to explain 
the evolution of the behavior
of ${\cal L}$ and ${\cal S}$ with $\beta_{pl}$.

\section{Damping coefficients}

\noindent
The damping coefficient of the modes
can be obtained by
the standard formula for the linear growth rate of the
modes in the weak damping approximation (e.g., Melrose 1968; BS73):

\begin{equation}
\Gamma =
- {\it i} \Big(
{{ E_i^* K^a_{ij} E_j }\over{ 16 \pi W }}
\Big)_{\omega_i =0}\omega_r
\label{damping_start}
\end{equation}

\noindent
where $ K^a_{ij}$ stands for the anti--Hermitian part of
the dielectric tensor which can be directly obtained from
Eqs.(\ref{dielectrictensor1}--\ref{VV}); here we closely follow the
approach in BS73.
From the {\it Sokhotskii--Plamelj} formula
(Eq.\ref{plamelj}), from Eq.(\ref{VV}) one has (e.g., BS73):

\begin{eqnarray}
E_i^*
\left(
{{ V_i V_j^*  }\over
{\omega - n \Omega - k_{\Vert} v_{\Vert} }}
\right)^a_{\alpha} E_j \,\,\,
\rightarrow \,\,
- {\it i} \pi {{\gamma}\over{k_{\Vert}}}
\delta \Big( \omega_r {{\gamma}\over{k_{\Vert}}}
- n \Omega_{o,\alpha} k_{\Vert}^{-1}
-{{p_{\Vert}}\over{m_{\alpha}}} \Big)
\Big\{v_{\Vert}^2 J_n(z_{\alpha})^2 |E_{\Vert}|^2
+{\it i} v_{\perp} v_{\Vert} J_n(z_{\alpha}) J^{\prime}_n(z_{\alpha})
\left( E_{\perp} E_{\Vert}^* - E_{\perp}^* E_{\Vert} \right)
\nonumber\\
+ v_{\perp}^2 (J^{\prime}_n(z_{\alpha}))^2 |E_{\perp}|^2
\Big \}
= - {\it i} \pi v_{\perp}^2 {{\gamma}\over{k_{\Vert}}}
\delta \Big( \omega_r {{\gamma}\over{k_{\Vert}}}
- n \Omega_{o,\alpha} k_{\Vert}^{-1}
-{{p_{\Vert}}\over{m_{\alpha}}} \Big)
\left| {\it i} J^{\prime}_n(z_{\alpha}) E_{\perp}
+ {{ p_{\Vert} }\over{p_{\perp}}} J_n(z_{\alpha}) E_{\Vert}
\right|^2
\label{evve}
\end{eqnarray}

\noindent
By making use of the properties of the
$\delta$--functions, 
from Eq.(\ref{damping_start}), Eq.(\ref{dielectrictensor1}),
and Eq.(\ref{evve}) one has :

\begin{equation}
\Gamma =
-{{ \pi}\over{{16} \omega_r W}}
{{k_{\Vert}}\over{|k_{\Vert}|}}
\sum_{\alpha , n}
\omega_{p,\alpha}^2
\int_0^{\infty}
dp_{\perp}
\int_{-\infty}^{\infty}
dp_{\Vert} p_{\perp}^2
\Psi_n^{\alpha}
\Big\{
\left(
{{\omega}\over{k_{\Vert}}}
- v_{\Vert} \right)
{{\partial \hat{f}_{\alpha}(p)
}\over{\partial p_{\perp}}}
+
v_{\perp}
{{\partial \hat{f}_{\alpha}(p)
}\over{\partial p_{\Vert}}}
\Big\} \,\,\,
\delta \left({{p_{\Vert} }\over{m_{\alpha}}}
+
{{ n \Omega_{o,\alpha} - \omega_r \gamma }\over
{k_{\Vert}}}
\right)
\label{damping_1}
\end{equation}

\noindent
where we define 

\begin{equation}
\Psi_n^{\alpha}=
2 \Big|
{\it i} J^{\prime}_n(z_{\alpha}) E_{\perp}
+ {{ p_{\Vert} }\over{p_{\perp}}} J_n(z_{\alpha}) E_{\Vert}
\Big|^2
\label{theta_n}
\end{equation}

\noindent
Here we focus on the TTD case, $n=0$,
which is the most important resonance of 
long wavelength ($z_{\alpha}<<1$) fast modes and 
magnetosonic waves.
In this case it is :

\begin{equation}
\Psi_o^{\alpha}
\buildrel {z_{\alpha}<<1} \over 
\longrightarrow
2 \Big|
{{ p_{\Vert} }\over{p_{\perp}}} E_{\Vert}
- {\it i} {{ k_{\perp} v_{\perp} \gamma }\over{ 2 \Omega_o}}
E_{\perp} \Big|^2
\label{theta_n=0}
\end{equation}

\noindent
and by using the properties of the $\delta$--functions 
in Eq.(\ref{damping_1}) we obtained a general formula for the
damping rate (TTD) with particles
of $\alpha$--species :

\begin{eqnarray}
\Gamma_{\alpha}(k)=
- {{\pi }\over{ 32 }}
{{|E_{\perp}|^2}\over{W }} 
{{ k_{\Vert} k_{\perp}^2}\over{ |k_{\Vert}|}}
{{\omega_{p,\alpha}^2 }\over{\omega}}
{\cal H}\left( 1 - | {{ \omega}\over{ k_{\Vert} c}} | \right)
{{ \sqrt{1-({{\omega}\over{k_{\Vert}c}})^2} }\over
{ m_{\alpha}^2 \Omega^2_{o,\alpha} }}
\int_0^{\infty}
{{ dp_{\perp} \, p_{\perp}^3 }\over{
\sqrt{ 1 + \left( {{p_{\perp}}\over{m_{\alpha} c }} \right)^2 }}}
\Big| {{p_{\perp}}\over{\sqrt{1 - (\omega/(k_{\Vert}c))^2 }}}
+ m_{\alpha} c \sigma_{\alpha}
\Big|^2
\left( {{\partial \hat{f}_{\alpha}(p)
}\over{\partial p_{\Vert}}}\right)_{p_{\Vert}({\rm res})}
\label{damping_2}
\end{eqnarray}

\noindent
where ${\cal H}(x)$ is the Heaviside step function (1 for $x>0$,
and 0 otherwise), the derivative of the particle
distribution function should be evaluated at :
 
\begin{equation}
p_{\Vert}({\rm res})=
m_{\alpha} c \left( {{\omega_r}\over{k_{\Vert} c}} \right)
\Big(
{{1 + \left( {{p_{\perp}}\over{m_{\alpha} c}}
\right)^2 }\over{
1 - \left( {{\omega}\over{k_{\Vert} c}} \right)^2}}
\Big)^{1/2}
\label{pres}
\end{equation}
 
\noindent
and $\sigma_{\alpha}$ is given by :

\begin{equation}
\sigma_{\alpha}=
2 {\it i} \left( {{\omega}\over{k_{\Vert} c}} \right)
\left( {{\Omega_o}\over{k_{\perp} c}} \right)_{\alpha}
{{ E_{\Vert} / E_{\perp}}\over{ 1 - (\omega/(k_{\Vert} c))^2 }}
\sqrt{ 1 - ( {{m_{\alpha} c}\over{p_{\perp}}} )^2 }
\label{sigma}
\end{equation}

\noindent
Since the fluctuations of the electric field are essentially
perpendicular to $B_o$ (Appendix B), it is
$E_{\Vert} \rightarrow 0$ and $\sigma_{\alpha} \rightarrow 0$ and thus
from the Faraday low (Eq.~\ref{B}) the expression for the
damping rate gets simplified :

\begin{equation}
\Gamma_{\alpha}(k)=
- {{\pi^2}\over{8}}
{{| B_k |^2}\over{B_o^2}} {{\omega }\over{W}}
\left(
{{k_{\perp}}\over{k}}
\right)^2
{{ k_{\Vert} }\over{ |k_{\Vert}|}}
{{ {\cal H}\left( 1 - \left|{{ \omega}\over{
k_{\Vert} c}} \right| \right)
N_{\alpha}/m_{\alpha} }\over
{\sqrt{1-(\omega/(k_{\Vert}c))^2} }}
\int_0^{\infty}
dp_{\perp} {{ p_{\perp}^5 }\over{
\sqrt{ 1 + \left( {{p_{\perp}}\over{m_{\alpha} c }} \right)^2 }}}
\left( {{\partial \hat{f}_{\alpha}(p)
}\over{\partial p_{\Vert}}}\right)_{p_{\Vert}({\rm res})}
\label{damping_2_trans}
\end{equation}

\noindent
Thus from Eq.(\ref{damping_2_trans})
one obtains formulae appropriate for the case of the ICM:

\begin{equation}
\Gamma_{e/p}(k,\theta) =
\sqrt{ {{\pi}\over 8} }
{{ |B_k |^2}\over{W(k,\theta)}}
{\cal H}
\Big(1- {{V_{\rm ph}}\over{c}} {{k}\over{ |k_{\Vert}| }} \Big)
{{ V_{\rm ph}^2}\over{B_o^2}}
\left( {{k}\over{|k_{\Vert}|}} \right)
\left( {{k_{\perp}}\over{k}} \right)^2
{{ \left( m_{e/p} k_B T \right)^{1/2}}\over{
1 - ( {{V_{\rm ph} k}\over{c k_{\Vert} }} )^2 }} N_{e/p}
\exp \Big\{ - {{ m_{e/p} V_{\rm ph}^2 }\over
{2 k_B T}} {{ \left( {{k / k_{\Vert}}} \right)^2 }\over{
1 - ( {{V_{\rm ph} k}\over{c k_{\Vert} }} )^2 }}
\Big\}
k
\label{damping_th_LFM}
\end{equation}

\noindent
in the case of thermal particles, and

\begin{equation}
\Gamma_{e/p}(k,\theta)=
- {{\pi^2}\over{8}}
{{
| B_k |^2 }\over{W(k,\theta)}}
\left(
{{k_{\perp}}\over{k}}
\right)^2
\left(
{{k}\over{|k_{\Vert}|}}
\right)
{\cal H}
\Big(1- {{V_{\rm ph}}\over{c}} {{k}\over{|k_{\Vert}|}} \Big)
{{ N_{e^{\pm}/p} \,\, V_{\rm ph}^2}\over
{B_o^2}} \,\, k
\big(
1 - ( {{V_{\rm ph} k}\over{c k_{\Vert} }} )^2 \big)^2
\int_0^{\infty}
p^4 dp
\left( {{\partial \hat{f}_{\alpha}(p)
}\over{\partial p}}\right)_{e/p}
\label{damping_ur}
\end{equation}

\noindent
in the case of relativistic particles (see also BS73), 
where in obtaining Eq.(\ref{damping_ur}) from Eq.(\ref{damping_2_trans}),
one takes :

\begin{equation}
p_{\Vert}(res) \rightarrow \left( {{\omega}\over{k_{\Vert} c}} \right)
{{ p_{\perp} }\over{\sqrt{ 1 - ( {{\omega}\over{k_{\Vert} c}} )^2 }}}
\label{p_res_ur}
\end{equation}

\noindent
which implies $p_{\perp} = p \sqrt{ 1 - ( \omega / k_{\Vert} c )^2 }$,
and the derivative of the distribution function is taken :

\begin{equation}
\left( {{\partial \hat{f}_{\alpha}(p)
}\over{\partial p_{\Vert}}}\right)_{p_{\Vert}({\rm res})}
= \left( {{\partial \hat{f}_{\alpha}(p)
}\over{\partial p}}
{{ p_{\Vert}}\over{p}}
\right)_{p_{\Vert}({\rm res})}=
\left( {{\omega}\over{k_{\Vert} c}} \right)
{{\partial \hat{f}_{\alpha}(p)
}\over{\partial p}}
\label{df/dp_ur}
\end{equation}


\begin{thebibliography}{}
\bibitem{}  Achterberg A., 1981, A\&A 97, 259
\bibitem{}  Akhiezer A.I., Akhiezer I.A., Polovin R.V., Sitenko A.G.,
Stepanov K.N., {\it Plasma Electrodynamics} (Pergamon, Oxford, 1975)
\bibitem{}  Barnes A., 1968, Phys. of Fluids 11, 2644
\bibitem{}  Barnes A., Scargle J.D., 1973, ApJ 184, 251
\bibitem{}  Baldwin D.E., Bernstein I.B., Weenink M.P.H., Advances in Plasma Phys. 3, 1
\bibitem{}  Becker P.A., Le T., Dermer C.D., 2006, ApJ 647, 539
\bibitem{}  Berezinsky V.S., Blasi P., Ptuskin V.S., 1997, ApJ 487, 529
\bibitem{}  Bertoglio J.P., Bataille F., Marion J.D.: 2001, Phys. Fluids 13, 290
\bibitem{}  Blandford R., Eichler D., 1987, PhR 154, 1B
\bibitem{}  Blasi P., 2001, APh 15, 223
\bibitem{}  Blasi P.: 2004, JKAS 37, 483
\bibitem{}  Blasi P., Colafrancesco S., 1999, APh 12, 169
\bibitem{}  Blasi P., Gabici S., Brunetti G., 2007, Int. Journal of Mod.
Phys. A 22, 681; astro-ph/0701545
\bibitem{}  Braginskii S.I., 1965, Rev. Plasma Phys. 1, 205
\bibitem{}  Brunetti G., 2003, in {\it Matter and Energy in Clusters of 
Galaxies}, Vol. 301, eds. S.Bowyer \& C.-Y.Hwang, San
Francisco: Astronomical Society of the Pacific, p.349
\bibitem{}  Brunetti G., 2004, JKAS 37, 493
\bibitem{}  Brunetti G., 2006, Astron. Nachr. 327, 615
\bibitem{}  Brunetti G., Setti G., Feretti L., Giovannini G., 2001, MNRAS 320, 365
\bibitem{}  Brunetti G., Blasi P., Cassano R., Gabici S., 2004, MNRAS 350, 1174
\bibitem{}  Brunetti G., Blasi P.: 2005, MNRAS 363, 1173
\bibitem{}  Buote D.A., 2001, ApJ 553, L15
\bibitem{}  Cassano R., Brunetti G., 2005, MNRAS 357, 1313
\bibitem{}  Chandran B.D.G., 2000, Phys. Rev. Lett., 85, 4656
\bibitem{}  Chandran B.D.G., 2003, ApJ 599, 1426
\bibitem{}  Chandran B.D.G., Maron J.L., 2004, ApJ 603, 23
\bibitem{}  Cho J., Lazarian A., 2002, ApJ 575, L63
\bibitem{}  Cho J., Lazarian A., Vishniac E.T., 2002, ApJ 564, 291
\bibitem{}  Cho J., Lazarian A., 2003, MNRAS 345, 325
\bibitem{}  Cho J., Lazarian A., 2004, ApJ 615, L41
\bibitem{}  Cho J., Lazarian A., Honein A., Knaepen B., Kassinos S., 
Moin P., 2003, ApJ 589, L77
\bibitem{}  Cho J., Lazarian A., 2006, ApJ 638, 811
\bibitem{}  Churazov E., Forman W., Jones C., Sunyaev R., B\"ohringer H., 2004, MNRAS 347, 29
\bibitem{}  Crawford C.S., Hatch N.A., Fabian A.C., Sanders J.S., 2005,
MNRAS 363, 216
\bibitem{}  De Gouveia dal Pino E.M., Lazarian A., 2005, A\&A 441, 845
\bibitem{}  Denisse J.F., Delcroix J.L., {\it Plasma Waves} (Interscience,
New York, 1963).
\bibitem{}  Dennison B.: 1980, ApJ 239, L93
\bibitem{}  Dogiel V.A., Colafrancesco S., Ko C.M., Kuo P.H., 
Hwang C.Y., Ip W.H., Birkinshaw M., Prokhorov D.A., 2007, A\&A 461, 433
\bibitem{}  Dolag K., Bartelmann M., Lesch H., 2002, A\&A 387, 383
\bibitem{}  Dolag K., Vazza F., Brunetti G., Tormen G., 2005, MNRAS 364, 753
\bibitem{}  Drury L., 1983, SSRv 36, 57
\bibitem{}  Eilek J.A., 1979, ApJ 230, 373
\bibitem{}  Eilek J.A., Henriksen R.N., 1984, ApJ 277, 820
\bibitem{}  Ensslin T.A., Biermann P.L., Kronberg P.P., Wu X.-P., 1997, ApJ 477, 560
\bibitem{}  Ensslin T.A., Wang Y., Nath B.B., Biermann P.L., 1998, A\&A 333, L47
\bibitem{}  Ensslin T.A., Vogt C., Pfrommer C., 2005, in {\it The Magnetized
Plasma in Galaxy Evolution}, eds. K.T. Chyzy, K. Otminowska--Mazur,
M. Soida, R.-J. Derrmar, Jagielonian University, Kracow, p. 231.
\bibitem{}  Fabian A.C., Sanders J.S., Crawford C.S., Conselice C.J.,
Gallagher J.S., Wyse R.F.G., 2003, MNRAS 344, L48
\bibitem{}  Farmer A.J., Goldreich P., 2004, ApJ 604, 671
\bibitem{}  Feretti L., 2005, Adv. Space Res. 36, 729 
\bibitem{}  Fermi E., 1949, Phys. Rev. 75, 1169 
\bibitem{}  Fisk L.A., 1976, Journal of Geophys. Research 81, 4633
\bibitem{}  Foote E.A., Kulsrud R.M., 1979, ApJ 233, 302
\bibitem{}  Fried B.D., Conte S.D., {\it The Plasma Dispersion Function}
(Academic, New York, 1961).
\bibitem{}  Fujita Y., Sarazin C.L., 2001, ApJ 563, 660
\bibitem{}  Fujita Y., Takizawa M., Sarazin C.L., 2003, ApJ 584, 190
\bibitem{}  Fusco-Femiano R., Orlandini M., Brunetti G., Feretti L.,
Giovannini G., Grandi P., Setti G., 2004, ApJ 602, L73
\bibitem{}  Gabici S., Blasi P., 2003, ApJ 583, 695
\bibitem{}  Gary S.P., McKean M.E., Winske D., Anderson B.J., Denton R.E.,
Fuselier S.A., 1994, Journal of Geoph. Research 99, 5903
\bibitem{}  Gastaldello F., Molendi S., 2004, ApJ 600, 670
\bibitem{}  Ginzburg V.L., {\it Propagation of Electromagnetic Waves in
Plasma} (New York: Gordon \& Breach, 1961).
\bibitem{}  Goldreich P., Sridhar S.: 1995, ApJ 438, 763
\bibitem{}  Govoni F., Feretti L., 2004, Int. Journal of Mod Phys. 13, 1549
\bibitem{}  Higdon J.C., 1984, ApJ 285, 109
\bibitem{}  Hwang C-Y., 2004, JKAS 37, 461
\bibitem{}  Jaffe W.J., 1977, ApJ 212, 1
\bibitem{}  Jokipii J.R., 1966, ApJ 146, 480
\bibitem{}  Jubelgas M., et al., 2006, MNRAS in press.
\bibitem{}  Kang H., Jones T.W., 1995, ApJ 447, 944
\bibitem{}  Kato S., 1968, PASJ 20, 59
\bibitem{}  Kowal G., Lazarian A., 2006, astro-ph/0611396
\bibitem{}  Krall N.A., Trivelpiece A.N., {\it Principles of Plasma
Physics} (McGraw--Hill, New York, 1973).
\bibitem{}  Lazarian A., 2006a, Astron. Nachr. 327, 609
\bibitem{}  Lazarian A., 2006b, ApJ 645, L25
\bibitem{}  Lazarian A., Vishniac E.T., 1999, ApJ 517, 700
\bibitem{}  Lazarian A., Beresnyak A., 2006, MNRAS, 373, 1195
\bibitem{}  Lithwick Y., Goldreich P., 2001, ApJ 562, 279
\bibitem{}  Liu C.S., Mok Y., 1977, Phys. Rev. Letters 38, 162
\bibitem{}  Melrose D.B., 1968, ApSS 2, 171
\bibitem{}  Miller J.A., 1991, ApJ 376, 342
\bibitem{}  Miller J.A., La Rosa T.N., Moore R.L., 1996, ApJ 461, 445
\bibitem{}  Miniati F., Jones T.W., Kang H., Ryu D., 2001, ApJ 562, 233
\bibitem{}  Moghaddam-Taaheri E., Vlahos L., Rowland H.L., Papadopoulos K.,
1985, Phys. of Fluids 28, 3356
\bibitem{}  Montgomery D., Brown M.R., Matthaeus W.H., 1987, Journal of Geophys. Research 92, 282
\bibitem{}  Narayan R., Medvedev M.V., 2001, ApJ 562, L129
\bibitem{}  Paesold G., Benz A.O., 1999, A\&A 351, 741
\bibitem{}  Percival D.J., Robinson P.A., 1998, Journal of Math. Phys. 39, 3678
\bibitem{}  Petrosian V., 2001, ApJ 557, 560
\bibitem{}  Petrosian V., 2003, in {\it Matter and Energy in Clusters of
Galaxies}, Vol. 301, eds. S.Bowyer \& C.-Y.Hwang, San
Francisco: Astronomical Society of the Pacific, p.337
\bibitem{}  Petrosian V., Liu S., 2004, ApJ 610, 550
\bibitem{}  Petrosian V., Yan H., Lazarian A., 2006, ApJ 644, 603
\bibitem{}  Pfrommer C., Ensslin T.A., 2004, MNRAS 352, 76
\bibitem{}  Pfrommer C., Springel V., Ensslin T.A., Jubelgas M., 2006, MNRAS 367, 113
\bibitem{}  Pilipp W., V\"olk H.J., 1971, Journal of Plasma Phys. 6, 1
\bibitem{}  Ptuskin V.S., 1988, Sov. Astr. Lett. 14, 255
\bibitem{}  Reimer A., Reimer O., Schlickeiser R., Iyudin A., 2004, A\&A 424, 773
\bibitem{}  Reimer O., 2004, JKAS 37, 307
\bibitem{}  Rephaeli Y., Gruber D., Arieli Y., 2006, ApJ 649, 673
\bibitem{}  Ricker P.M., Sarazin C.L., 2001, ApJ 561, 621
\bibitem{}  Roettiger K., Burns J.O., Loken C., 1996, ApJ 473, 651
\bibitem{}  Roettiger K., Loken C., Burns J.O., 1997, ApJS 109, 307 
\bibitem{}  Ryu D., Kang H., Hallman E., Jones T.W., 2003, ApJ 593, 599
\bibitem{}  Sarazin C.L., 1999, ApJ 520, 529
\bibitem{}  Schlickeiser R., 2002, {\it Cosmic Ray Astrophysics}, Springer-Verlag, Berlin Heidelberg
\bibitem{}  Schlickeiser R., Miller J.A., 1998, ApJ 492, 352
\bibitem{}  Schlickeiser R., Sievers A., Thiemann H.: 1987, A\&A 182, 21
\bibitem{}  Schuecker P., Finoguenov A., Miniati F., Boehringer H., Briel U.G., 2004, A\&A 426, 387
\bibitem{}  Schekochihin A.A., Cowley S.C., Kulsrud R.M., Hammett G.W.,
Sharma P., 2005, ApJ 629, 139
\bibitem{}  Schekochihin A.A., Cowley S.C., 2006, PhPl 13, 56501
\bibitem{}  Shafranov V.D., {\it Reviews of Plasma Physics Vol.~3}
(New York: Consultants Bureau, 1967).
\bibitem{}  Shebalin J.V., Matthaeus W.H., Montgomery D., 1983, Journal of Plasma Phys. 29, 525
\bibitem{}  Shebalin J.V., Montgomery D., 1988, Journal of Plasma Phys. 39, 339
\bibitem{}  Simon A., 1955, Phys. Rev. 100, 1557
\bibitem{}  Subramanian K., Shukurov A., Haugen N.E.L.: 2006, MNRAS 366, 1437
\bibitem{}  Sun M., Jones C., Forman W., Nulsen P.E.J., Donahue M., Voit
G.M., 2006, ApJ 637, L81
\bibitem{}  Sunyaev R.A., Norman M.L., Bryan G.L., 2003, Astron. Lett. 29, 783
\bibitem{}  Tsytovich V.N., {\it An Introduction to the Theory of Plasma
Turbulence} (Pergamon Press, Oxford, 1972).
\bibitem{}  Tsytovich V.N., {\it Theory of Turbulent Plasma} (Consultant
Bureau, New York, 1977). 
\bibitem{}  Yan H., Lazarian A., 2002, Phys. Rev. Lett. 89, 1102
\bibitem{}  Yan H., Lazarian A., 2004, ApJ 614, 757
\bibitem{}  Vazza F., Tormen G., Cassano R., Brunetti G., Dolag K., 2006, MNRAS 369, L14
\bibitem{}  Vogt C., Ensslin T.A., 2005, A\&A 434, 67
\bibitem{}  Voigt L.M., Fabian A.C., 2004, MNRAS 347, 1130
\bibitem{}  V\"{o}lk H.J., Aharonian F.A., Breitschwerdt D., 1996, SSRv 75, 279
\bibitem{}  V\"olk H.J., Atoyan A.M., 1999, APh 11, 73
\bibitem{}  Zank G.P., Matthaeus W.H., 1992, Journal of Geoph. Research 97, 189
\bibitem{}  Zank G.P., Matthaeus W.H., 1993, Physics of Fluids A 5, 257
\bibitem{}  Zhou Y., Matthaeus W.H., 1990, Journal of Geoph. Research 95,
14881
\end{thebibliography}
\end{document}